\documentclass[11pt,final,a4paper,oneside,twocolumn,nofonttune]{IEEEtran}
\usepackage{moreverb,url}
\usepackage{natbib}
\usepackage[colorlinks,bookmarksopen,bookmarksnumbered,citecolor=red,urlcolor=red]{hyperref}

\usepackage{orcidlink}
\usepackage{amsmath}
\usepackage[font=small,labelfont=bf]{caption}
\usepackage{subcaption}
\usepackage{tabularray}
\usepackage{tikz}
\usetikzlibrary{positioning}

\definecolor{resampling}{rgb}{0.12156862745098039, 0.4666666666666667, 0.7058823529411765}
\definecolor{replication}{rgb}{1.0, 0.4980392156862745, 0.054901960784313725}
\definecolor{extent}{rgb}{0.17254901960784313, 0.6274509803921569, 0.17254901960784313}

\graphicspath{{./images/}}

\title{Evaluating and Improving Weak Scalability Analysis of Visualization Algorithms}

\author{Marvin Petersen\,\orcidlink{0000-0003-2324-9661}, Jonas Lukasczyk\,\orcidlink{0000-0001-6650-770X}, and Christoph Garth\,\orcidlink{0000-0003-1669-8549}%
\thanks{Marvin Petersen, Jonas Lukasczyk and Christoph Garth are with the University of Kaiserslautern-Landau (RPTU)}}

\begin{document}
\widowpenalty=10000
\clubpenalty=10000
\brokenpenalty=10000

\maketitle
    
\begin{abstract}
    Research on visualizing large-scale datasets traditionally relies on empirical evaluation of scalability, determining the effectiveness of specific computation methods, algorithmic strategies, or implementations. 
    Weak scalability, which assesses the algorithm's performance as problem size and computing resources increase, is a valuable indicator for a method's applicability at scale. 
    However, sufficiently large data sets with increasing size are needed for weak scalability studies. To this end, it is customary to use simple scaling techniques to increase problem size by generating larger input data sets from a base data set.
    Nevertheless, many visualization algorithms' workload depends on factors beyond input size, such as input data complexity or output size, leading to inaccuracies in the attributed weak scalability. 
    In this work, we highlight different common data scaling methods on multiple algorithms and data sets, recognizing that the suitability of scaling approaches varies across algorithms and data sets. We present a method that effectively mitigates the observed inconsistencies in workload increases for the different scaling methods in a shared-memory setting. With this work, we aim to further the discussion on how to evaluate and report scalable visualization research.
\end{abstract}

\begin{IEEEkeywords}
weak-scalability, shared-memory parallelism, scientific visualization, empirical performance analysis, data scaling methods, workload estimation
\end{IEEEkeywords}    

\section{Introduction}

    Scalability studies are used to investigate the overhead of parallel applications and their ability to handle larger data sets and use additional computing resources. In visualization, the parallel scalability of visualization algorithms is of special interest. Decreasing time-to-solution via parallelism to guarantee interactivity and thus allowing for better data exploration can be beneficial for analysis tasks. 
    Depending on scenario, visualization algorithms must adapt to a prescribed compute budget efficiently.

    Due to growing simulation data and the need for visualization in the analysis, ensuring a computation method's fitness for future data sets becomes crucial. To this end, weak scalability benchmarks are an important measuring tool for present and future visualization algorithms, since they describe the performance development of algorithms and methods on growing data sets with aligned increases in computing resources. Basically, the assumption is that a weakly scalable algorithm can be used on problems of growing size if sufficient resources are available to it. Due to the high complexity of many visualization algorithms and their data dependent runtime behavior, empirical studies are used to demonstrate weak scalability on practically relevant problem classes. 
    
    However, the data sets used to study algorithm runtimes are not typically available at multiple sizes, and thus differently-sized data sets have to be generated for an effective weak scalability demonstration. While datasets can be scaled in many different ways, it is customary to increase the size of an algorithm's input data set by the same factor as the amount of available computational resources. Since the execution time of some visualization algorithms depend on more parameters than raw input size, this practice is not always sufficient to consistently and accurately represent the algorithm's scalability. Therefore, in this paper, we investigate the impact of different data scaling methods on weak scalability studies in visualization. Further, we present an approach that can provide more accurate insight into weak scalability than is afforded by data size scaling alone. To this end, we perform experiments with several scaling techniques that suggest that in addition to data size, algorithmic complexity has to be considered as well, necessitating a scaling approach that is tailored to the visualization algorithm under examination. 

    Our work is intended to further the discussion on how contributions in scalable visualization research are evaluated. 
    We wish to bring to attention nuances of scalability that are not always considered in adequate manner when evaluating visualization algorithms. 
    The work we present here is not aimed at providing a broad approach that can be used to tease out the details of scalability for a majority of visualization algorithms. 
    To simplify our arguments, we mainly consider uniformly gridded data on shared-memory systems, and speculate on how our insights generalize.

    This paper is structured as follows. First, we provide some background information on parallel scalability and data scaling methods and provide exemplary evidence that for specific combinations of algorithms and problem scaling methods, the workload per core does not stay constant. Afterwards, we discuss the implications of these findings. In \autoref{sec:workscaling}, we describe a method that mitigates this problem and investigate its performance on several visualization algorithms in \autoref{sec:experiments}. 
    We discuss the results of our method, its limitations and implications for weak scalability benchmarks in visualization in general in \autoref{sec:discussion}.

\section{Background}\label{sec:background}

    The foundation for scalability analyses is formed by the laws of \citet{amdahl1967validity} and \cite{gustafson1988reevaluating}, describing upper bounds on the theoretical possible speed up by parallelism for programs with a parallelizable fraction. The main difference between the two laws lies in the examination of fixed and scaled problem sizes in relation to the amount of computational resources. Due to the dependence of the parallel fraction on the considered problem size, as described by Gustafson, a less strictly bound \emph{scaled speedup} can be described, as opposed to Amdahl's law. Since we will cover some implications of Gustafson's law in more detail, we provide a short summary.

    \vspace{0.5\baselineskip}
    \begin{quote}        
        If $N$ is the number of cores and, respectively, $s$ and $p$ the serial and parallel part of the execution time on $N$ cores, then the \emph{scaled speedup} can be written as:
        \begin{align}
            \text{scaled speedup} &= \frac{s+Np}{s+p} \label{q:gustafson}
        \end{align}
    \end{quote}
    This formula is based upon the assumption that only the parallel part of a program scales with problem size. It is further assumed that, under perfect scaling, the scaled workload for $N$ cores does therefore take $(s+Np)$ time on a single core if it takes $(p+s)$ time on $N$ cores. This last assumption is a central part of the presented work here.

\section{Related Work}\label{sec:relatedwork}
    The laws of Amdahl and Gustafson have been the topic of several scientific discussions, leading to more refined models that include a multitude of different factors like I/O, available memory and communication overheads \citep{sun1993,alhayanni2020}. Although more complex models are available, this work focuses on Gustafson's law, for two reasons. Firstly, Gustafson's work is a common reference and foundation for many weak scalability studies in visualization algorithm engineering \citep{zhang2003, richer2022}. Secondly, the problem discussed here is a direct consequence of Gustafson's law, since it requires scaled problem sizes. 

    It is common practice to scale the input size for weak scalability studies in visualization. 
    \cite{zhang2003} and \cite{petersen2022prototype} studied weak scalability of contour extraction with accordingly scaled up data sets. 
    Similarly, \cite{werner2021} and \citet{carr2022} investigate the scalability of contour tree computation. 
    For volume rendering, weak scalability on input size scaled data sets was investigated by \cite{biedert2017task,petersen2022prototype} and \cite{shih2016}.
    \cite{howison2011hybrid} also investigated volume rendering, but reported weak scalability both with respect to input data size, upscaled with trilinear interpolation, and with respect to output image size.
    Upscaling via interpolation was compared to replication of a dataset in the context of contour extraction scalability by \cite{childs2010extreme}.
    
    Scaling behavior is investigated with potentially many different target measures, execution time being only one of them. Also, energy consumption and CPU usage efficiency receives growing attention \citep{cassidy2012}. Another alternative for execution time is discussed by \cite{moreland2015formal}. They suggest to use \emph{rate} and \emph{efficiency} in place of execution time for scalability studies and propose best practices for reporting algorithm scaling behavior. The results presented here also apply to their findings if the \emph{rate} represents the input size. However, finding a \emph{rate} that mirrors the development of execution time best, is closely related to the scaling of the problem size and the included work. This is investigated here.

    The observation that parallel scalability also depends on the algorithmic complexity with respect to the problem size was described by \cite{snyder1986modest}, which plays an important role in keeping the work per core constant.
    Scaling problem size in a meaningful way was already investigated by \cite{grama1993}. They provide a definition of the problem size that does not derive directly from the input size, as they define the problem size as the total number of basic operations for this problem. They also describe the inconsistency arising from a parameter describing the input size and not the actual work, since it is also application dependent. However, their goal is to scale the problem size such that a fixed efficiency is achieved. This work, in contrast, uses proportional problem size scaling to observe the actual efficiency. Furthermore, their work did not explicitly consider problem size to be data dependent.
    
    In this work, only problems generated via input size scaling for uniform rectilinear grids are considered. However, there are algorithms where alternative parameters are investigated. This is often the case with integral curves where the number of seed points is increased proportionally \citep{zhang2018}. Another example is the scaling of image resolution and thus number of pixel for ray casting methods \citep{howison2011hybrid,wu2018}.
    
    The state of scalability studies in visualization in a broader sense was investigated by \cite{richer2022}. The scalability discussed here fits their description of parallel computing scalability, which focuses on strong and weak scalability based on the laws by Amdahl and Gustafson.

\section{Scaling Input Data Size Only}\label{sec:inputscaling}

    Weak scaling experiments are often conducted using proportionally scaled data with respect to the amount of available parallel computing resources. For structured grids this generally means that the number of cells are adjusted. 
    Importantly, for algorithms that are output sensitive or otherwise data dependent apart from the input size, this scaling does not necessarily scale the problem size or \textit{work} of the data for that algorithm.

    As described, this proportionality of the problem size to the resources is a foundation of Gustafson's law that defines the theoretically possible parallel speedup. 
    Nonetheless, the actual execution time on single core machines is rarely checked. Since the optimal \emph{scaled speedup} assumes perfect scaling, the single core time will likely be lower than the described $(s+Np)$. However, depending on the scaled data set, this effect can be counteracted by potentially overestimating the work in the scaled data sets, leading to lower parallel execution time $(s+p)$ and thus a lower hypothetical execution time on single cores. This can let the weak scalability appear better than it is. This means that for data dependent algorithms weak scaling experiment results could be skewed if no care is taken.
    
\subsection{Data Scaling Methods}\label{sec:scalingmethods}

    \begin{figure}
        \centering
        \includegraphics[width=0.8\linewidth]{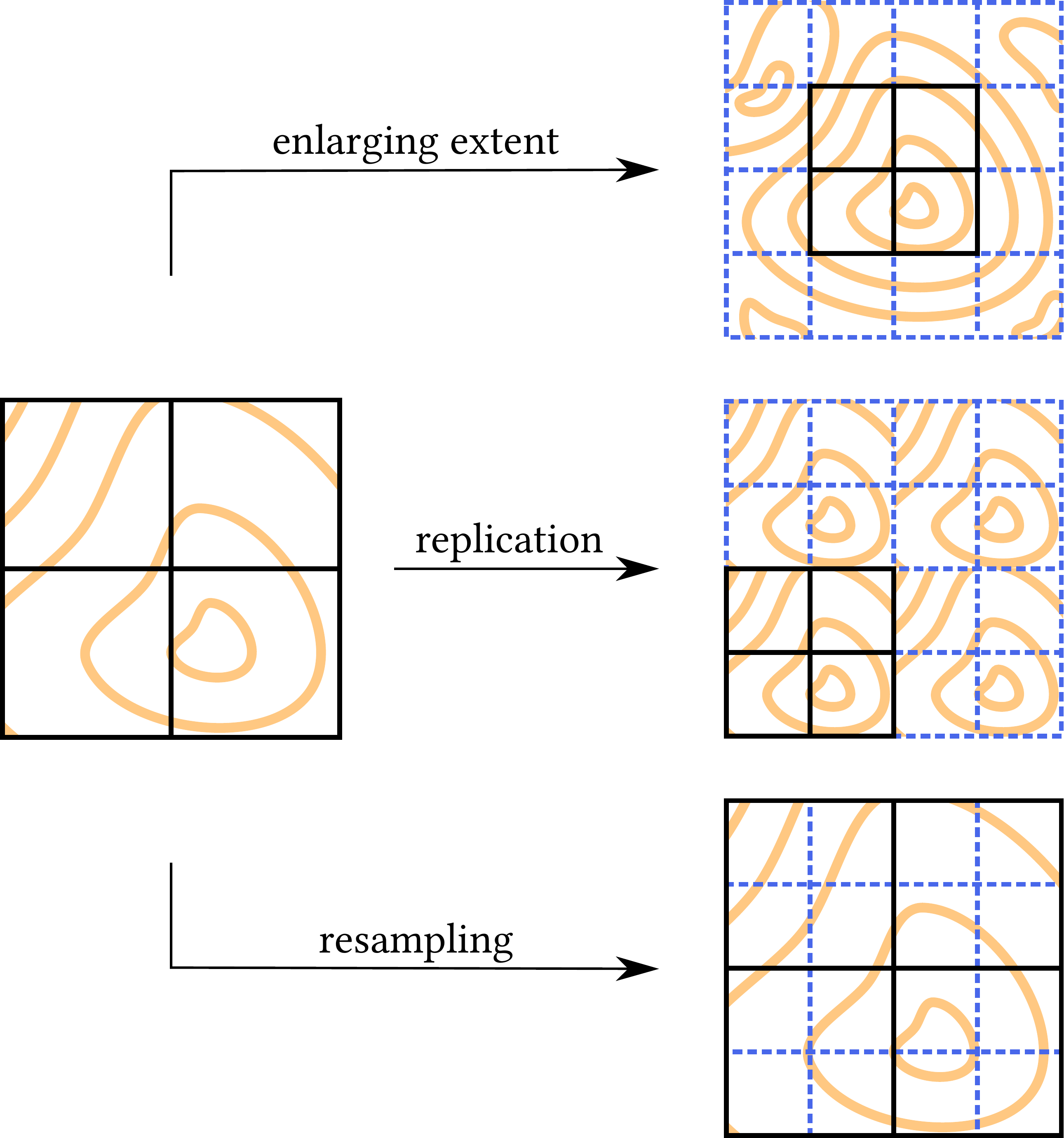}
        \caption{The three different scaling methods described in \autoref{sec:scalingmethods} on a two-dimensional example. New cells are shown in blue dotted lines. Note the additional patterns added with \emph{enlarging extent}. The scaling factor for this example is $4$.}
        \label{fig:scaling-methods}
    \end{figure}

    \begin{figure}
        \centering
        \includegraphics[width=\columnwidth]{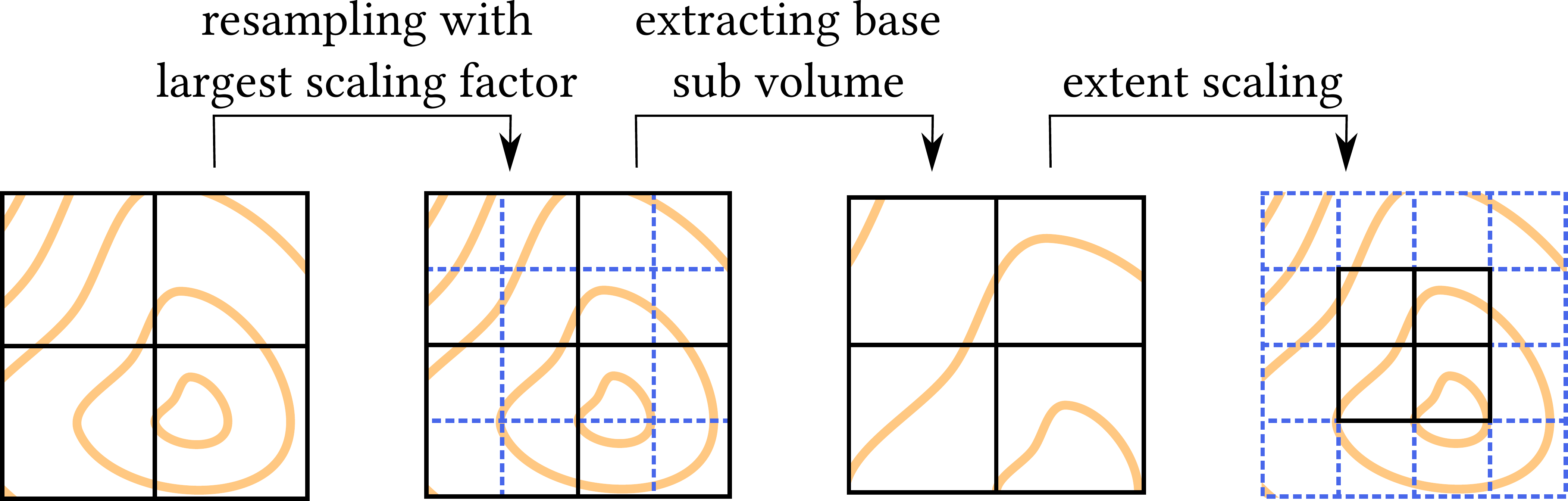}
        \caption{Process of generating extent scaled data sets for small base data sets where no high resolution data is available, as used in the experiments.}
        \label{fig:extent-scaling}
    \end{figure}

    \begin{figure*}
        \centering
        \includegraphics[width=0.33\textwidth]{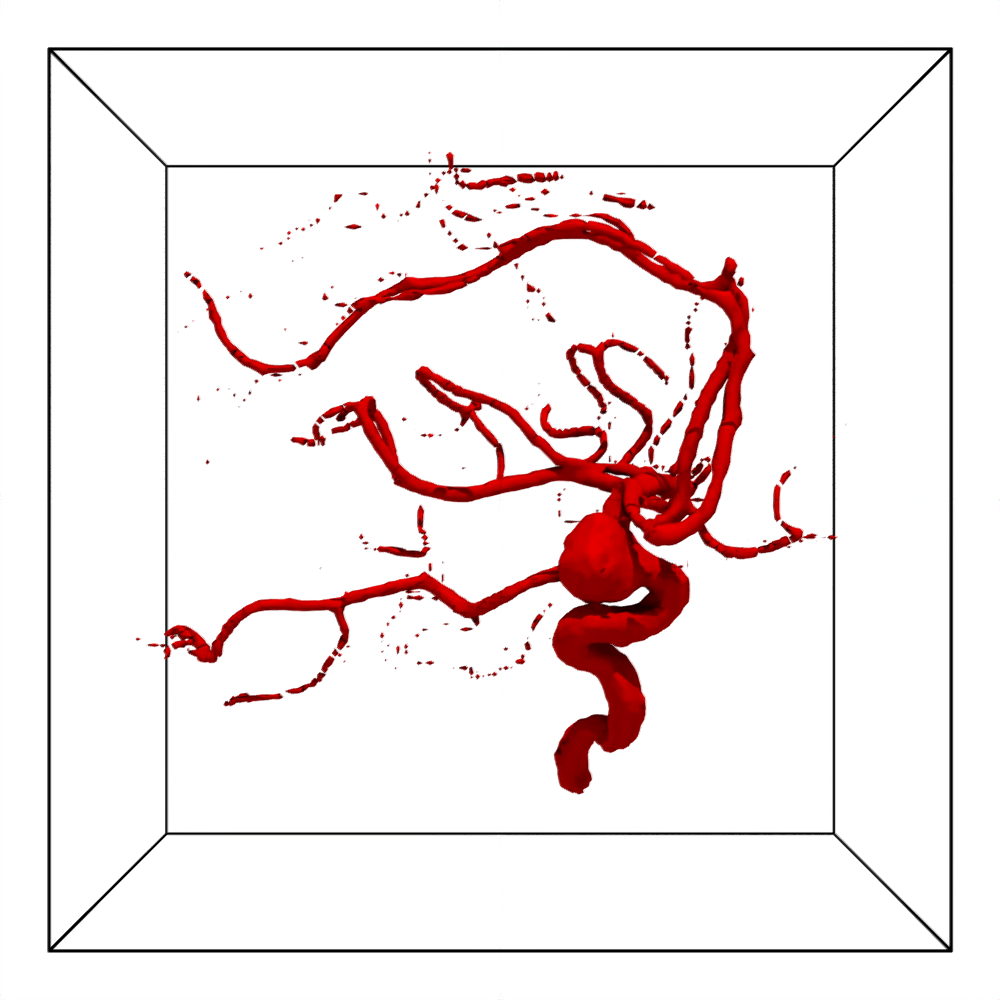}\hfill
        \includegraphics[width=0.33\textwidth]{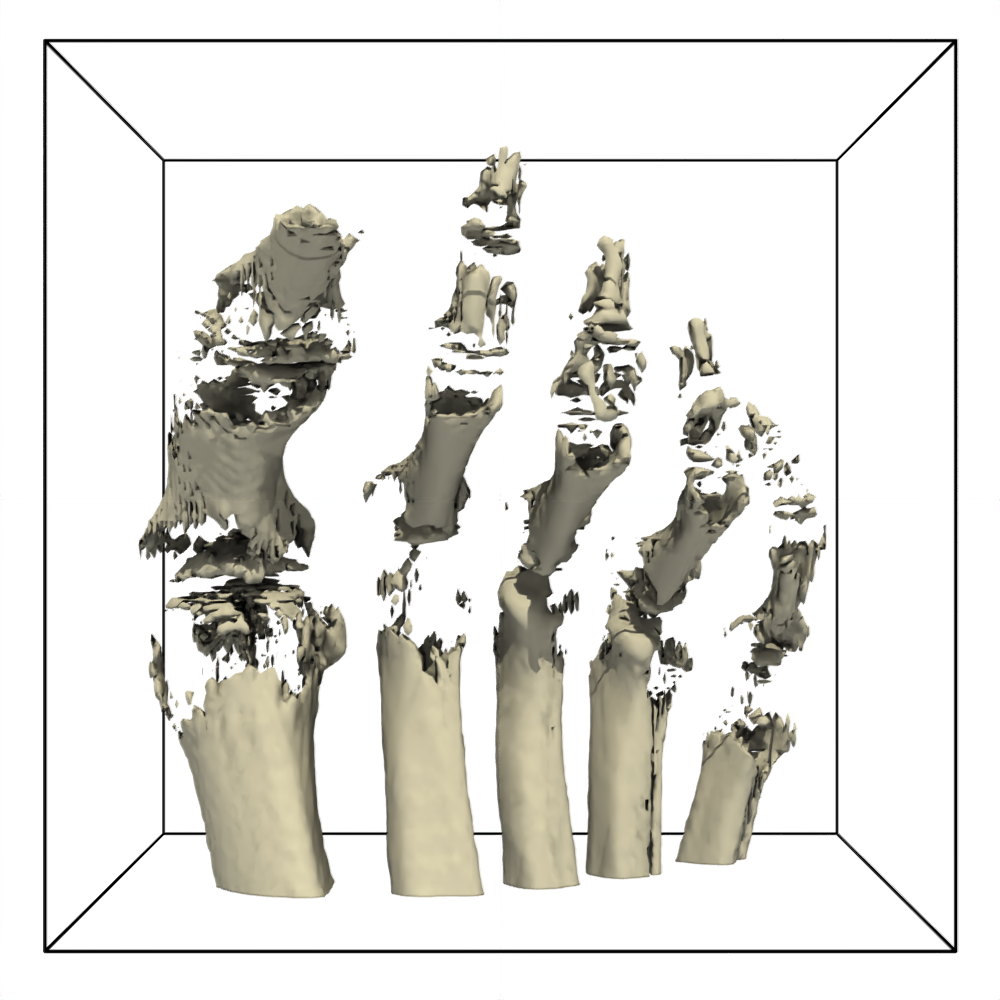}\hfill
        \includegraphics[width=0.33\textwidth]{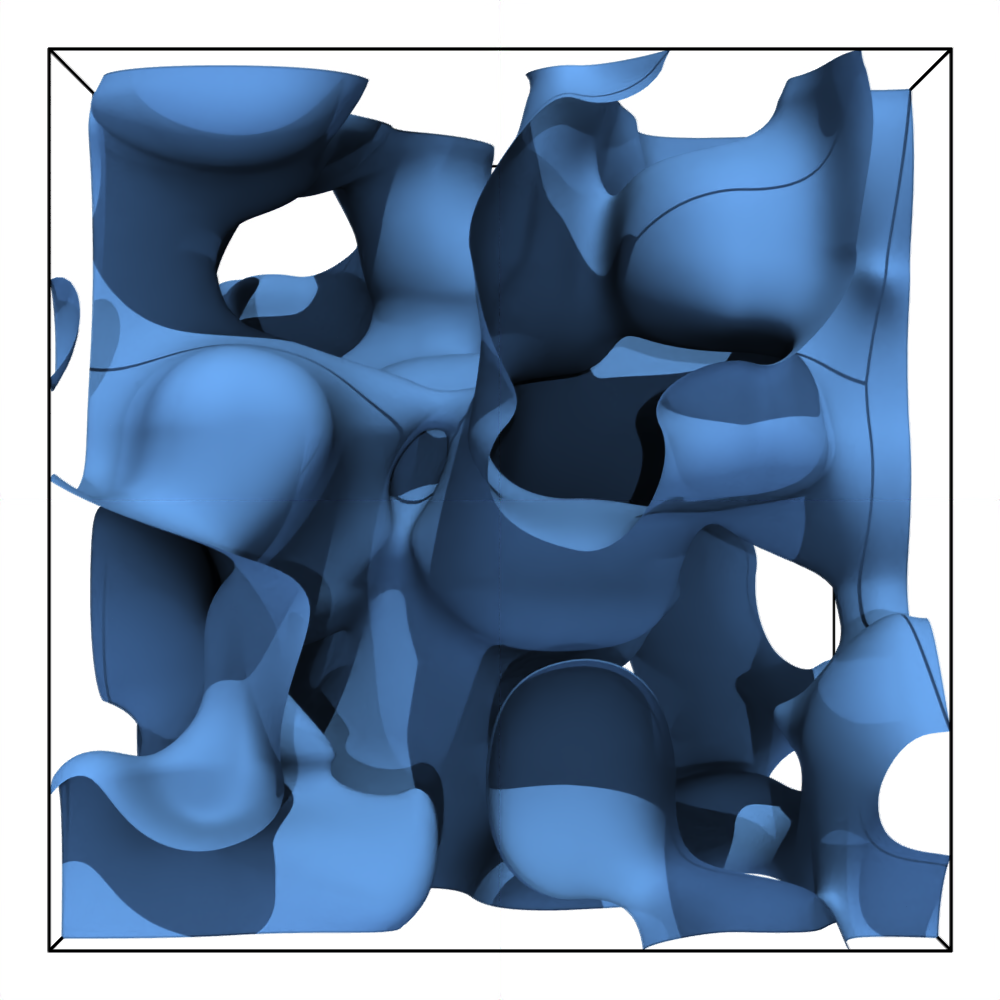}
        \caption{Iso-contour renderings of the data sets used in this work. From left to right: aneurysm, foot, Perlin noise.}
        \label{fig:datasets}
    \end{figure*}

    We will shortly introduce and discuss the most prevalent data scaling methods for uniform rectilinear grids on which the examples here and the method presented in \autoref{sec:workscaling} is based upon. A two dimensional example for each scaling method discussed here is shown in \autoref{fig:scaling-methods}. In our case, the scaling factor $f$ with which a data set is scaled, describes the factor that is multiplied with the data set's size. For example, a data set that should be scaled with a scaling factor of $f=2$, would result in a data set with twice the original size.    
    Among other, more problem or domain specific procedures, the following methods are commonly used:
\paragraph*{Resampling} 
    Resampling is done by sampling in a different resolution from the original in the bounds of the original data, depending on the scaling factor. For values between original data points interpolation is used. In this work, we restrict ourselves to linear interpolation. There are also more sophisticated resampling methods that preserve certain characteristics of the data, which are not considered here.
\paragraph*{Enlarging Extent} 
    Size scaling can also be done by enlarging the extent and including additional data in the newly added cells. This is only possible if there is a larger original data set to get this additional data from or the data can be procedurally generated, e.g., Perlin noise. 
    For the experiments in this paper, we first create larger base data sets for usage in extent scaling via resampling, since the base data sets are not available at higher resolution.
    From this up sampled base data set we then extract subvolumes according to the scaling factor. A scaling factor of $f=1$ would lead to the same number of cells as in the original data set. Furthermore, the subvolumes are extracted around the center of the up sampled original. This process is exemplified in \autoref{fig:extent-scaling}.
\paragraph*{Replication}
    Replication, similarly to extent scaling, increases the extent, but fills it with periodic repetitions of the original data. If the scaling factor does not yield a multiple of the original data, only parts are appended. In this work we use two replication modes. The first replicates the data only along one axis, meaning with a scaling factor of $2$ the output would be two times the whole data set. The second mode replicates the data in all three dimensions. This means that the resulting data will contain partial replications in each direction if the scaling factor's third root is not an integer. We use the second mode only for the ray tracing example in order to keep the same viewing frustum and an equal amount of image space occupied by data.

    These problem size scaling methods have different influences on the data itself and thus also the algorithms that are investigated on them. For example, \emph{resampling} with linear interpolation does not introduce new local maxima or minima, while growing the \emph{extent} and \emph{replicating} the data can. In the following, we exemplify the influence of different problem size scaling methods in three different scenarios on multiple data sets.

\subsection{Examples}\label{sec:examples}

    Execution times for weak scalability with scaling factor equal to core count and the corresponding measured execution time on a single core are shown here for different visualization algorithms on three different data sets each scaled with the three previously mentioned scaling methods. Each data point shows the mean over 5 executions on a shared-memory parallel system and a confidence interval of $0.95$, with the exception of the reported execution times for volume rendering where only the median of $12$ evenly spaced camera angles is reported. All timings were gathered on a machine with 512 GB of memory and two AMD EPYC 7453 with 28 physical cores each. SMT was disabled for all measurements.

\paragraph*{Data Sets}

    The considered data sets are a scan of a foot \citep{openscivisdata}, an aneurysm data set \citep{openscivisdata}, and some Perlin noise \citep{perlin2002} with a frequency of $0.02$. All base data sets are of size $256^3$ for the iso-contour example and $128^3$ for the augmented contour tree and the volume rendering. We choose these data sets as their features range from localized to evenly distributed inside the domain. An iso-contour rendering for each of the three data sets is shown in \autoref{fig:datasets} with an iso-value in the middle of the scalar range. In the examples, these data sets represent the baseline and are the input of the algorithm on one core, while up-scaled versions are used for executions on multiple cores. The data sets are scaled such that the number of input cells increases by the same factor as the number of cores.

\paragraph*{Iso-Contour Extraction}

\begin{figure}
        \centering
        \includegraphics[width=.95\columnwidth]{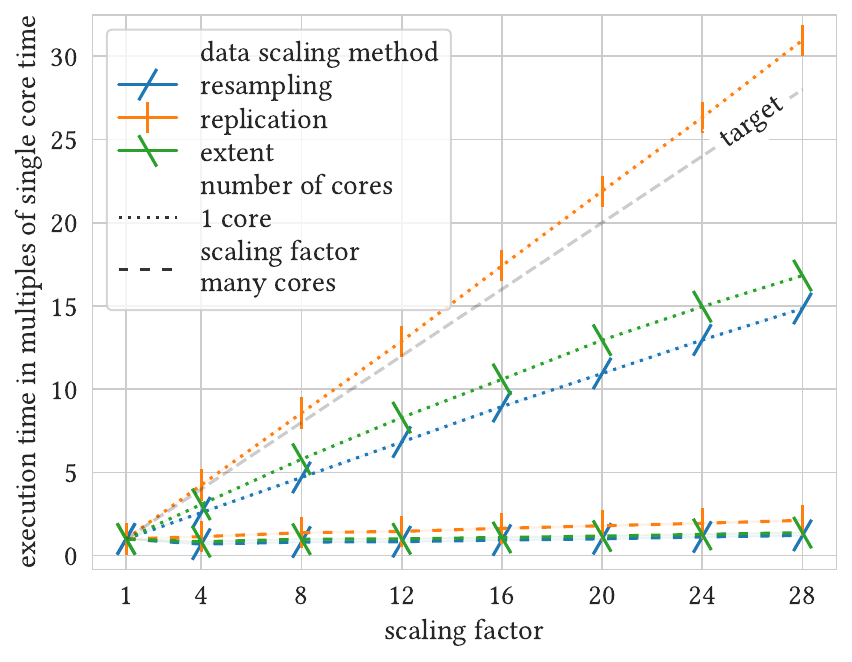}
        \caption{Weak scaling timings of contour extraction with the flying edge algorithm on the foot data set. The dotted lines show the corresponding execution timings with the same data on a single core. It shows a discrepancy in the scaling of the workload from one to many cores for the \emph{resampling} and \emph{extent} scaling techniques.}
        \label{fig:ic_input_scaling}
    \end{figure}

    For this example, the flying edge implementation by \cite{schroeder2015} in VTK \citep{vtk} was used and applied to the foot data set. 
    Execution time measurement for one and $N$ cores can be seen in \autoref{fig:ic_input_scaling}. We are able to observe that \emph{replication} scales the work on a single core almost perfectly on the foot data set with respect to the measured execution time on a single core. However, \emph{resampling} and \emph{extent} scaling introduce substantially less work. All scaling approaches seem to result in good weak scalability considering the execution time on \emph{scaling factor} many cores.

    The size of an iso-contour extraction's output depends largely on the selected iso-value and its occurrence in the data set. Generating the geometry for the respective cells requires additional work that is reflected in the execution time. Depending on the implementation, preprocessing steps and optimizations, determining the work for a specific input becomes even more involved. To this end, it is not necessarily the case that the number of output cells, and thus work overall, scales proportionally to the number of input cells depending on the scaling method.

\paragraph*{Direct Volume Rendering via Ray Casting}

    \begin{figure}
        \centering
        \includegraphics[width=.95\columnwidth]{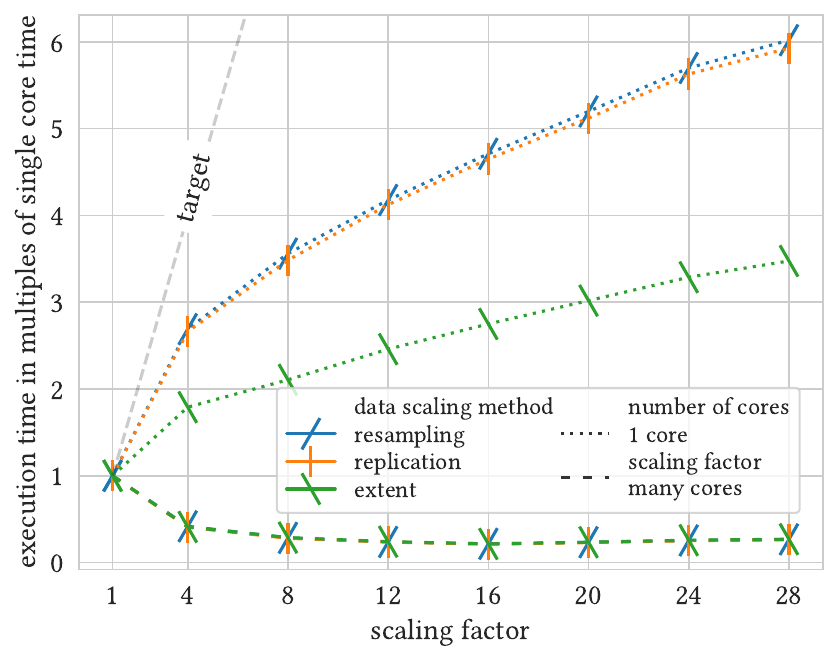}
        \caption{Weak scaling timings of volume rendering via ray casting on the foot data set. The dotted lines show the corresponding execution timings with the same data on a single core. Here it can be seen that all scaling methods do not increase the workload by the scaling factor from one to many cores, \emph{extent} scaling increasing least.}
        \label{fig:vr_input_scaling}
    \end{figure}

    For this example we use VTK's OpenGLGPUVolumeRayCastMapper with OSMesa on the foot data set.
    One can observe that the execution time on the scaling number of processors improves at first and then keeps nearly constant, seeming to exhibit good weak scalability. However, looking at the single core execution timings, all data scaling methods introduce substantially less work than needed for a one-to-one scaling of work and cores (cf. \autoref{fig:vr_input_scaling}).
    
    In ray casting, a query for each sampling position along each ray is made. Depending on the data structure this query can be dependent on the size of the whole data set. But the number of queries is largely dependent on the number of rays and the sampling distance between the points. For more complex techniques, including optimizations like early-ray-termination and empty-space-skipping, or path tracing, there are many more factors which depend on the data that can influence the execution time. For example, the opacity transfer function has a large influence on early-ray-termination. Furthermore, if a 3-dimensional uniform rectilinear grid is increased in each direction by a factor of two, the data size increases by a factor of eight while the number of points sampled per ray only roughly doubles. Thus, the actual work does not increase by the same factor as the number of input cells.

\paragraph*{Augmented Contour Tree Computation}

    \begin{figure}
        \centering
        \includegraphics[width=.95\columnwidth]{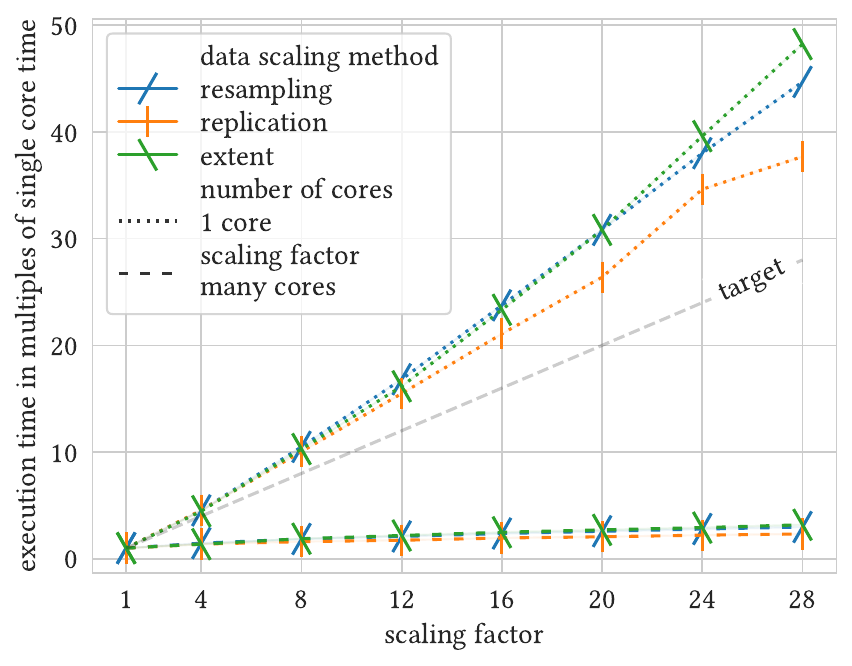}
        \caption{Weak scaling timings of augmented contour tree computation via parallel peak pruning on the Perlin noise data set. The dotted lines show the corresponding timings with the same data computed on a single core. It can be seen that all scaling methods do increase the workload by some amount that is larger than the scaling factor from one to many cores.
        }
        \label{fig:ct_input_scaling}
    \end{figure}
    
    This example shows the execution timings of the parallel peak pruning algorithm by \cite{carr2016,carr2021} available in VTK-m \citep{vtkm}.
    In \autoref{fig:ct_input_scaling}, showing the actual work on a single core for each scaling method on the Perlin noise data set, each scaling method introduces more work than the one-to-one ratio. 
    The computation of the augmented contour tree does not have linear complexity. Its time complexity is in $\mathcal{O}(E\log t\log E)$, with $E$ being the number of edges in the input and $t$ the number of critical points. Here, independent of the data, too much work is generated by scaling the input and there is an additional data dependency via the number of critical points.

\subsection{Implications}\label{sec:implications}

    Let $T(N,|d|)$ denote the execution time of some algorithm on $N$ cores with data of size $|d|$. Then $T(N,N|d|)$ is the execution time of that algorithm on a with a factor $N$ scaled up data set $d$. $T(1,N|d|)$ denotes the execution time of the algorithm with the same data on a single core. For Gustafson's law we also need a hypothetical execution time on a single core $\hat{T}(1,N|d|)$, which differs to $T(1,N|d|)$, since it is derived from $T(N,N|d|)$, taking the parallel fraction into account but otherwise assuming perfect scalability (see \autoref{q:gustafson}). For simplicity we set ${T(N,N|d|) = 1}$.
    From Gustafson's law follows under the assumed ideal scaling conditions that an algorithm shows good weak scalability via $\hat{T}(1, N|d|)/T(N,N|d|) \simeq N$ for small serial fractions.
    With ${T(N,N|d|) = 1}$, it follows that $\hat{T}(1, N|d|) \simeq N$.
    However, when looking at the actual $T(1,N|d|)$, sometimes $T(1,N|d|) < N$ and even $T(1,N|d|) > N$ can be observed.
    $T(1,N|d|) < N$ either means additional parallel overhead, larger serial fraction or the problem size is not scaled correctly.
    $T(1,N|d|) > N$ indicates that the problem size is too large.
    If we actually compare $T(1,N|d|)$ to $T(1,|d|)$ on smaller data we can identify that the problem size does indeed scale differently to the number of cores and input size. Since the single core execution time is not affected by actual parallelization overhead, it should hold for algorithms with linear complexity that $T(1,N|d|) \simeq N T(1,|d|)$.
    We can see that this is not the case for all scaling methods. It seems to depend on the algorithm, the data itself and thus also on the scaling method (\autoref{fig:ic_input_scaling}, \ref{fig:vr_input_scaling} and \ref{fig:ct_input_scaling}).

    Although this might have extensive consequences for the consistency and general applicability of weak scalability studies, it is common practice to only scale the input size. 

\section{Scaling Problem Size by Work}\label{sec:workscaling}

    As shown, the na\"ive scaling of the number of input cells does not lead to a commensurate increase in work contained in the data set depending on the algorithm. In the following, we describe a method that mitigates this problem by using the mentioned scaling methods introducing as many cells as needed to increase the work proportionally. 

    The goal of this method is to find a corrected scaling factor for the data so that the algorithm on the scaled data takes $N$ times longer than on the base data set on a single core. Then, the resulting data set can be used as the input for the execution on $N$ cores as part of the weak scalability benchmark.

    In order to scale the problems according to the amount of work for a specific algorithm, first, the work on the original data has to be identified. Under the assumption of constant core speeds, the amount of work can be approximated by the execution time $T_o$ of said algorithm on one core. Given a target factor $F$ for the problem size, we scale the data with the respective method and a scaling factor $f$ as a parameter, for which, as a first guess, we use the target factor $F$. The work of this now scaled data is again approximated by executing the algorithm with this data on a single core, meaning we can describe $T$ as a function of $f$. Now, we can compute the achieved work scaling factor $T(f)/T_o$. So we are looking for an $f$ as a parameter for the data scaling such that $T(f)/T_o = F$. With this and the following equations, we can iteratively update our scaling factor using a quasi-Newton's method. 
    \begin{align*}
        f_0 &= \frac{T_o}{T_o} = 1 \\
        f_1 &= \frac{T(F)}{T_o} \\
        e(f) &= \frac{T(f)}{T_o} - F \\
        e'(f_i) &= \frac{e(f_i) - e(f_{i-1})}{f_i-f_{i-1}} &\forall i>0 \\
        f_{i+1} &= f_i - \frac{e(f_i)}{e'(f_i)} &\forall i>0
    \end{align*}
    This iteration is terminated if the error is small enough $e(f_i) < 0.01$, the approximated derivative is zero $e'(f_i) = 0$, a maximum number of iterations is reached $i > 10$, the number of cells in the scaled data does not change, or the scaling factor exceeds twice the target factor and would still grow $f_i>2F_\text{t}, f_{i+1} > f_i$.

    Getting accurate scaling factors with this approach, highly depends on $T(f)$. However, this method can be tweaked via the termination conditions and starting points, number of repetitions for each step and other parameters to yield acceptable scaling factors. 
    On that note, if the presented approach does not converge, other techniques can be employed, like adding damping.
    Nonetheless, we achieved satisfactory accuracy in our experiments. Whenever the targeted execution time with the \emph{work scaled} data set has a relative error of more than five percent $\left((T(f_{last})/T_o - F)/F > 0.05\right)$, it is highlighted in the result plots in the following section by a square marker. Similar to the accuracy, the costs of creating the \emph{work scaled} data sets may vary depending on the algorithm. Of course, executing algorithms on a single core for large data is not ideal. However, we get some valuable insights in return, demonstrated in the next section.

\section{Experiments}\label{sec:experiments}

    In the experiments, we investigate the same algorithms for which we provided examples in \autoref{sec:inputscaling}. We compare the traditional scaling approach, which we will call \emph{input scaling}, to the approach presented in \autoref{sec:workscaling}, which we will call \emph{work scaling} in a weak scaling experiment.

    The goal in this section is to assess whether \emph{work scaling} methods yield more meaningful results than \emph{input scaling} methods. The results presented here are not meant to discuss whether a method exhibits good weak scalability.
    Since \emph{input scaling} was shown to yield varying results across scaling approaches, we consider the \emph{work scaling} approach to perform well if the execution time measurements coincide across data scaling techniques or are at least closer together as the corresponding \emph{input scaled} execution time measurements, meaning more consistent. This also means that the \emph{work scaling} approach should yield similar results to \emph{input scaling}, when \emph{input scaling} works well.
    
    Analogous to \autoref{sec:inputscaling}, the benchmarks were executed 5 times each. The \emph{work scaled} data was computed based upon the median of the work estimation of seven runs on one core representing $T_0$. Every iteration of the scaling factor is based on the median of five runs ($T(f)$). For the \emph{extent scaled} data sets, the original data set was resampled with a scaling factor of $32$ and the resulting data set was used to create the \emph{extent scaled} subvolumes. These experiments are conducted on the same machine used in \autoref{sec:examples}.

    \paragraph*{How to read the plots}
    For each algorithm we report experiment results on the three data sets and the three scaling methods from \autoref{sec:inputscaling}. Each combination of data set and algorithm results in three plots. The first shows weak-scalability time measurements for \emph{work scaling} and \emph{input size scaling}. 
    In this plot we can observe whether there are large differences in execution time between differently \emph{input scaled} data sets by comparing the dashed lines with different colors corresponding to the scaling method (cf. \autoref{fig:fe_aneurism_work_vs_input},\ref{fig:fe_bones_work_vs_input},\ref{fig:fe_pn_work_vs_input}). Further, we can asses whether the execution timings of \emph{work scaled} data sets are closer together, i.e., more consistent, by comparing the spread of the dashed lines with the spread of the solid lines. 
    The second type of plot displays the \emph{work scaled} data size in multiples of the base data size---and thus also the scaling factor---that is used for the \emph{work scaling} approach in comparison to the line with slope one, representing the \emph{input scaling} approach. In these plots it can be observed in which directions and by how much the \emph{work scaling} method corrects the data size to meet the target work requirement (cf. \autoref{fig:fe_aneurism_sf},\ref{fig:fe_bones_sf},\ref{fig:fe_pn_sf}). 
    The last plot compares the execution time of \emph{input scaled} data sets on a single core with the execution time of \emph{input scaled} data sets on the number of cores corresponding to the scaling factor. This is the same kind of plot as in \autoref{sec:inputscaling}. These plots show the deviation of the actual amount of work from the target amount in the sense of Gustafson's scaled speedup for \emph{input scaling} (\autoref{fig:fe_aneurism_input_scaling},\ref{fig:fe_bones_input_scaling},\ref{fig:fe_pn_input_scaling}).

\subsection{Iso-Contouring}\label{sec:ex_ic}

    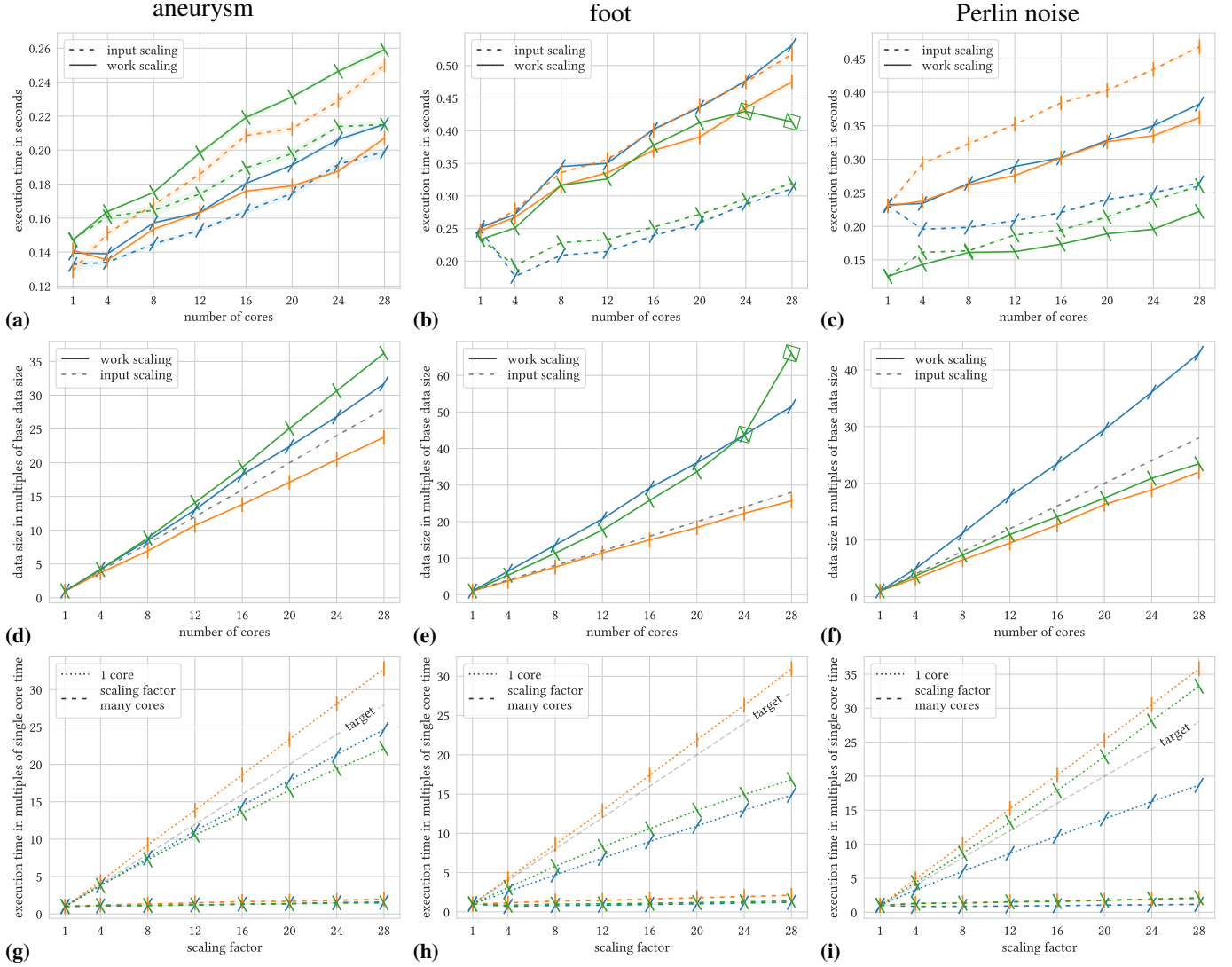
\begin{figure*}[tp]
        \centering
        \begin{tikzpicture}[ampersand replacement=\&]
            \matrix [outer sep=0pt] {
              \node[]{data scaling method:}; \&
              \draw[color=resampling, very thick] (0.1,-0.1) -- (0.3,0.1);
              \draw[color=resampling, very thick] (0.0,0.0) -- (+0.4,0.0) node[right,black] {resampling}; \&
              \draw[color=replication, very thick] (0.2,-0.1) -- (0.2,0.1);
              \draw[color=replication, very thick] (0.0,0.0) -- (+0.4,0.0) node[right,black] {replication}; \&
              \draw[color=extent, very thick] (0.3,-0.1) -- (0.1,0.1);
              \draw[color=extent, very thick] (0.0,0.0) -- (+0.4,0.0) node[right,black] {extent}; \\
            };
        \end{tikzpicture}
        \vspace{0.5em}
        
        \begin{subfigure}{0.33\textwidth}
            \centering
            aneurysm
            \vspace{2pt}
            
            \includegraphics[width=\textwidth]{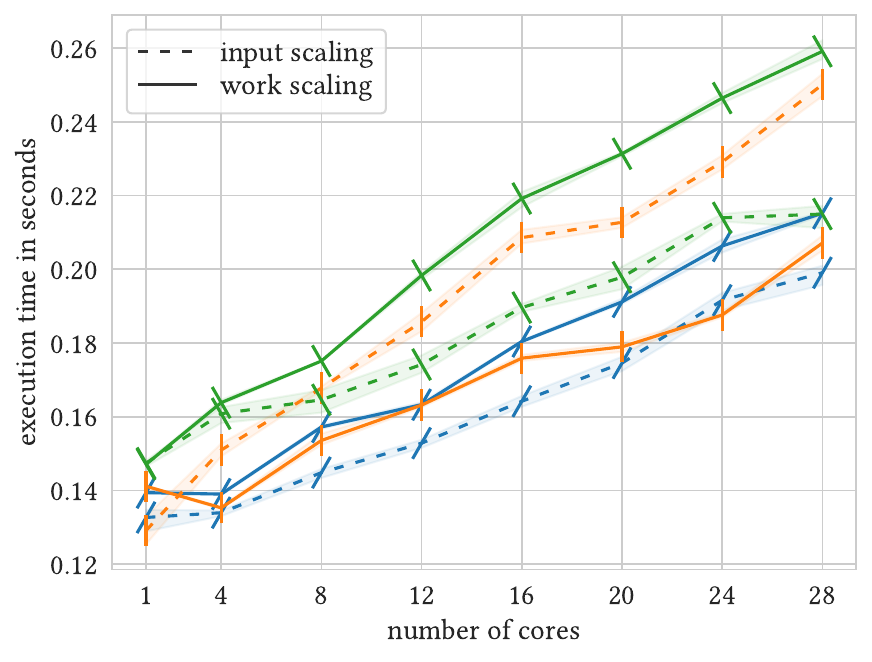}
            \vspace{-1cm}
            \caption{}\label{fig:fe_aneurism_work_vs_input}
        \end{subfigure}
        \begin{subfigure}{0.33\textwidth}
            \centering
            foot
            \vspace{2pt}
            
            \includegraphics[width=\textwidth]{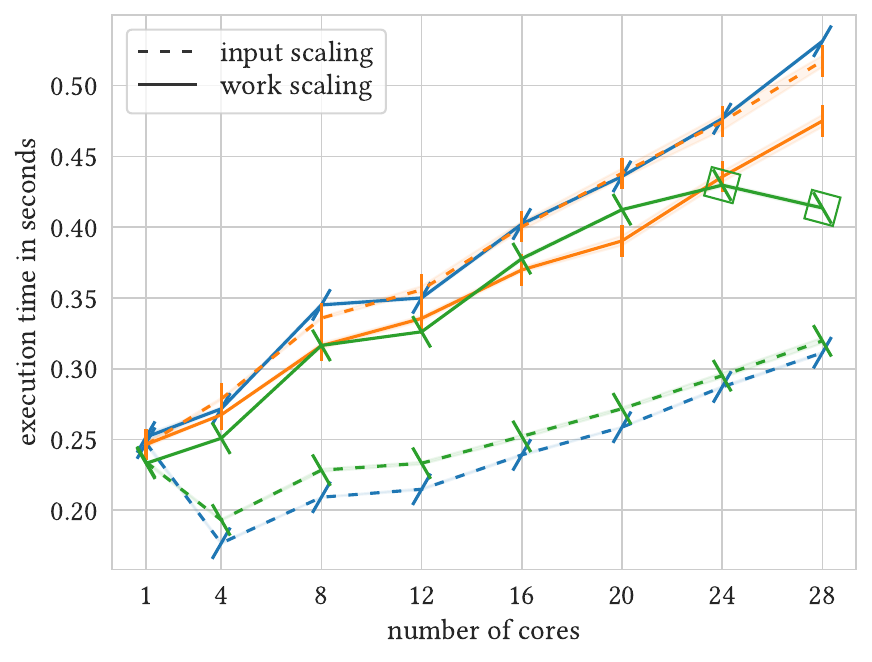}
            \vspace{-1cm}
            \caption{}\label{fig:fe_bones_work_vs_input}
        \end{subfigure}
        \begin{subfigure}{0.33\textwidth}
            \centering
            Perlin noise
            \vspace{2pt}
            
            \includegraphics[width=\textwidth]{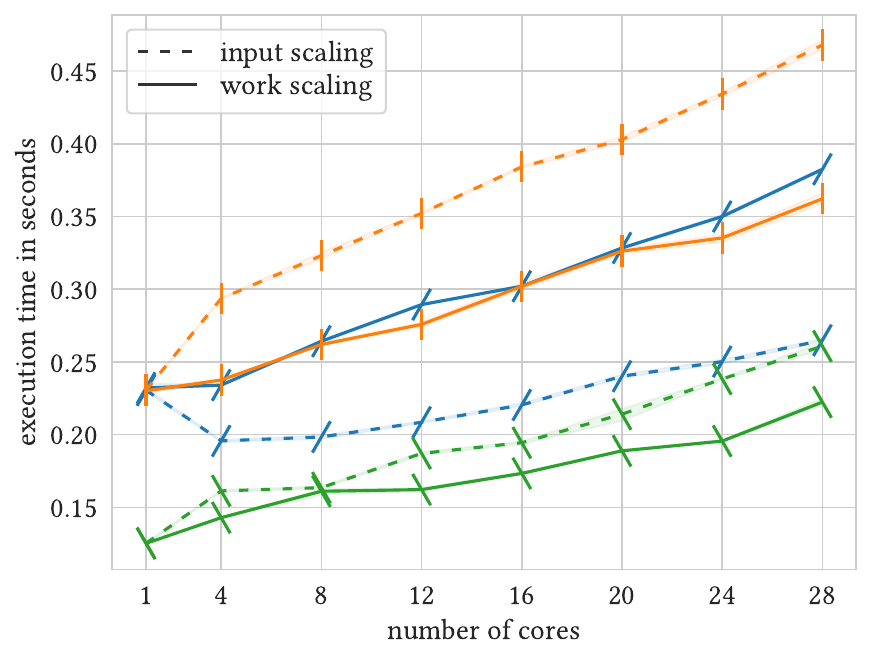}
            \vspace{-1cm}
            \caption{}\label{fig:fe_pn_work_vs_input}
        \end{subfigure}
        \begin{subfigure}{0.33\textwidth}
            \centering
            \includegraphics[width=\textwidth]{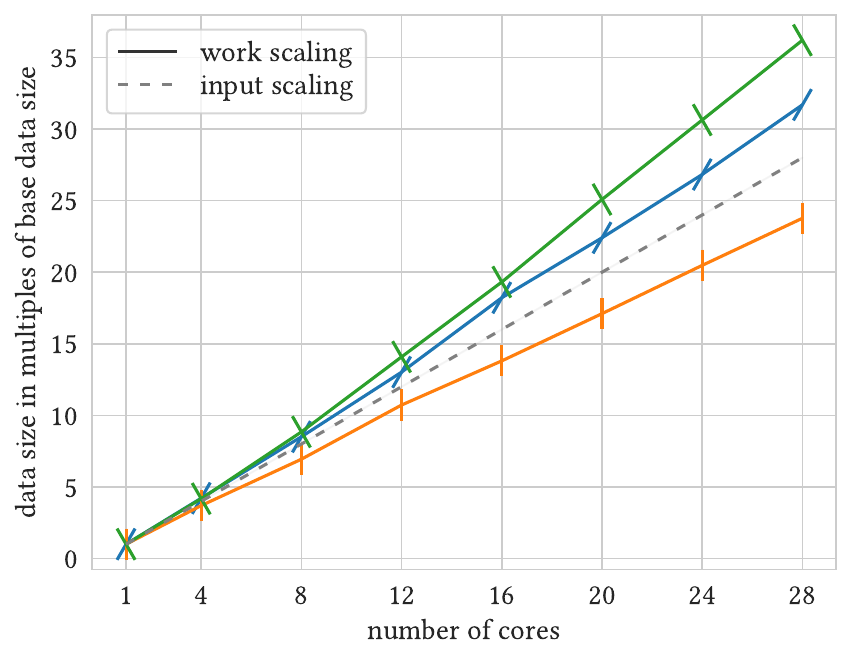}
            \vspace{-1cm}
            \caption{}\label{fig:fe_aneurism_sf}
        \end{subfigure}
        \begin{subfigure}{0.33\textwidth}
            \centering
            \includegraphics[width=\textwidth]{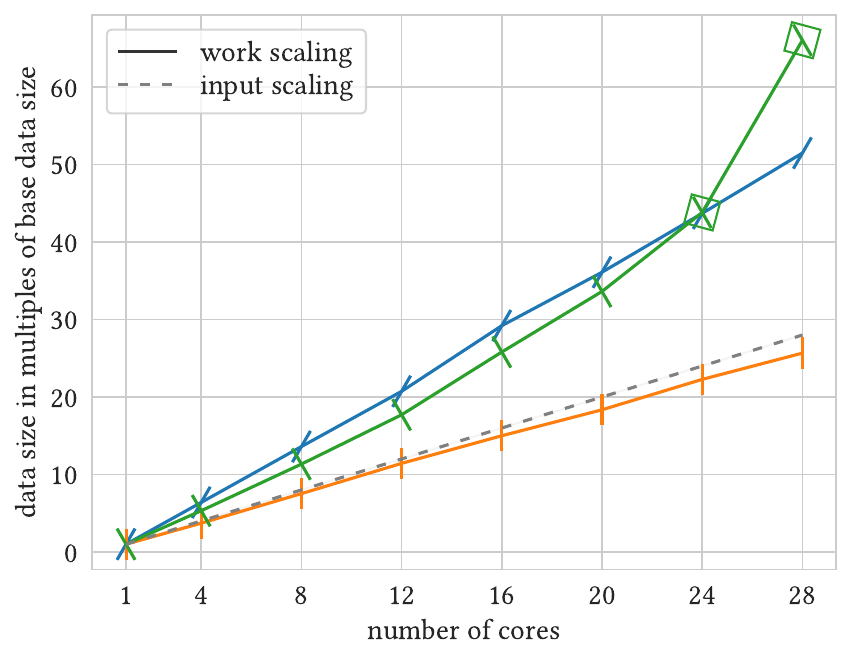}
            \vspace{-1cm}
            \caption{}\label{fig:fe_bones_sf}
        \end{subfigure}
        \begin{subfigure}{0.33\textwidth}
            \centering
            \includegraphics[width=\textwidth]{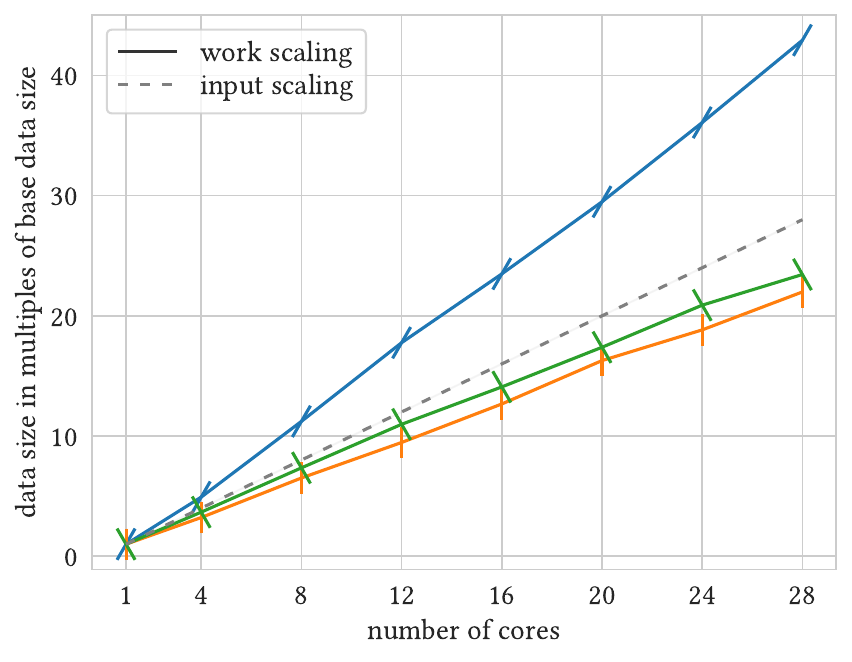}
            \vspace{-1cm}
            \caption{}\label{fig:fe_pn_sf}
        \end{subfigure}
        \begin{subfigure}{0.33\textwidth}
            \centering
            \includegraphics[width=\textwidth]{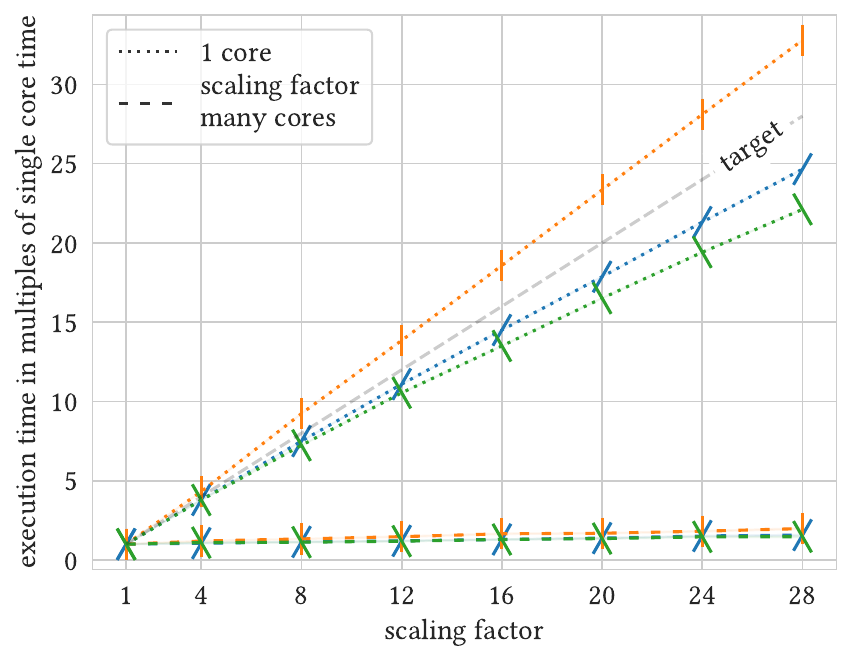}
            \vspace{-1cm}
            \caption{}\label{fig:fe_aneurism_input_scaling}
        \end{subfigure}
        \begin{subfigure}{0.33\textwidth}
            \centering
            \includegraphics[width=\textwidth]{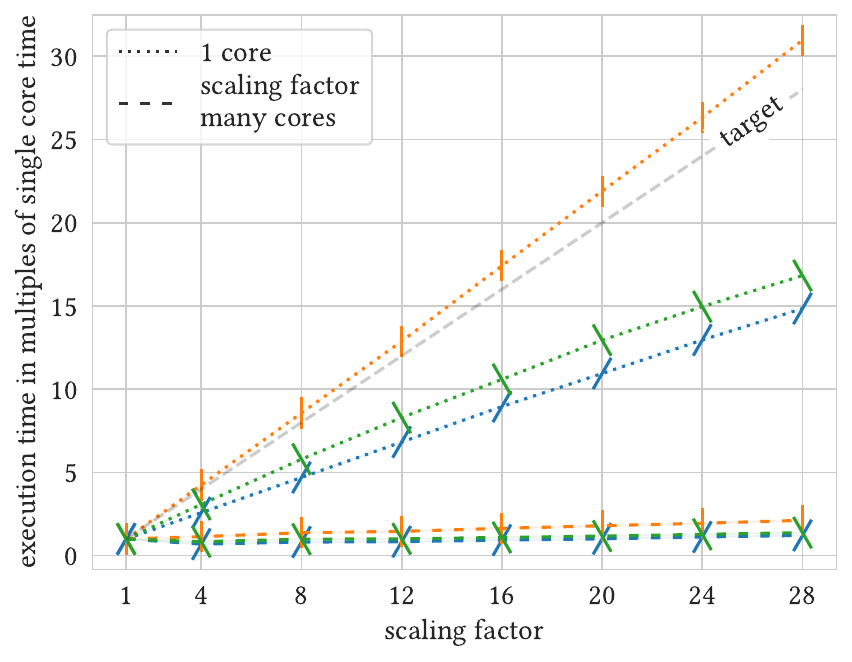}
            \vspace{-1cm}
            \caption{}\label{fig:fe_bones_input_scaling}
        \end{subfigure}
        \begin{subfigure}{0.33\textwidth}
            \centering
            \includegraphics[width=\textwidth]{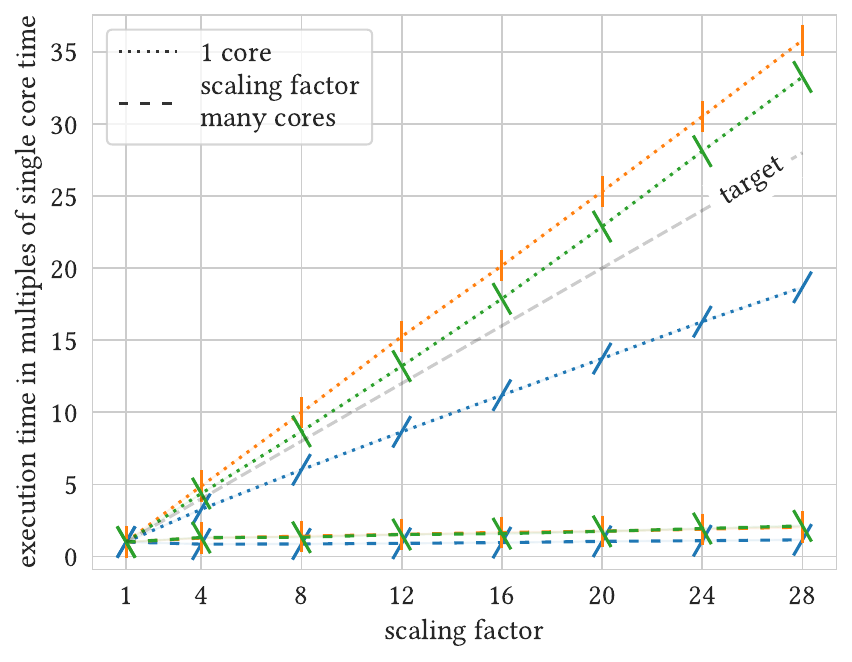}
            \vspace{-1cm}
            \caption{}\label{fig:fe_pn_input_scaling}
        \end{subfigure}
        \caption{Flying edge plots showing \emph{work scaling} vs. \emph{input scaling} (top row), the corresponding scaling factor (middle row) and the work approximation of input-size scaling via execution time on a single core (bottom row). The columns correspond to the different data sets aneurysm, foot and Perlin noise from left to right. Timings on \emph{work scaled} data sets are comparatively consistent across \emph{resampling} and \emph{replication}. \emph{Input scaling} introduces insufficient amounts of work for \emph{resampling} and \emph{extent} scaling. \emph{Extent} scaling deviates from the other methods depending on the data set, due to the different reference data for single core execution.
        }
        \label{fig:contour_flyingedge}
    \end{figure*}

    \begin{figure*}
        \vspace{1em}
        
        \centering
        \begin{tikzpicture}[ampersand replacement=\&]
            \matrix [outer sep=0pt] {
              \node[]{data scaling method:}; \&
              \draw[color=resampling, very thick] (0.1,-0.1) -- (0.3,0.1);
              \draw[color=resampling, very thick] (0.0,0.0) -- (+0.4,0.0) node[right,black] {resampling}; \&
              \draw[color=replication, very thick] (0.2,-0.1) -- (0.2,0.1);
              \draw[color=replication, very thick] (0.0,0.0) -- (+0.4,0.0) node[right,black] {replication}; \&
              \draw[color=extent, very thick] (0.3,-0.1) -- (0.1,0.1);
              \draw[color=extent, very thick] (0.0,0.0) -- (+0.4,0.0) node[right,black] {extent}; \\
            };
        \end{tikzpicture}
        \vspace{0.5em}
        
        \begin{subfigure}{0.33\textwidth}
            \centering
            aneurysm
            \vspace{2pt}
            
            \includegraphics[width=\textwidth]{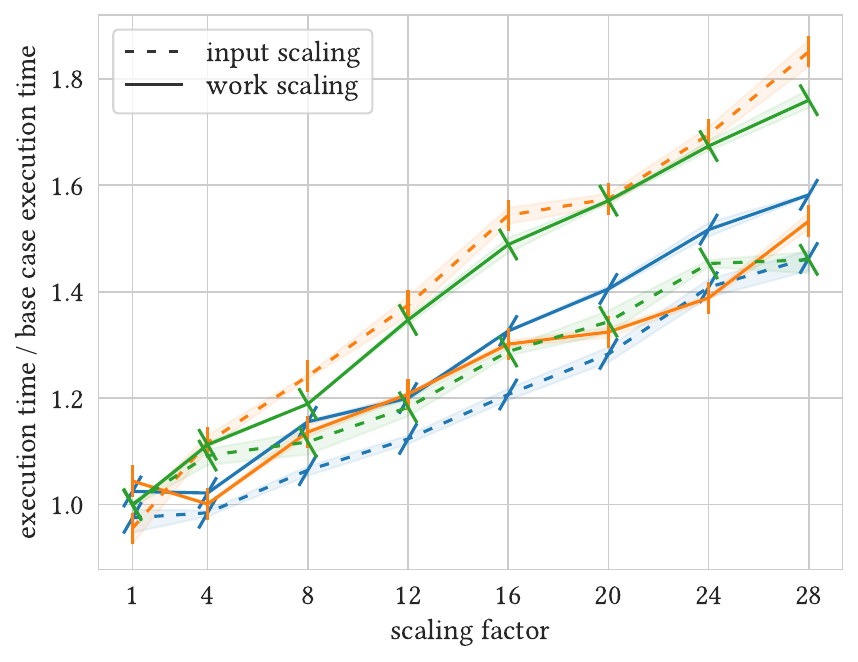}
            \vspace{-1cm}
            \caption{}\label{fig:contour_aneurysm_normalized}
        \end{subfigure}
        \begin{subfigure}{0.33\textwidth}
            \centering
            foot
            \vspace{2pt}
            
            \includegraphics[width=\textwidth]{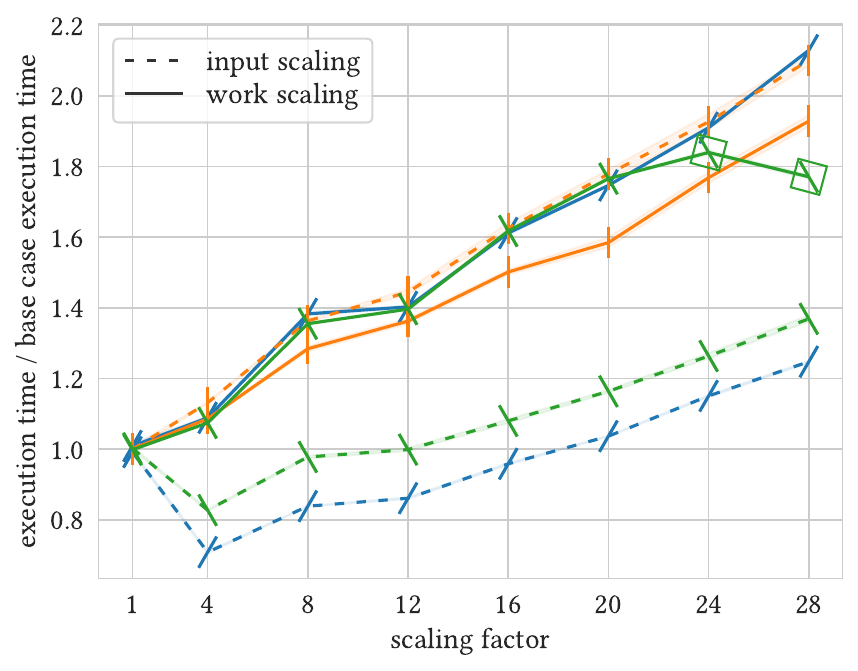}
            \vspace{-1cm}
            \caption{}\label{fig:contour_foot_normalized}
        \end{subfigure}
        \begin{subfigure}{0.33\textwidth}
            \centering
            Perlin noise
            \vspace{2pt}
            
            \includegraphics[width=\textwidth]{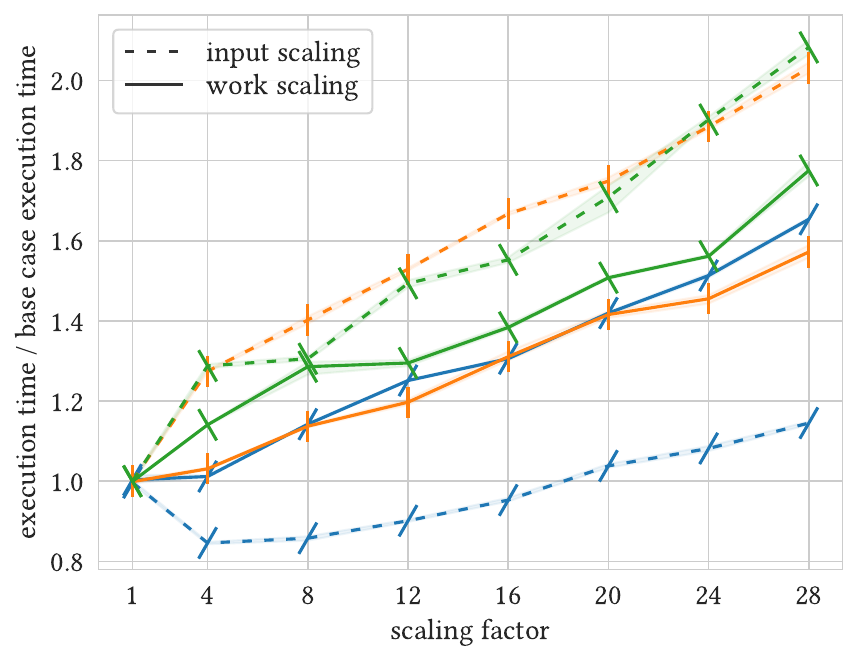}
            \vspace{-1cm}
            \caption{}\label{fig:contour_perlinnoise_normalized}
        \end{subfigure}
        \caption{Flying edge plots showing execution time in multiples of the execution time for the base case on a single core. The plots correspond to the different data sets aneurysm, foot and Perlin noise from left to right. 
        Here we can observe that the relative change in execution time of \emph{extent} scaling also aligns with the other methods, most notably in the Perlin noise data set (right).
        }
        \label{fig:contour_flyingedge_normalized}
    \end{figure*}
    
    Again, we consider the flying edge algorithm \citep{schroeder2015} for iso-contour extraction as provided by VTK \citep{vtk}. The test cases each compute five different iso-values evenly spread across the range in the data. The results are depicted in \autoref{fig:contour_flyingedge}. Creating the 72 data sets---8 per scaling method and data set---took about 9 hours.

    In the aneurysm and Perlin noise data set, the \emph{extent} scaled version deviates from the other methods in \emph{input scaling} mode as well as in \emph{work scaling} mode. We can observe a difference in the execution time of its base case on a single core for these two data sets (see \autoref{fig:fe_aneurism_work_vs_input} and \autoref{fig:fe_pn_work_vs_input}). In the aneurysm data set it exhibits longer execution times than the other base case and for Perlin noise it is faster. In the aneurysm data set, most of the features are next to the center of the data and become fewer when moving to the boundary. In the Perlin noise data set, the features are equally distributed over the whole data set (cf. \autoref{fig:datasets}). This means that for \emph{extent} scaling, the base case of the aneurysm data set contains proportionally more features and the Perlin noise base case proportionally fewer features than the base cases for \emph{resampling} and \emph{replication}. Since the presented method determines the scaled work with respect to its base case, the results for \emph{extent} scaling do not align with the other methods. 
    However, if we look at the execution time relative to the respective base case execution time, we find that \emph{extent scaling} also aligns (see \autoref{fig:contour_flyingedge_normalized}).

    \emph{Input scaling} with \emph{replication} works well on foot and aneurysm data sets, meaning it creates roughly target factor times more work (see \autoref{fig:fe_bones_input_scaling} and \autoref{fig:fe_aneurism_input_scaling}).
    However, it introduces too much work on the Perlin noise data set (\autoref{fig:fe_pn_input_scaling}). This could be caused by discontinuities at the borders between replications.
    \emph{Input scaling} via \emph{resampling} and \emph{extent} scaling introduce similar amount of work albeit not enough to meet the target factor (\autoref{fig:fe_bones_input_scaling}, \ref{fig:fe_aneurism_input_scaling}, \ref{fig:fe_pn_input_scaling}).
    In the foot data set, where traditional \emph{replication} introduces roughly scaling factor times more work, our method is able to correct the scaling of \emph{resampling} and enlarging the \emph{extent} so that the weak scalability execution time measurements of all three approaches are closer together (cf. \autoref{fig:fe_bones_work_vs_input}).
    For the other two data sets, the \emph{work scaled} \emph{resampling} and \emph{replication} measurements are closer most of the time, indicating that weak scalability via \emph{work scaling} is more consistent across data scaling methods than \emph{input scaling}.

\subsection{Direct Volume Rendering}

    \begin{figure*}
        \centering
        \begin{tikzpicture}[ampersand replacement=\&]
            \matrix [outer sep=0pt] {
              \node[]{data scaling method:}; \&
              \draw[color=resampling, very thick] (0.1,-0.1) -- (0.3,0.1);
              \draw[color=resampling, very thick] (0.0,0.0) -- (+0.4,0.0) node[right,black] {resampling}; \&
              \draw[color=replication, very thick] (0.2,-0.1) -- (0.2,0.1);
              \draw[color=replication, very thick] (0.0,0.0) -- (+0.4,0.0) node[right,black] {replication}; \&
              \draw[color=extent, very thick] (0.3,-0.1) -- (0.1,0.1);
              \draw[color=extent, very thick] (0.0,0.0) -- (+0.4,0.0) node[right,black] {extent}; \\
            };
        \end{tikzpicture}
        \vspace{0.5em}
        
        \begin{subfigure}{0.33\textwidth}
            \centering
            \includegraphics[width=\textwidth]{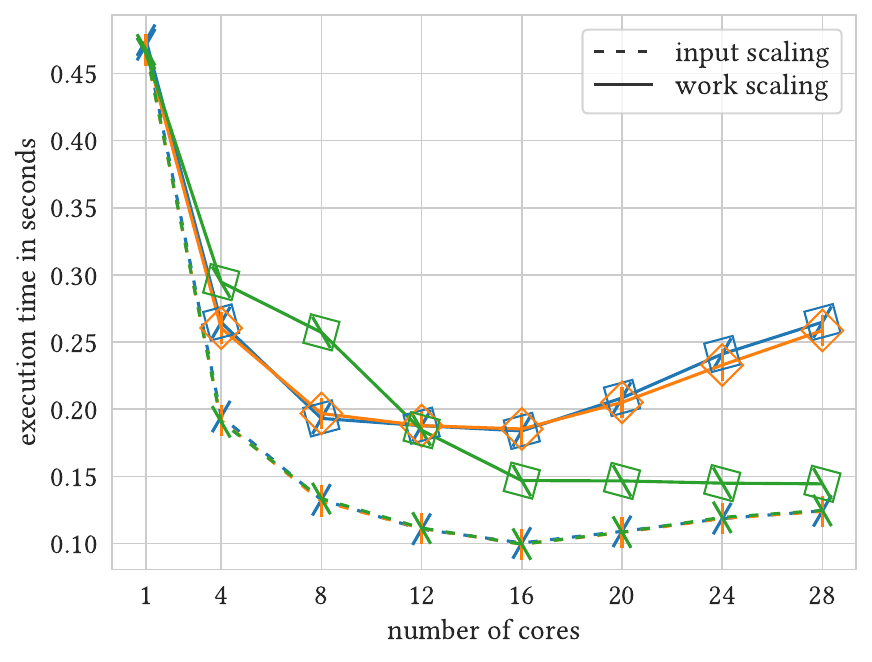}
            \vspace{-1cm}
            \caption{}\label{fig:vr_bones_work_vs_input}
        \end{subfigure}
        \begin{subfigure}{0.33\textwidth}
            \centering          
            \includegraphics[width=\textwidth]{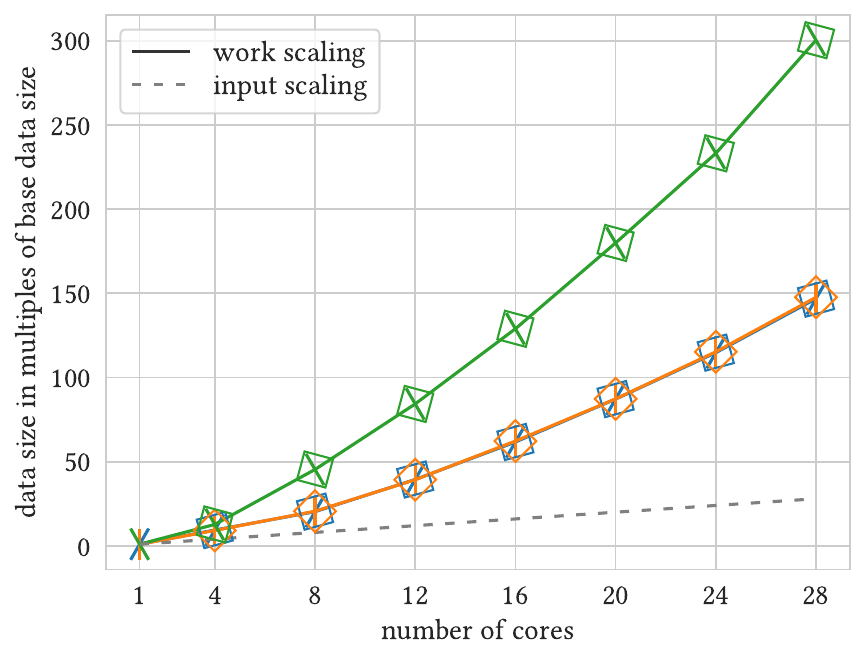}
            \vspace{-1cm}
            \caption{}\label{fig:vr_bones_sf}
        \end{subfigure}
        \begin{subfigure}{0.32\textwidth}
            \centering
            \includegraphics[width=\textwidth]{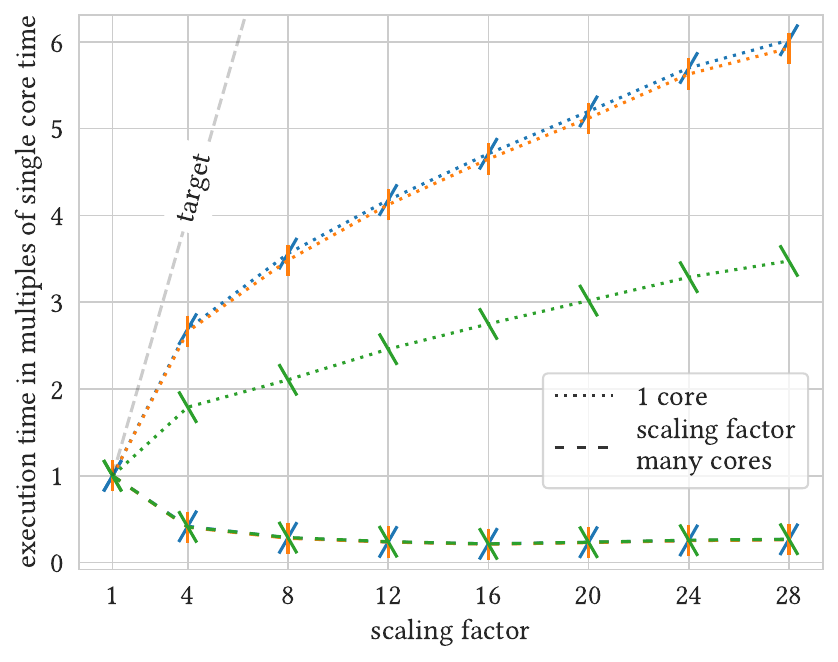}
            \vspace{-1cm}
            \caption{}\label{fig:vr_bones_input_scaling}
        \end{subfigure}
        \caption{Volume rendering plots showing \emph{work scaling} vs. \emph{input scaling} (left), the corresponding scaling factor (middle) and the work approximation of input-size scaling via execution time on a single core (right). The results displayed are for executions on the foot data set only. The results on the other data sets are omitted for brevity due to their similarity to plots shown here. It can be observed that \emph{input scaling} introduces insufficient work for weak scalability for all scaling methods. This was not entirely corrected by our method due to termination when data sets grew too large. This indicates that the execution time of the volume rendering algorithm does not depend to a larger degree on the size of the input data set.
        }
        \label{fig:volumerender}
    \end{figure*}

    Here we are looking at the effects of \emph{work scaling} on volume rendering with VTK’s OpenGLGPUVolumeRayCastMapper with OSMesa. For the test cases a single opacity transfer function was used that increased linearly over the whole data range from $0$ to $0.02$ and early ray termination is enabled. The size of the rendered image is $800\times800$ pixel. The results are depicted in \autoref{fig:volumerender}. Executions on other data sets are omitted due to being nearly identical to the ones on the foot data set. Creating all 72 data sets---8 per scaling method and data set---took about 9 hours.

    The weak scalability plot (\autoref{fig:vr_bones_work_vs_input}) shows more than optimal weak scalability behavior with the negative slope showing a decrease in execution time on more cores. As anticipated in \autoref{sec:inputscaling}, the execution times on a single core reveal the disproportional scaling of work (\autoref{fig:vr_bones_input_scaling}). This behavior can only slightly be mitigated by our \emph{work scaling} approach, since the data size increases drastically and would exceed memory capabilities very fast (\autoref{fig:vr_bones_sf}). Nonetheless, \autoref{fig:vr_bones_sf} showing the size of \emph{work scaled} data sets and \autoref{fig:vr_bones_input_scaling} displaying single core execution time, indicate a sub-optimal choice for the parameter scaling the work, being the input size in this case. What we observe here in the execution time plots (\autoref{fig:vr_bones_work_vs_input}) is more the general complexity of the algorithm than the overhead of parallelization.

\subsection{Augmented Contour Tree Computation}

    \begin{figure*}
        \centering
        \begin{tikzpicture}[ampersand replacement=\&]
            \matrix [outer sep=0pt] {
              \node[]{data scaling method:}; \&
              \draw[color=resampling, very thick] (0.1,-0.1) -- (0.3,0.1);
              \draw[color=resampling, very thick] (0.0,0.0) -- (+0.4,0.0) node[right,black] {resampling}; \&
              \draw[color=replication, very thick] (0.2,-0.1) -- (0.2,0.1);
              \draw[color=replication, very thick] (0.0,0.0) -- (+0.4,0.0) node[right,black] {replication}; \&
              \draw[color=extent, very thick] (0.3,-0.1) -- (0.1,0.1);
              \draw[color=extent, very thick] (0.0,0.0) -- (+0.4,0.0) node[right,black] {extent}; \\
            };
        \end{tikzpicture}
        \vspace{0.5em}
        
        \begin{subfigure}{0.33\textwidth}
            \centering
            aneurysm
            \vspace{2pt}
            
            \includegraphics[width=\textwidth]{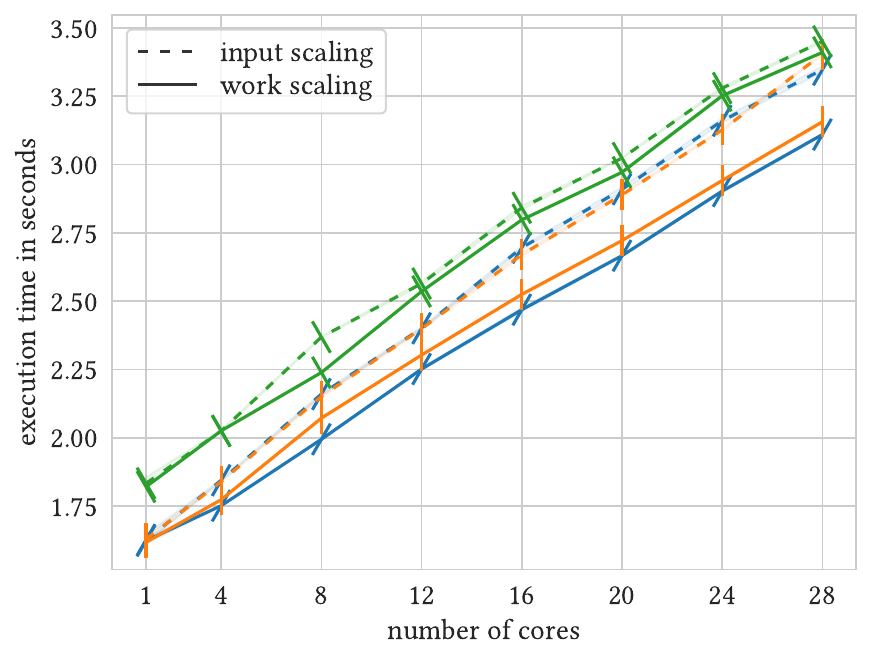}
            \vspace{-1cm}
            \caption{}\label{fig:vtkmct_aneurism_work_vs_input}
        \end{subfigure}
        \begin{subfigure}{0.33\textwidth}
            \centering
            foot
            \vspace{2pt}
            
            \includegraphics[width=\textwidth]{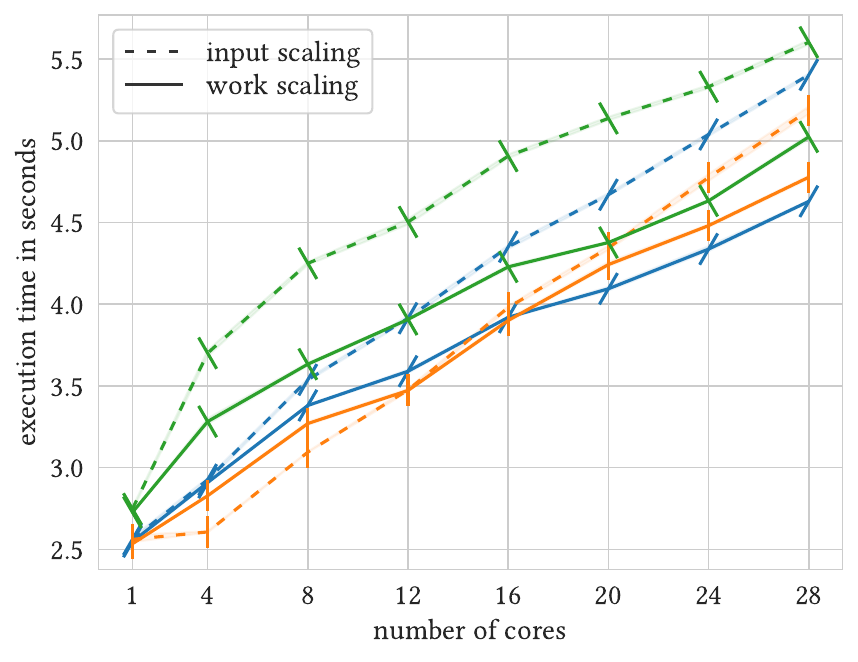}
            \vspace{-1cm}
            \caption{}\label{fig:vtkmct_bones_work_vs_input}
        \end{subfigure}
        \begin{subfigure}{0.33\textwidth}
            \centering
            Perlin noise
            \vspace{2pt}
            
            \includegraphics[width=\textwidth]{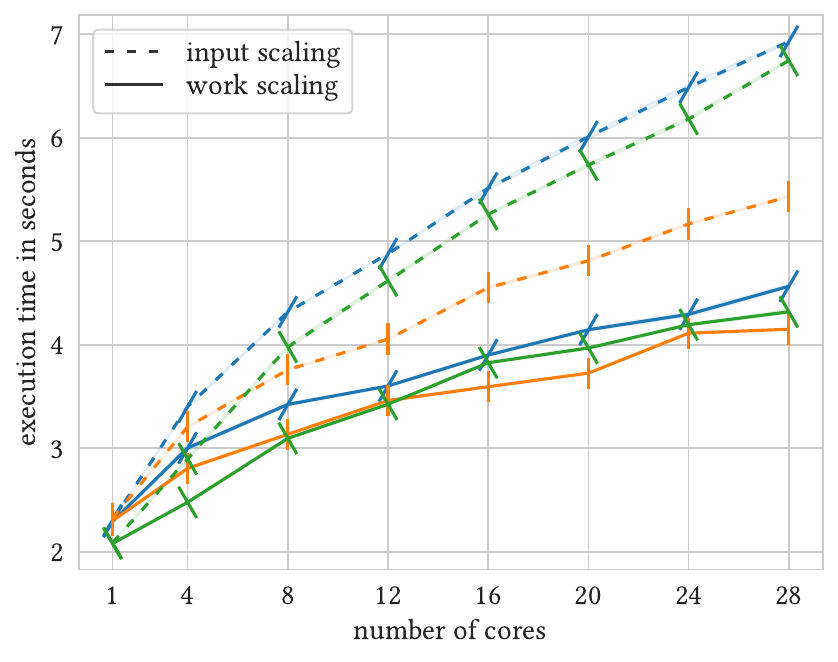}
            \vspace{-1cm}
            \caption{}\label{fig:vtkmct_pn_work_vs_input}
        \end{subfigure}
        \begin{subfigure}{0.33\textwidth}
            \centering
            \includegraphics[width=\textwidth]{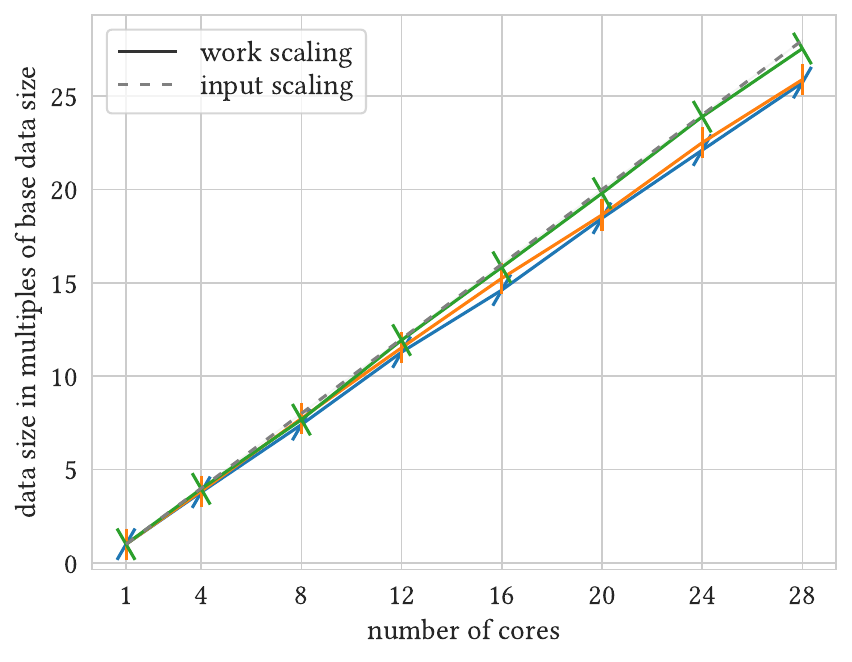}
            \vspace{-1cm}
            \caption{}\label{fig:vtkmct_aneurism_sf}
        \end{subfigure}
        \begin{subfigure}{0.33\textwidth}
            \centering
            \includegraphics[width=\textwidth]{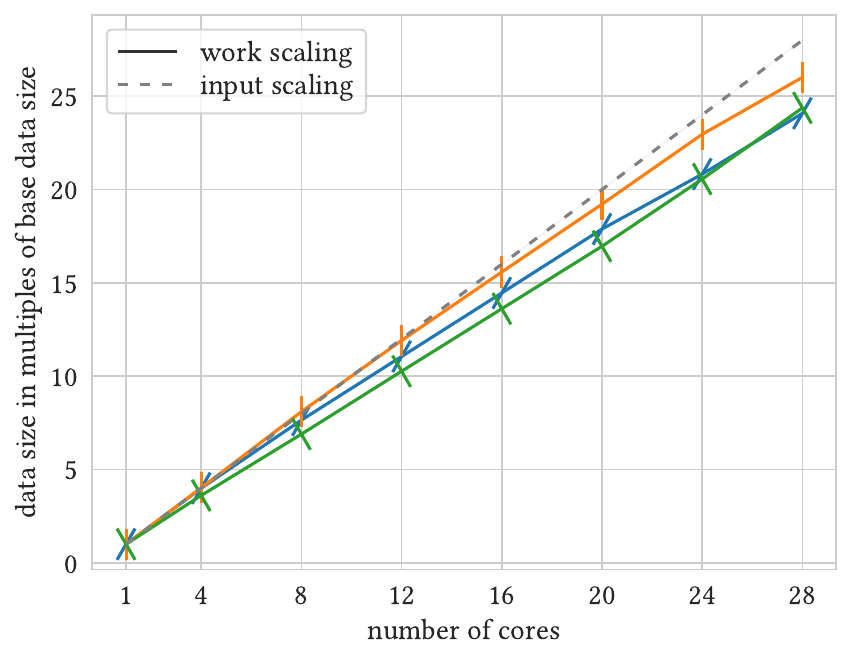}
            \vspace{-1cm}
            \caption{}\label{fig:vtkmct_bones_sf}
        \end{subfigure}
        \begin{subfigure}{0.33\textwidth}
            \centering
            \includegraphics[width=\textwidth]{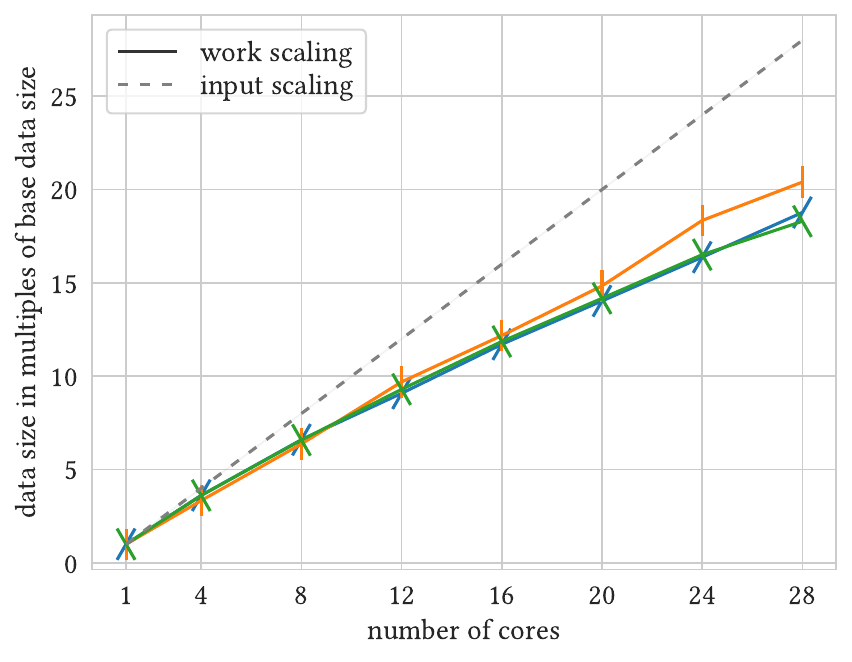}
            \vspace{-1cm}
            \caption{}\label{fig:vtkmct_pn_sf}
        \end{subfigure}
        \begin{subfigure}{0.33\textwidth}
            \centering
            \includegraphics[width=\textwidth]{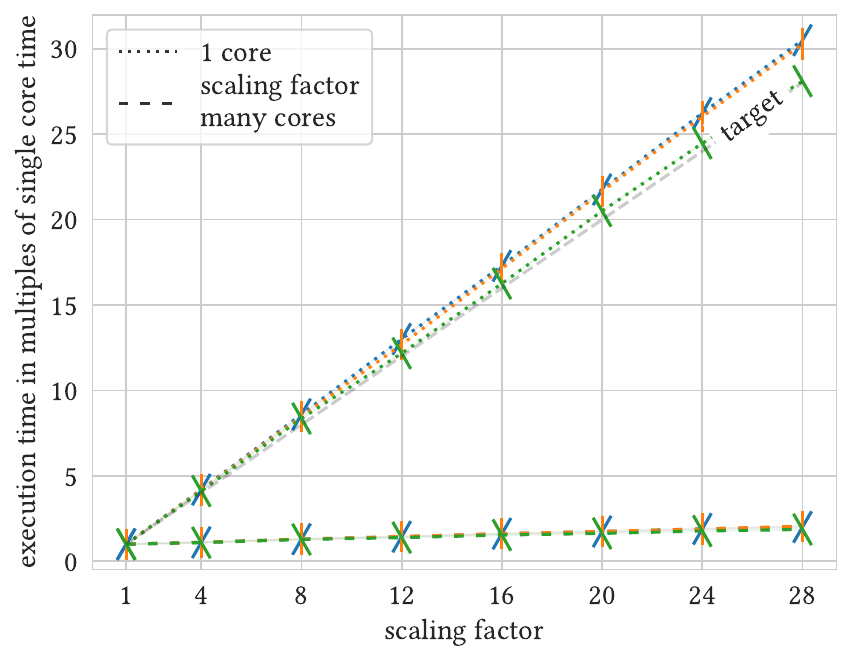}
            \vspace{-1cm}
            \caption{}\label{fig:vtkmct_aneurism_input_scaling}
        \end{subfigure}
        \begin{subfigure}{0.33\textwidth}
            \centering
            \includegraphics[width=\textwidth]{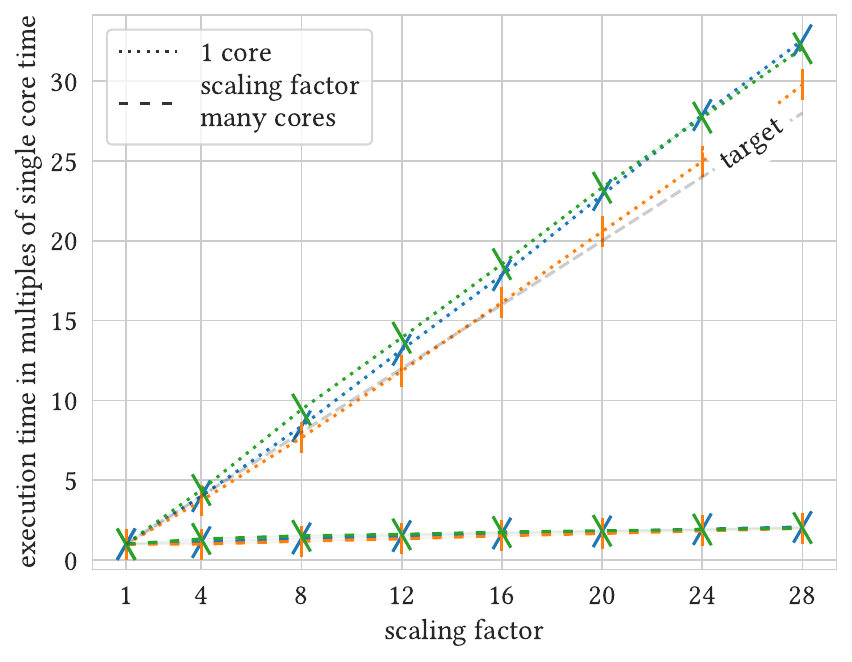}
            \vspace{-1cm}
            \caption{}\label{fig:vtkmct_bones_input_scaling}
        \end{subfigure}
        \begin{subfigure}{0.33\textwidth}
            \centering
            \includegraphics[width=\textwidth]{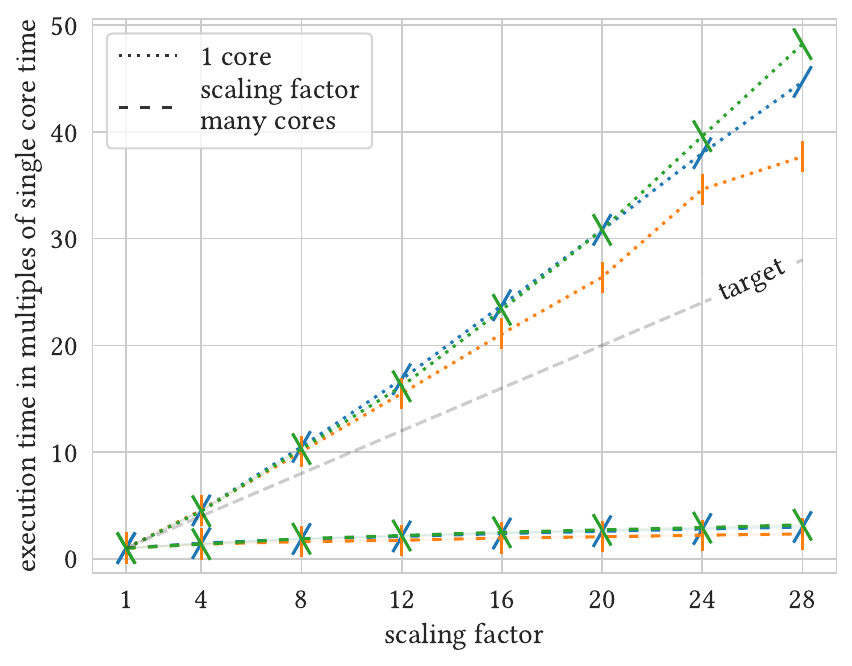}
            \vspace{-1cm}
            \caption{}\label{fig:vtkmct_pn_input_scaling}
        \end{subfigure}
        \caption{Plots showing \emph{work scaling} vs. \emph{input scaling} (top row), the corresponding scaling factor (middle row) and the work approximation of input-size scaling via execution time on a single core (bottom row) for augmented contour tree computation with parallel peak pruning. The columns correspond to the different data sets aneurysm, foot and Perlin noise from left to right. For the aneurysm data set all methods behave similar and only minor corrections can be observed for the scaling factor. Executions on the foot data set show a slight upward deviation in work for \emph{extent} and \emph{resampling} \emph{input scaling}. This can also be observed in the timings, with \emph{extent} scaling showing the longest execution timings. For Perlin noise the \emph{input scaling} methods exhibit differing behavior, while \emph{work scaling} stays consistent across all methods.
        }
        \label{fig:contourtree_ppp}
    \end{figure*}
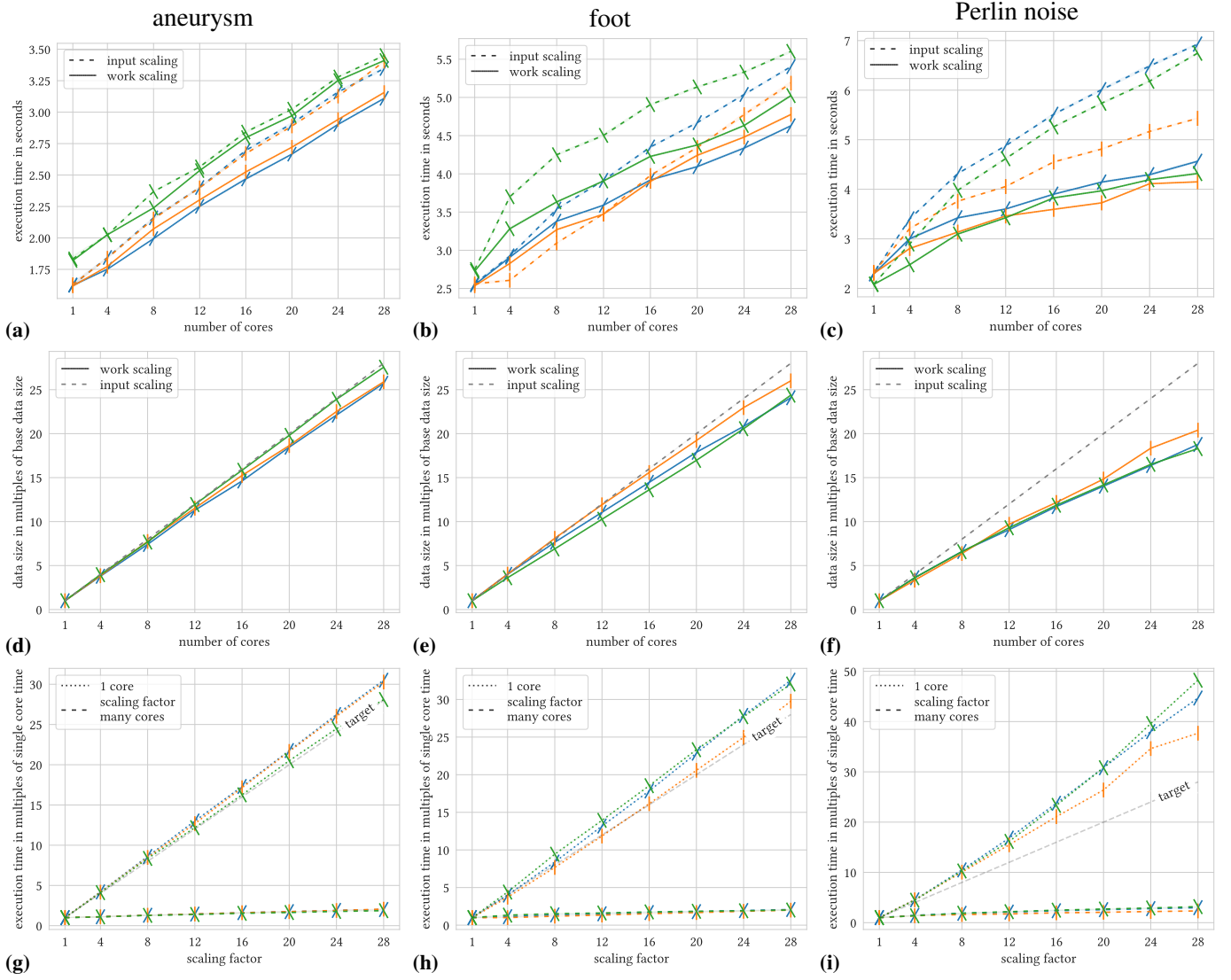

    Here we investigate the \emph{work scaling} method with the parallel peak pruning algorithm computing the augmented contour tree. Creating the 72 data sets---8 per scaling method and data set---took about 20 hours for the parallel peak pruning algorithm. 

    It can be observed that the scaling methods introduce more work than aimed for depending on the data set. For the aneurysm data set near perfect scaling of work is achieved by all three methods with \emph{input scaling} (\autoref{fig:vtkmct_aneurism_input_scaling}), while scaling the foot data set exhibits slightly too much work (cf. \autoref{fig:vtkmct_bones_input_scaling}) and in the Perlin noise data set, the work is roughly $1.5$ times the target amount  (\autoref{fig:vtkmct_pn_input_scaling}).
    Similarly, the weak scaling execution time measurements tend from a linear behavior in the aneurysm data set to a presumably logarithmic behavior with the Perlin noise data set (\autoref{fig:vtkmct_aneurism_work_vs_input}, \ref{fig:vtkmct_pn_work_vs_input}).
    With the Perlin noise data set we can also observe the largest difference between the \emph{input scaled} methods, where \emph{replication} shows better weak scalability than the other two methods (\autoref{fig:vtkmct_pn_work_vs_input}).
    Especially in the Perlin noise data set and similar to the volume rendering case, we measure not only the parallel overhead but also the algorithmic complexity of the algorithm itself, since the algorithm does not linearly depend on the input data size.
    We find that \emph{work scaled} \emph{resampling} and \emph{replication} results for weak scaling are again close together. Furthermore, they show a better weak scalability, due to mitigating the super-linear increase in execution time arising from the general algorithmic complexity of parallel peak pruning by reducing the problem size (\autoref{fig:vtkmct_pn_sf}).

\subsection{On Distributed Weak Scalability}

    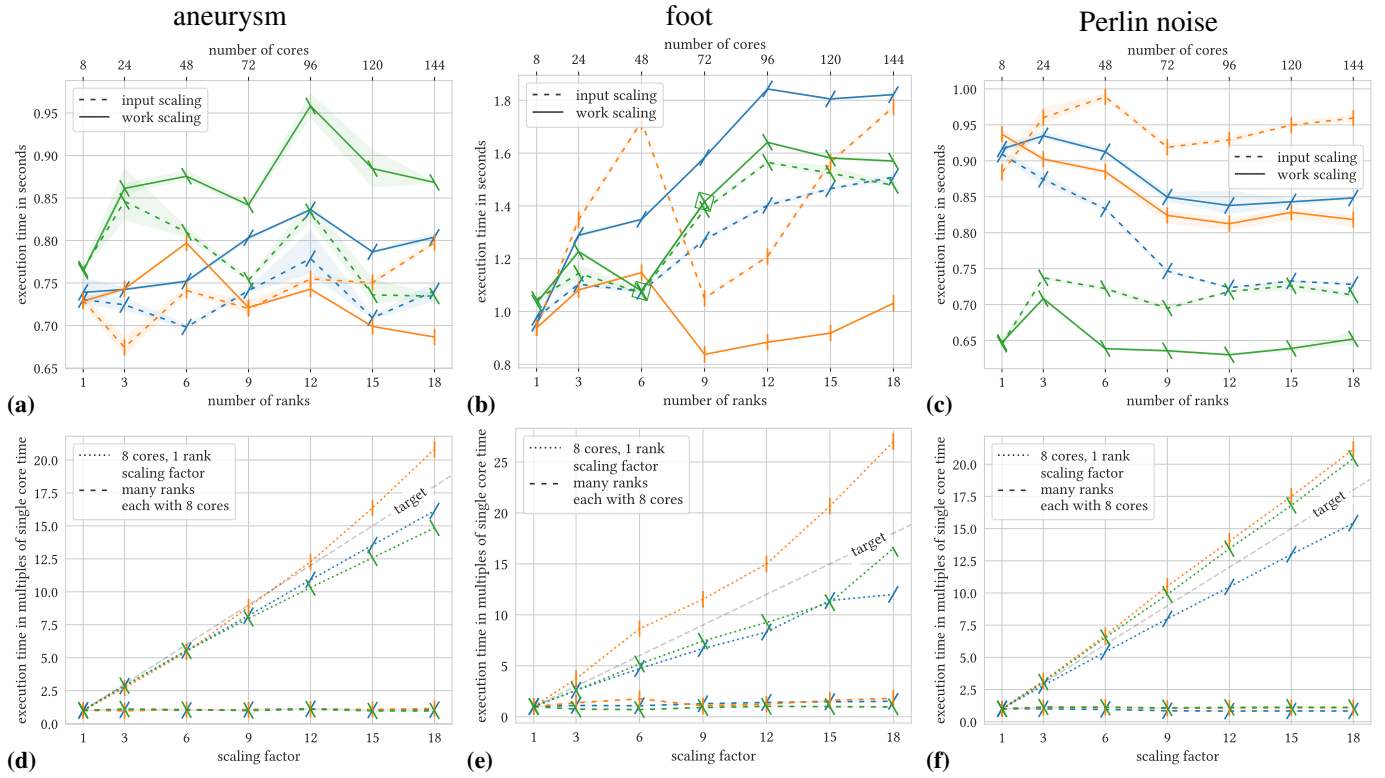
\begin{figure*}
        \centering
        \begin{tikzpicture}[ampersand replacement=\&]
            \matrix [outer sep=0pt] {
              \node[]{data scaling method:}; \&
              \draw[color=resampling, very thick] (0.1,-0.1) -- (0.3,0.1);
              \draw[color=resampling, very thick] (0.0,0.0) -- (+0.4,0.0) node[right,black] {resampling}; \&
              \draw[color=replication, very thick] (0.2,-0.1) -- (0.2,0.1);
              \draw[color=replication, very thick] (0.0,0.0) -- (+0.4,0.0) node[right,black] {replication}; \&
              \draw[color=extent, very thick] (0.3,-0.1) -- (0.1,0.1);
              \draw[color=extent, very thick] (0.0,0.0) -- (+0.4,0.0) node[right,black] {extent}; \\
            };
        \end{tikzpicture}
        \vspace{0.5em}
        
        \begin{subfigure}{0.33\textwidth}
            \centering
            aneurysm
            \vspace{2pt}
            
            \includegraphics[width=\textwidth]{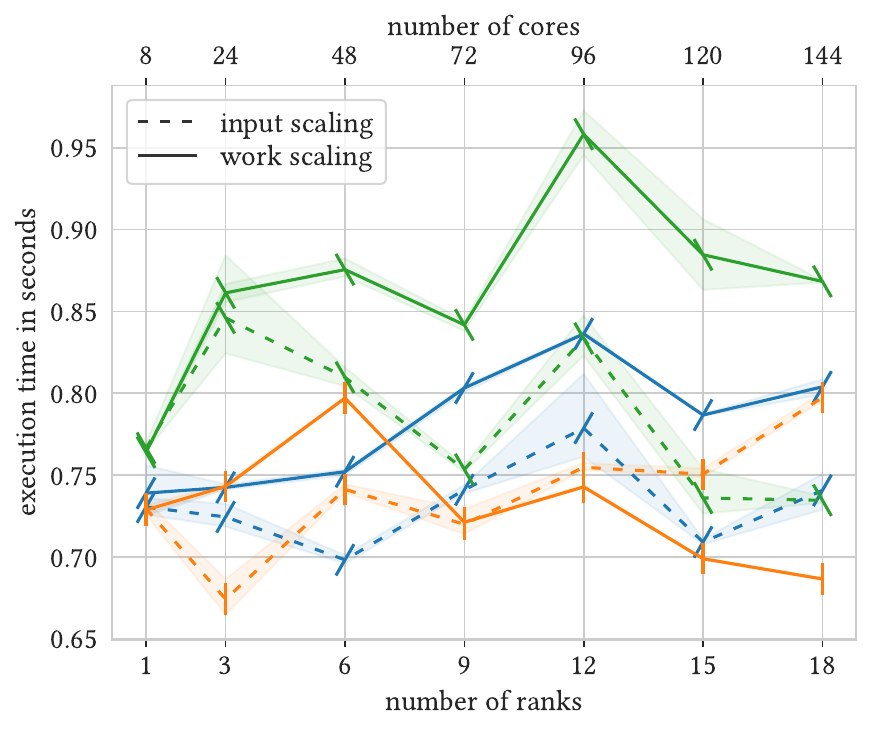}
            \vspace{-1cm}
            \caption{}\label{fig:dist_fe_aneurism_work_vs_input}
        \end{subfigure}
        \begin{subfigure}{0.33\textwidth}
            \centering          
            foot
            \vspace{2pt}
            
            \includegraphics[width=\textwidth]{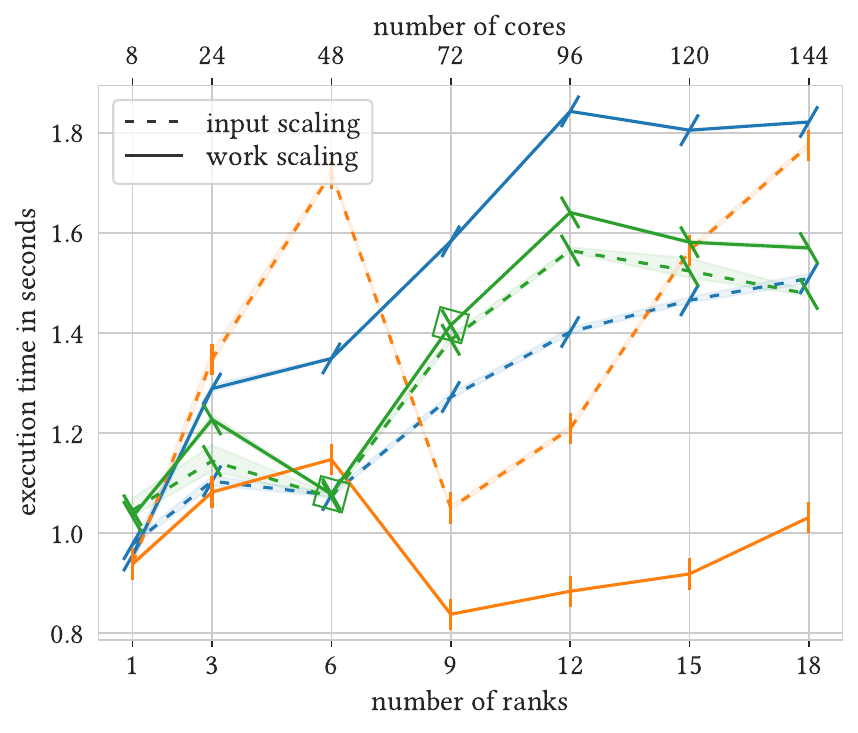}
            \vspace{-1cm}
            \caption{}\label{fig:dist_fe_bones_work_vs_input}
        \end{subfigure}
        \begin{subfigure}{0.33\textwidth}
            \centering
            Perlin noise
            \vspace{2pt}
            
            \includegraphics[width=\textwidth]{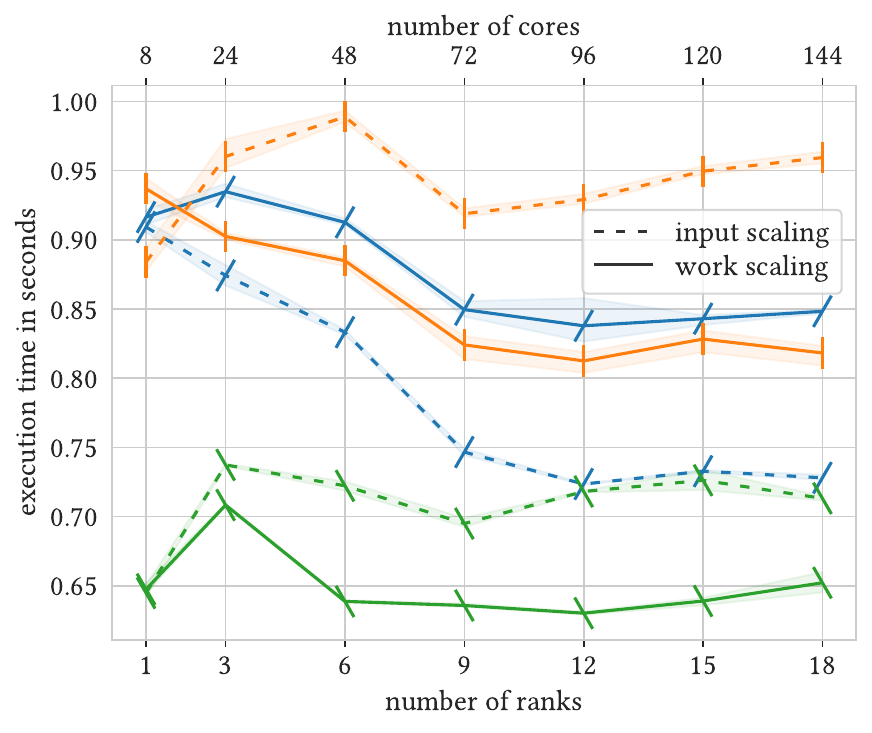}
            \vspace{-1cm}
            \caption{}\label{fig:dist_fe_pn_work_vs_input}
        \end{subfigure}
        \begin{subfigure}{0.33\textwidth}
            \centering
            \includegraphics[width=\textwidth]{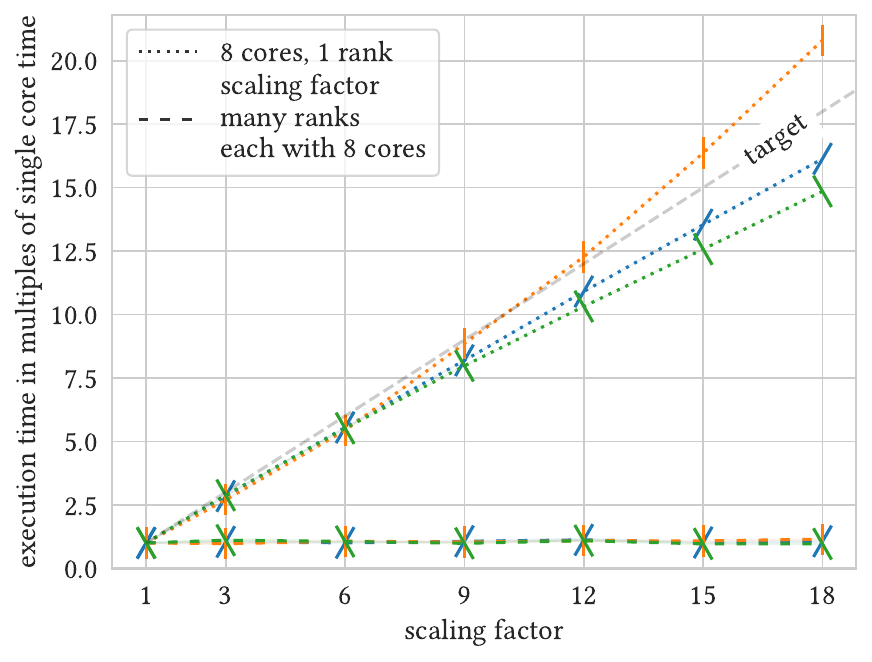}
            \vspace{-1cm}
            \caption{}\label{fig:dist_fe_aneurism_input_scaling}
        \end{subfigure}
        \begin{subfigure}{0.33\textwidth}
            \centering
            \includegraphics[width=\textwidth]{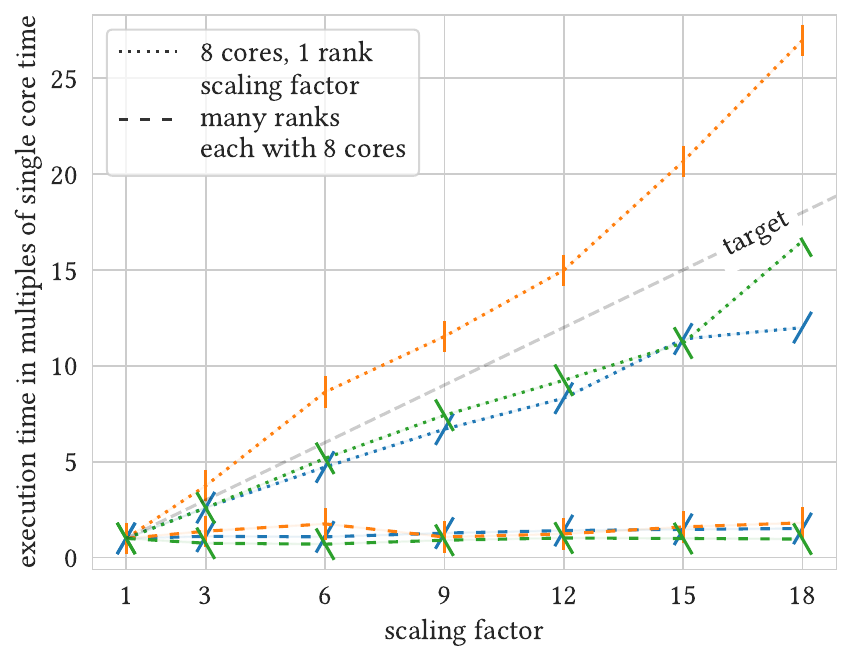}
            \vspace{-1cm}
            \caption{}\label{fig:dist_fe_bones_input_scaling}
        \end{subfigure}
        \begin{subfigure}{0.33\textwidth}
            \centering
            \includegraphics[width=\textwidth]{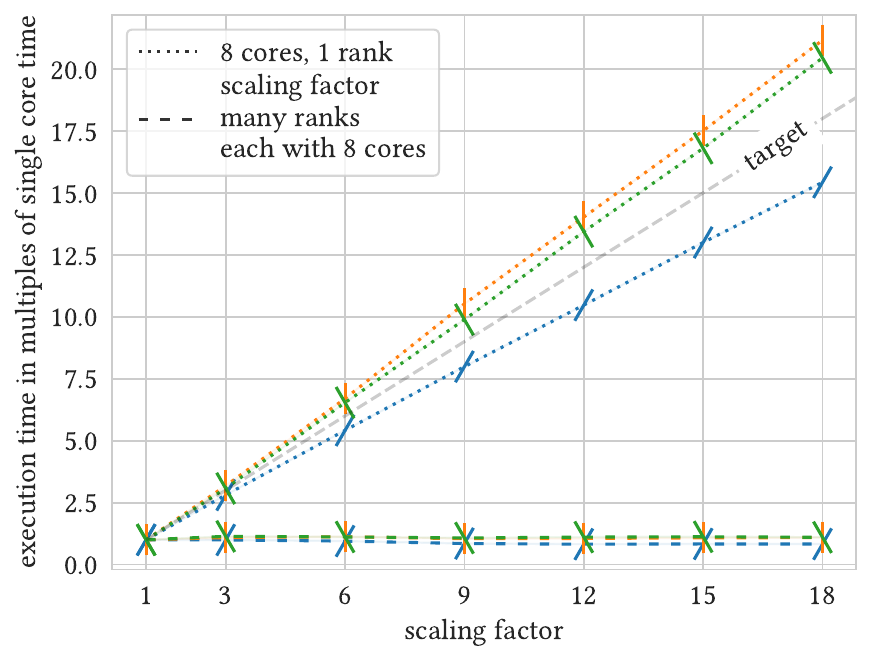}
            \vspace{-1cm}
            \caption{}\label{fig:dist_fe_pn_input_scaling}
        \end{subfigure}
        \caption{Plots showing \emph{work scaling} vs. \emph{input scaling} (top row) and the work approximation of input-size scaling via execution time on a single rank with 8 cores (bottom row) for distributed iso-contour computation. The columns correspond to the different data sets aneurysm, foot and Perlin noise from left to right. In contrast to the shared-memory execution (cf. \autoref{fig:contour_flyingedge}) the execution timings in the aneurysm and foot data set exhibit larger variations across the number of MPI ranks. For Perlin noise this is not as evident.
        }
        \label{fig:dist_fe}
    \end{figure*}

    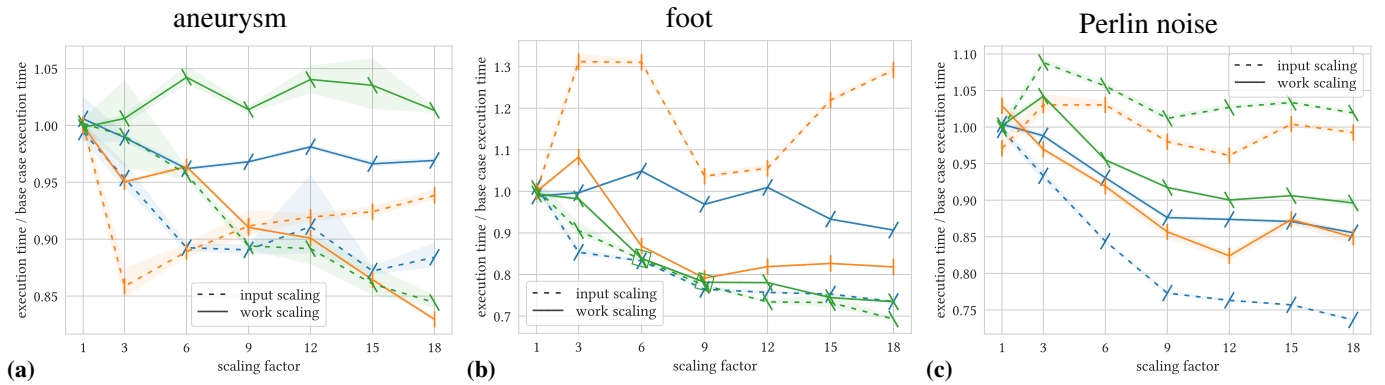
\begin{figure*}
        \centering
        \begin{tikzpicture}[ampersand replacement=\&]
            \matrix [outer sep=0pt] {
              \node[]{data scaling method:}; \&
              \draw[color=resampling, very thick] (0.1,-0.1) -- (0.3,0.1);
              \draw[color=resampling, very thick] (0.0,0.0) -- (+0.4,0.0) node[right,black] {resampling}; \&
              \draw[color=replication, very thick] (0.2,-0.1) -- (0.2,0.1);
              \draw[color=replication, very thick] (0.0,0.0) -- (+0.4,0.0) node[right,black] {replication}; \&
              \draw[color=extent, very thick] (0.3,-0.1) -- (0.1,0.1);
              \draw[color=extent, very thick] (0.0,0.0) -- (+0.4,0.0) node[right,black] {extent}; \\
            };
        \end{tikzpicture}
        \vspace{0.5em}
        
        \begin{subfigure}{0.33\textwidth}
            \centering
            aneurysm
            \vspace{2pt}
            
            \includegraphics[width=\textwidth]{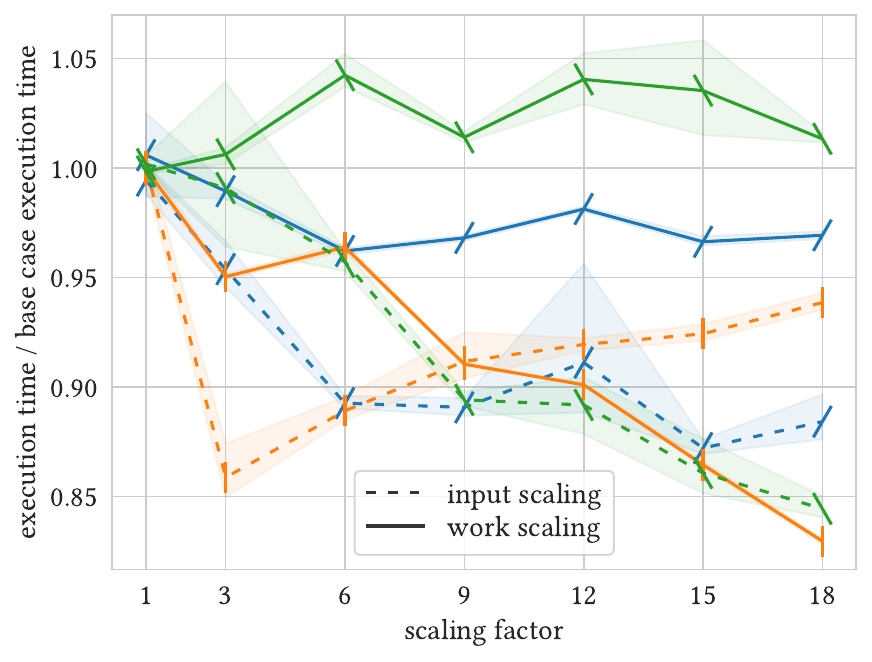}
            \vspace{-1cm}
            \caption{}\label{fig:dist_fe_aneurism_work_vs_input_lbc}
        \end{subfigure}
        \begin{subfigure}{0.33\textwidth}
            \centering          
            foot
            \vspace{2pt}
            
            \includegraphics[width=\textwidth]{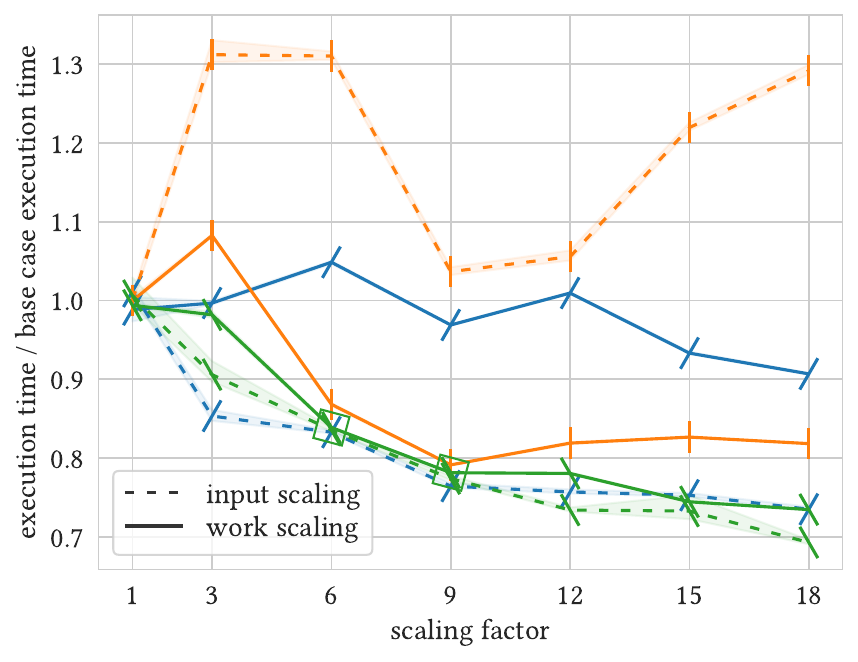}
            \vspace{-1cm}
            \caption{}\label{fig:dist_fe_bones_work_vs_input_lbc}
        \end{subfigure}
        \begin{subfigure}{0.33\textwidth}
            \centering
            Perlin noise
            \vspace{2pt}
            
            \includegraphics[width=\textwidth]{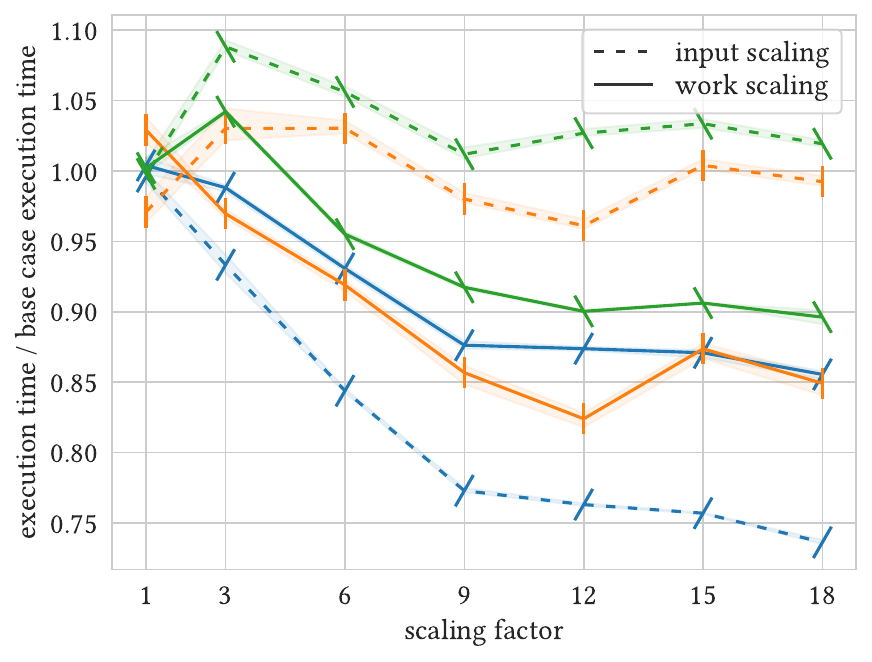}
            \vspace{-1cm}
            \caption{}\label{fig:dist_fe_pn_work_vs_input_lbc}
        \end{subfigure}       
        \caption{
        Plots showing \emph{work scaling} vs. \emph{input scaling} with the average execution time of MPI ranks in multiples of the base case for distributed iso-contour computation execution. The columns correspond to the different data sets aneurysm, foot and Perlin noise from left to right. Averaging over the MPI ranks execution timings seems to have a larger effect on the aneurysm and foot data set (cf. \autoref{fig:dist_fe}). Overall, the averaging leads to fewer execution time variations over different number of MPI ranks. The \emph{work scaling} method appears more consistent for the Perlin noise and foot data set in the case of excluding differences in execution time of MPI ranks.
        }
        \label{fig:dist_fe_lbc}
    \end{figure*}

    In this work we largely excluded distributed execution. Here we report on preliminary findings of the \emph{work scaling} method in a distributed execution environment for iso-contour extraction and further give reasons on the difficulties and the potential for future work.

    The experiment setup is similar to the shared-memory experiments. The size of the base data set is resampled to $512^3$ and in the case of extent scaling this base data set was up sampled with a scaling factor of $22$ before extracting the \emph{extent scaled} subvolumes. Replication was applied in all three dimensions. We repeated every measurement five times. The used machines are 1-6 nodes each with two Intel XEON 6126 processors and 96GB of memory.
    The base execution time is not computed on a single core but on eight, albeit on a single MPI rank. This allows to scale to higher core counts due to the decreased base case execution time. There is a small overhead due to the parallelization on eight cores instead of one, but the impact is considerably small compared to distribution across multiple nodes requiring communication. In addition to the total execution time of a method we also gather the execution time per rank.

    The first observation is a similarly skewed work increase of the \emph{input-scaling} method as in the shared-memory setting (cf.\ \autoref{fig:dist_fe_aneurism_input_scaling},\ref{fig:dist_fe_bones_input_scaling},\ref{fig:dist_fe_pn_input_scaling} with \autoref{fig:fe_aneurism_input_scaling},\ref{fig:fe_bones_input_scaling},\ref{fig:fe_pn_input_scaling}). Due to the similar means of computation with no communication overhead and in the distributed case low parallel overhead, this is expected. However, when applying the method to the three exemplary data sets we find that the methods efficacy varies strongly. For Perlin noise the result is similar to shared memory, leading to more consistent weak scalability results across data scaling methods. The same is not true for the \emph{foot} and \emph{aneurysm} data set. The results are not more consistent and overall seem to differ strongly depending on the amount of resources. For these cases larger load imbalances across the MPI ranks can be noticed. 

    When comparing the plots for the total execution time (see \autoref{fig:dist_fe}) with plots showing the averaged execution time per rank (see \autoref{fig:dist_fe_lbc}), one can observe a large difference for the \emph{foot} and \emph{aneurism} data set, but not for the Perlin noise data set. This partially explains the seemingly noisy execution times in \autoref{fig:dist_fe}. Moreover, this also reveals that data dependent execution times can induce load imbalances even if the input data is evenly distributed. Since cells containing the iso-value are evenly spread in the Perlin noise data set, this effect is only small. But for the other two data sets with uneven dsitribution of features (cf. \autoref{fig:datasets}) this has a large effect.

\section{Discussion}\label{sec:discussion}

    It is our interpretation that there are two main motivators for weak scalability studies. The first motivation is to observe and analyze the behavior of an algorithm on specific data sets akin to a case study. The second motivation is to quantify the parallel overhead of a specific approach and observe its development in a weak scaling setting, ideally as general as possible. Both approaches aim to investigate the behavior of a specific approach as data set size and the amount of computing resources increase. Although seemingly similar, these approaches differ in important aspects.

    In order to quantify the  parallel overhead, which is characterized by the additional work and latencies introduced by parallel computation, the actual work on the problem has to be known, when observing total execution time.
    This implies that input scaling is only suitable for investigations following the first motivation, since the actual work of the problem can be unknown or uncertain.
    The experimental results demonstrate that traditional weak scaling experiments of data dependent algorithms can differ largely depending on the choice of data and the scaling method to generate variable size data sets. The scaling behavior depends on the actual work which in turn depends on the data for specific algorithms. Thus different scaling methods are suited better or worse for different data and algorithms. Unfortunately, none of the investigated data scaling methods are superior, meaning more consistent across data and algorithms, to the others in all cases. \emph{Resampling} suffers from smoothing, but produces continuous fields in contrast to \emph{replication}. \emph{Extent} scaling generates continuous fields and does not change the original data, however, it may suffer from uneven distribution of features, which is not the case with \emph{resampling} and \emph{replication}.
    Due to this complexity, it is necessary to reassure that the scaling is a valid comparison between workloads to get generally applicable results.

    One large unknown in this discussion and other contributions in this topic is, whether data actually scales in the directions investigated by weak scalability studies. This depends, among other things, on the area of application and data sources. 
    And while the results of this study indicate that some weak scalability benchmarks in literature were not executed on proportionally scaled problem sizes with respect to work, this does not invalidate them, since they describe a possible scenario of data set development. This is especially the case if a reasonable use case or prediction on development of future data sets is provided. Furthermore, comparisons of weak scalability between algorithms on the same data sets can still showcase algorithmic improvements of one over another.

    Nonetheless, they are not necessarily suitable for quantifying the parallel overhead.
    In order to provide relevant analysis of parallelization overhead in weak scalability studies, the amount of work for each measurement should be incorporated into the analysis. 
    With the method presented in \autoref{sec:workscaling}, we can observe that \emph{work scaling} via execution time on a single core adequately scales the problem size, leading to more consistent results in a shared memory setting. It can compensate for scaling methods introducing too much or not enough work, with the caveat that scaling methods introducing not enough work possibly lead to very large data sets. 
    Although, most weak scalability benchmarks with our methods were more consistent across data scaling methods, \emph{extent} scaling poses an exception.
    As explained in \autoref{sec:ex_ic}, the observed differences with \emph{extent} scaling presumably stem from the differing base case.
    
    With this method, it is easier to distinguish general algorithmic improvements or the influence of a specific data set from a reduction in parallelization overhead. 
    However, we also noticed that this method does not always yield meaningful results. This is the case for methods with execution times that depend on the input size by only a small degree, like ray casting. And exemplarily, it also does not lead to more consistent results in distributed executions.

    Keeping the work consistent across scaling approaches is necessary but not sufficient for consistent weak scalability benchmarks. Some work in a serial computation can induce different amounts of parallel overhead, although it yields equivalent execution time measurements on a single core. This leads to different weak scalability results with base problems of the same size. 
    This is not as apparent in shared-memory applications, but is emphasized for distributed applications with static load distribution.
    Unfavorable data distribution leads to data parts that contain vastly more work than others. Since this is not apparent on a single node or core it is not accounted for in the scaling.
    However, the costs of efficient data distribution, load balancing or the absence of it are a part of parallel overhead. The measurements are still meaningful, however in many cases they will not be more consistent, since the inconsistency stems from the data dependence of the method and its parallelization itself and not the scaling.

    As a guideline, we propose a classification of algorithms, that helps to contextualize scalability results and supports the creation of more robust assessments.
    \textbf{Strictly input dependent algorithms with linear complexity} with respect to the input data size are the first and simplest class. The traditional weak scalability approach should work in assessing parallel overhead and ability to scale.
    For \textbf{strictly input dependent algorithm with non-linear complexity} with respect to the input data size, already special care has to be taken. The observed traditional weak scalability will also be dependent on the general algorithmic complexity, not only parallel overhead if the problem size is scaled linearly. In this case, careful scaling of the problem size according to the algorithmic complexity could be a solution, when aiming to analyze the parallel overhead.
    \textbf{Data or output dependent algorithms} like iso-contour extraction and volume rendering via ray casting may show varying results across data sets. The observed traditional weak scalability may contain general algorithmic complexity that depends on the composition of the data, not only parallel overhead. Special care must be taken when artificially creating data sets for higher core counts. Finding a meaningful rate as replacement for execution time could also pose a solution.
    \textbf{Data or output dependent algorithms with data dependent parallelization overhead} are even more intricate. In this class the parallel overhead is not only dependent on the algorithm itself and the amount of available resources but additionally on the composition of the input data. A prominent example is the distributed computation of streamlines. In a parallelize-over-data approach, streamlines that exit the local domain induce communication overhead. How often this occurs also depends on the underlying vector field, leading to a dependency of the parallel overhead on the data. Unfortunately, also algorithms of the previous class can exhibit a data dependent parallelization overhead by using a static data distribution scheme, as discussed before.
    For these problems it seems that traditional weak scalability might be too simplistic and more nuanced performance models should be employed that incorporate the individual algorithm's sources of parallel overhead.

    For now, the different data dependent aspects of any algorithm should ideally be part of the algorithms scalability or performance assessment. At least using multiple data sets with varying feature density across the domain is recommended. Future work could explore general approaches to the aforementioned classes.

    Although we only considered uniform grids, inconsistent workload increases can generally appear with other data structures too. For example in surface rendering, if added cells are occluded by others, presumably less work has to be done in comparison to visible cells. Whether added cells are occluded or not, again depends on the method of scaling the data, meaning adding those cells. For data structures that do not have constant access time, but depend on the size of the data, like octrees or BVH, this would have to be taken into account too. So the observations made here, are only a step towards scaling the workload for different data types and algorithms in visualization. Since the findings presented here are only exemplified through scientific visualization methods, they may also be transferable to other domains with data dependent algorithms.

\subsection{Limitations}\label{sec:limitations}

    The iterative computation in \autoref{sec:workscaling} does not precisely compute the scaling factor for a work-scaled data set in the sense of Gustafson. Ideally, we want to compute $s+Np$ with $s+p$ being the execution time on a single core. Actually, we scale $X$ until $\frac{s+Xp}{s+p} = N$ for a given $N$. The error can be described by:
    \begin{align*}
        \frac{s+Xp}{s+p} &= N \\
        s+Xp &= N(s+p) \\
        Xp-Ns-Np &= -s \\
        X-N-\frac{Ns}{p} &= \frac{-s}{p} \\
        X-N &= \frac{s(N-1)}{p}
    \end{align*}
    This means that $X$ is \emph{not} equal to $N$, but for small values of $s$ this error is negligible for our purpose. To overcome this inaccuracy, one would have to determine $s$ and $p$ beforehand and use them to replace $N$ with $N'$ such that $X=N$. However, evaluating $s$ and $p$ empirically proves difficult due to their dependency on the problem size as described by \cite{gustafson1988reevaluating}. 

    The observed challenge in comparability for \emph{extent} scaling to the other methods reveals another limitation.
    Comparing different algorithms scalability with \emph{work scaled} data is not directly possible. While a difference in execution time between two algorithms for the base cases could be counteracted by reporting execution time in multiples of the execution time for the base case instead, \emph{work scaling} may lead to differently scaled problem sizes for each algorithm. 
    However, the main motivation behind this method is to investigate the parallelism induced overhead of a single algorithm more consistently across different data sets and scaling methods. Also parallel overhead can be compared, but problem size for the instances have to be considered in conjunction. As demonstrated, the problem size depends on more than just input size, so one could argue that comparisons of algorithms on different \emph{input scaled} data sets are also not as easily comparable as they seem.
    
    A common challenge with scaling data on one core is that data often exceeds memory limits with larger scaling factors. This is less of a problem for shared-memory weak scalability analysis, as done here, since every instance has to fit into the memory anyway. But it can pose a challenge when extending this method to distributed executions. A possible solution could be to compute the data on the smallest machine possible and scale it there. This would lead to shorter strips of weak scaling experiments (e.g. 1-4, 4-16, 16-64 nodes and so on), where a base timing for scaling is retrieved from the lowest core count of the current strip. A somewhat similar approach has been proposed by \cite{moreland2015formal} as a replacement for weak scaling, where several strong scaling experiments are executed at different data sizes.

    Another limitation is time. Executing an algorithm on a single core on a data set that is supposed to be processed by multiple cores, can take a long time. The approach for memory exceeding data sets could also be used here with the same caveat that some parallel overhead is included in the base cases. Ideally, alternative, less computation intensive methods for determining the work on scaled data sets could be devised as future work.

    Other relevant factors in practice, like cache behavior, are not considered here. Although they can influence the weak scalability results to a large degree, taking them into account in scalability studies is orthogonal to the presented challenge and approach.

    Given these restrictions and additional considerations, we still believe this method can be useful for estimating the increase in work for specific data sets and scaling approaches, leading to more consistent weak scalability studies in shared-memory environments, even if they are only applied to smaller core counts.

\section{Conclusion}
    
    In this paper we revealed that current studies of weak scalability often do not cover the nuances of workload increases in data for data dependent algorithms. 
    The typical approach of increasing the size of the input does not suffice to lead to an increase in the workload for all methods. 
    Since weak scalability is investigated by keeping the work per computing resource constant while increasing the amount thereof, using the input size of a problem as a stand-in for workload, might lead to less meaningful results.
    
    We were able to show on three examples that the choice of data scaling method and the data set has a large impact on the assessed weak scalability for data dependent algorithms. 
    In order to mitigate this we have presented a method that yields more consistent weak scalability results over the different data scaling methods. 
    The method achieves this by estimating the workload contained in a data set by measuring execution time on a single core and comparing it to the execution time on a base data set.
    The conducted experiments revealed shortcomings of this method for different kinds of algorithms. 
    From this we derived a classification of algorithms to guide future scalability research and improve the expressiveness of weak scalability studies.
    
    We hope to have sparked the discussion on improving scalability assessments for future visualization algorithms and methods.

\section*{Supplemental Material}
    In addition to the algorithms discussed in the main experiments section, we evaluated several others.
    For iso-contour extraction we investigate a marching cubes implementation \citep{lorensen1987} in addition to the flying edge implementation. 
    We also include all benchmark plots related to volume rendering.
    Furthermore, we present benchmarks for an alternative augmented contour tree extraction given by the FTM algorithm by \cite{gueunet2017} available in TTK \citep{ttk}.
    All additional results and comparisons are provided in the supplemental material.

    The source code for the experiments is available at \url{https://github.com/scivislab/weak-scaling-visualization-algorithms} with instructions on how to replicate the experiments.
    A suitable execution environment is provided via docker container.

\section*{Acknowledgment}
    The authors thank Florian Wetzels for helpful discussions and gratefully acknowledge the German Federal Ministry of Education and Research (BMBF) and the state government of RP for supporting this work as part of the NHR funding.

\bibliographystyle{IEEEtranN}
\bibliography{references}

\clearpage

\onecolumn 

\setcounter{section}{0}
\setcounter{figure}{0}
\setcounter{table}{0}
\setcounter{page}{1}

\renewcommand{\thesection}{S.\Roman{section}}
\renewcommand{\thefigure}{S\arabic{figure}}
\renewcommand{\thetable}{S\arabic{table}}
\renewcommand{\thepage}{S\arabic{page}}

\section{Supplemental Material: Additional Experiments}

    \begin{figure*}[h]
        \centering
        \begin{tikzpicture}[ampersand replacement=\&]
            \matrix [outer sep=0pt] {
              \node[]{data scaling method:}; \&
              \draw[color=resampling, very thick] (0.1,-0.1) -- (0.3,0.1);
              \draw[color=resampling, very thick] (0.0,0.0) -- (+0.4,0.0) node[right,black] {resampling}; \&
              \draw[color=replication, very thick] (0.2,-0.1) -- (0.2,0.1);
              \draw[color=replication, very thick] (0.0,0.0) -- (+0.4,0.0) node[right,black] {replication}; \&
              \draw[color=extent, very thick] (0.3,-0.1) -- (0.1,0.1);
              \draw[color=extent, very thick] (0.0,0.0) -- (+0.4,0.0) node[right,black] {extent}; \\
            };
        \end{tikzpicture}
        \vspace{0.5em}

        \begin{subfigure}[h]{0.33\textwidth}%
            \centering
            aneurysm
            \vspace{2pt}

            \includegraphics[width=\textwidth]{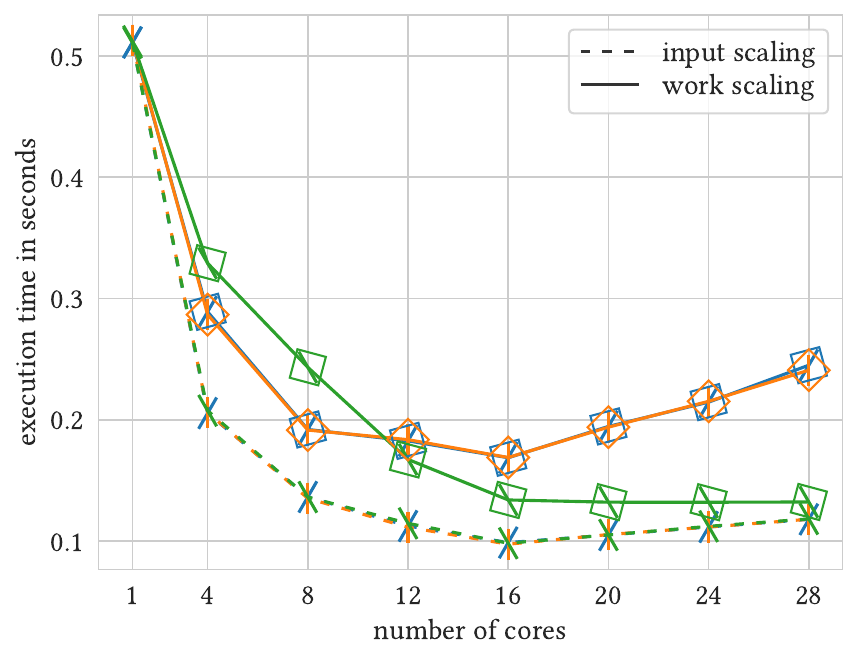}
            \vspace{-1cm}
            \caption{}\label{fig:apdx_vr_aneurism_work_vs_input}
        \end{subfigure}
        \begin{subfigure}[h]{0.33\textwidth}%
            \centering
            foot
            \vspace{2pt}

            \includegraphics[width=\textwidth]{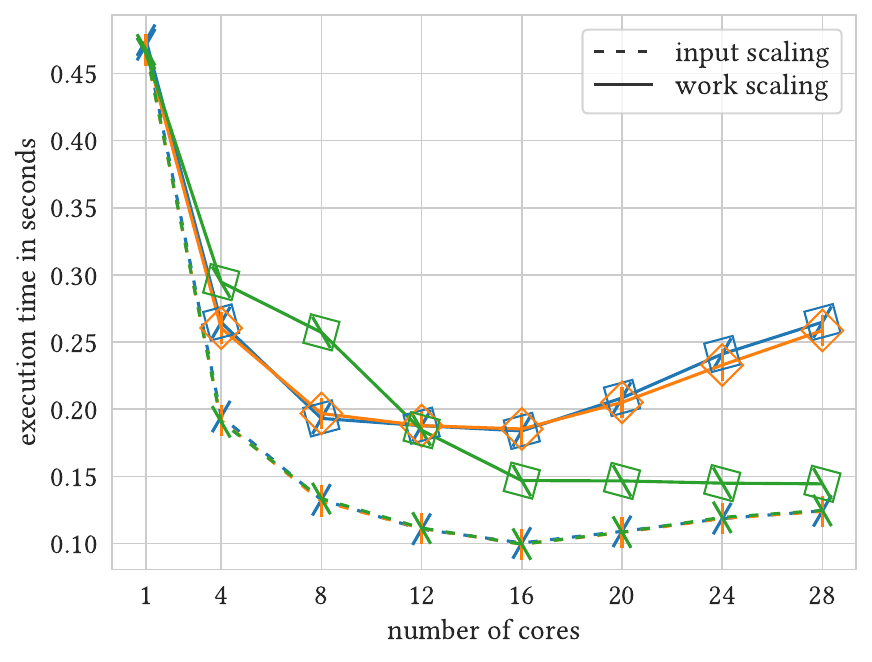}
            \vspace{-1cm}
            \caption{}\label{fig:apdx_vr_bones_work_vs_input}
        \end{subfigure}
        \begin{subfigure}[h]{0.33\textwidth}%
            \centering
            Perlin noise
            \vspace{2pt}

            \includegraphics[width=\textwidth]{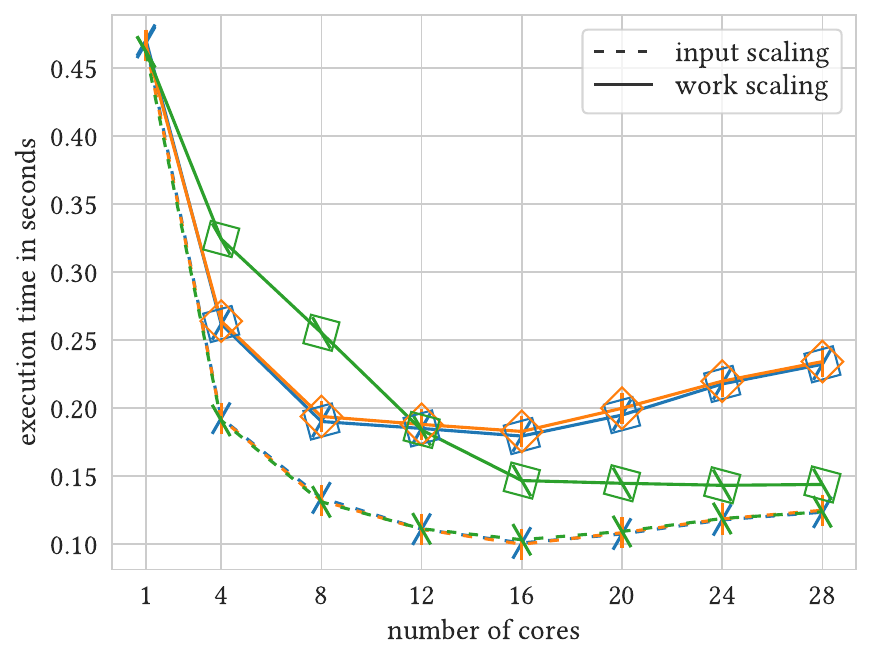}
            \vspace{-1cm}
            \caption{}\label{fig:apdx_vr_pn_work_vs_input}
        \end{subfigure}
        \begin{subfigure}[h]{0.33\textwidth}%
            \centering
            \includegraphics[width=\textwidth]{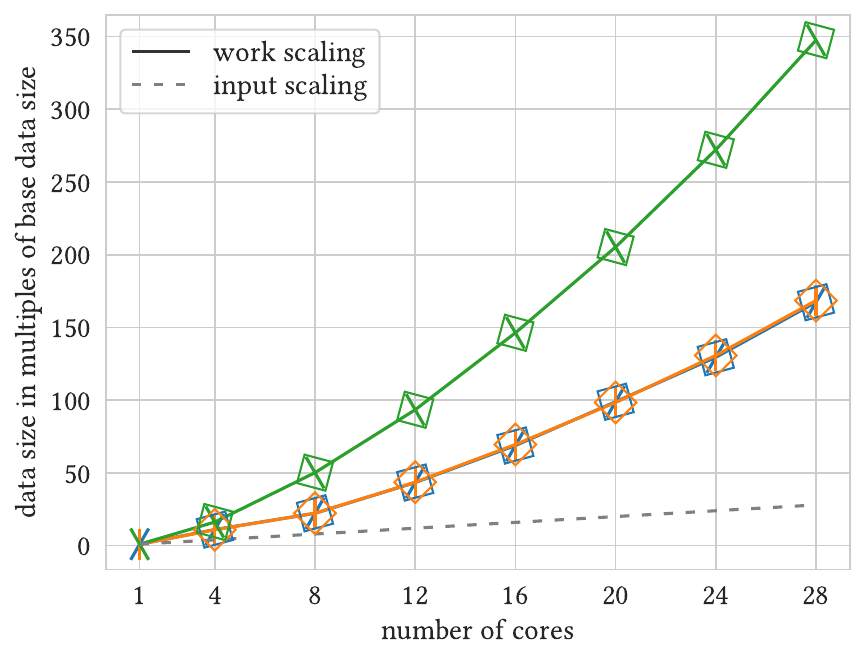}
            \vspace{-1cm}
            \caption{}\label{fig:apdx_vr_aneurism_sf}
        \end{subfigure}
        \begin{subfigure}[h]{0.33\textwidth}%
            \centering
            \includegraphics[width=\textwidth]{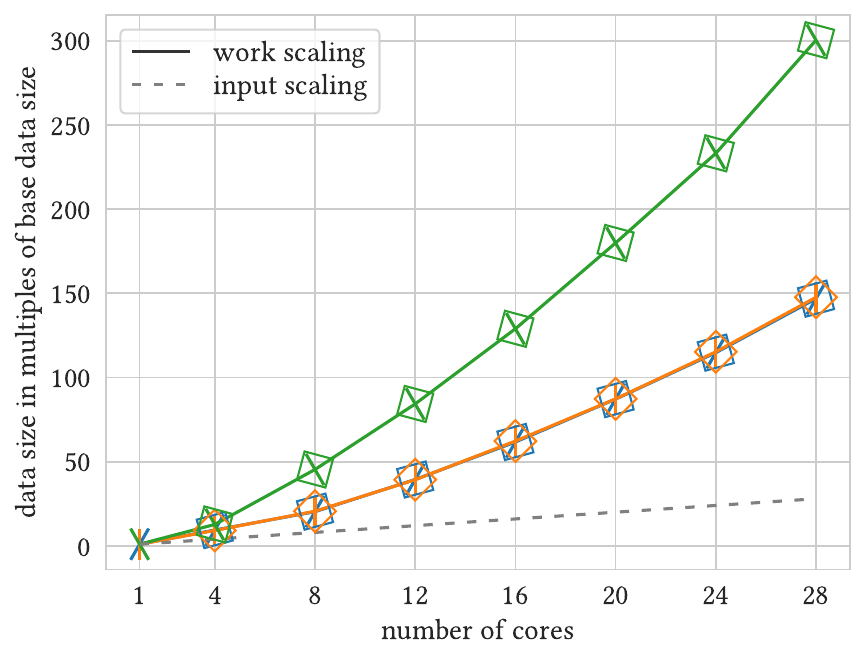}
            \vspace{-1cm}
            \caption{}\label{fig:apdx_vr_bones_sf}
        \end{subfigure}
        \begin{subfigure}[h]{0.33\textwidth}%
            \centering
            \includegraphics[width=\textwidth]{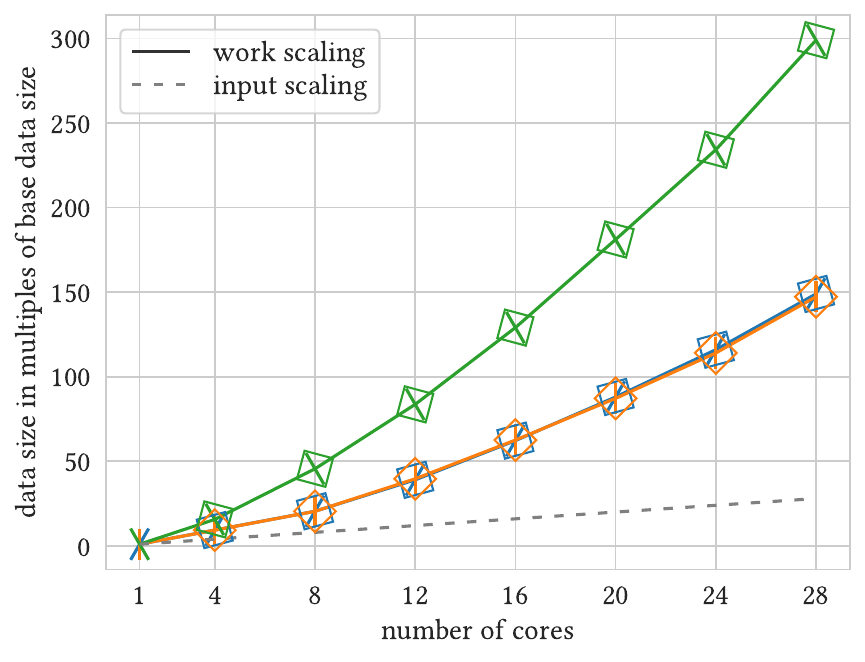}
            \vspace{-1cm}
            \caption{}\label{fig:apdx_vr_pn_sf}
        \end{subfigure}
        \begin{subfigure}[h]{0.33\textwidth}%
            \centering
            \includegraphics[width=\textwidth]{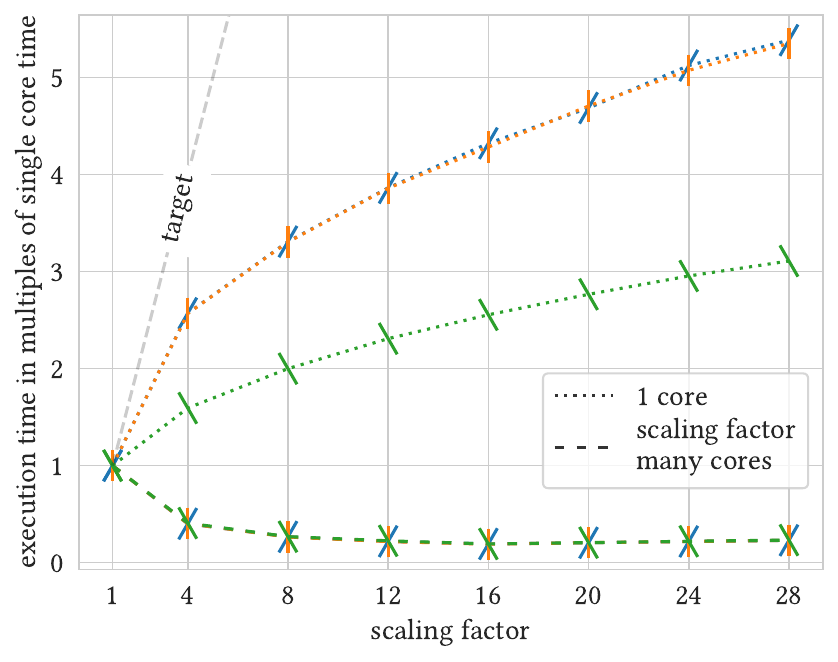}
            \vspace{-1cm}
            \caption{}\label{fig:apdx_vr_aneurism_input_scaling}
        \end{subfigure} 
        \begin{subfigure}[h]{0.33\textwidth}%
            \centering
            \includegraphics[width=\textwidth]{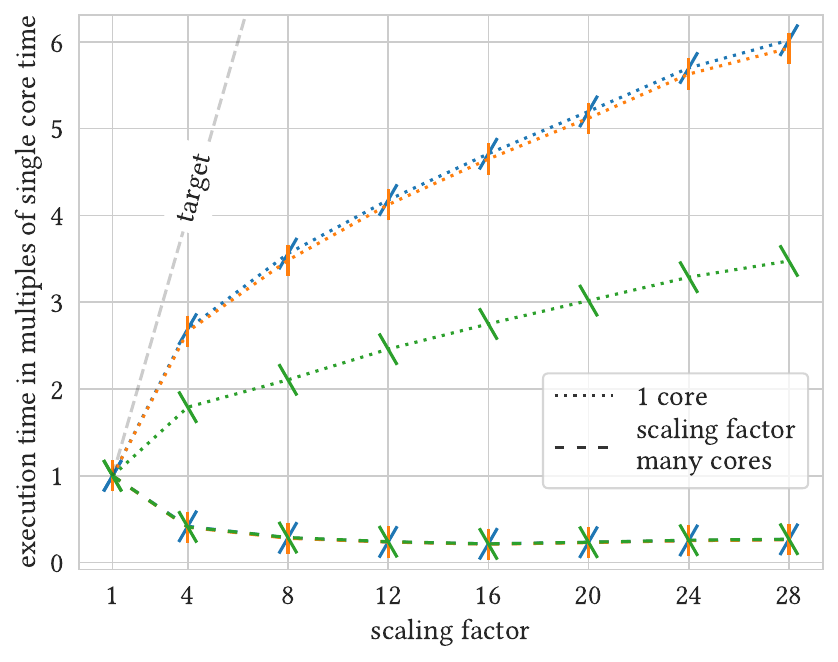}
            \vspace{-1cm}
            \caption{}\label{fig:apdx_vr_bones_input_scaling}
        \end{subfigure}
        \begin{subfigure}[h]{0.33\textwidth}%
            \centering
            \includegraphics[width=\textwidth]{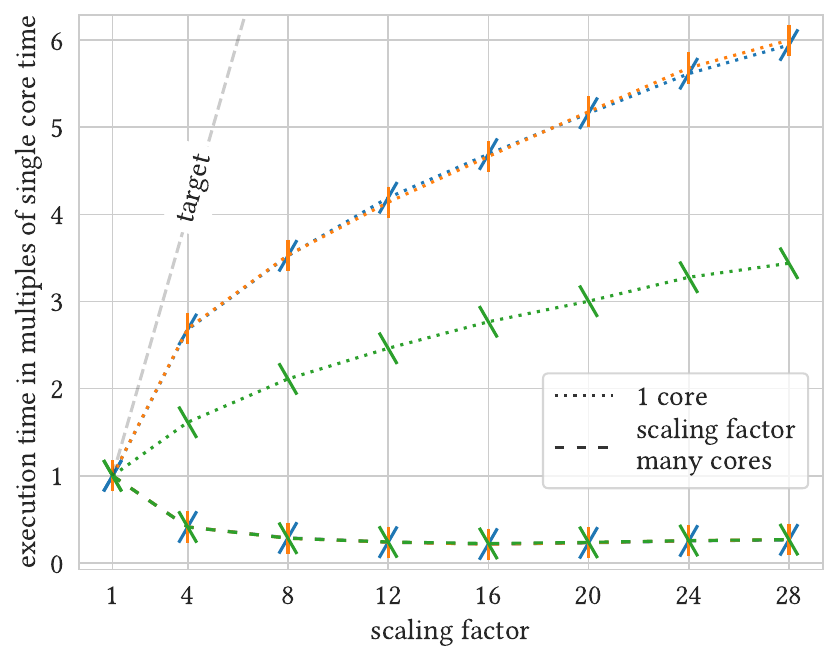}
            \vspace{-1cm}
            \caption{}\label{fig:apdx_vr_pn_input_scaling}
        \end{subfigure} 
        \caption{All volume rendering plots showing \emph{work scaling} vs. \emph{input scaling} (left), the corresponding scaling factor (middle) and the work approximation of \emph{input scaling} via execution time on a single core (right). The columns correspond to the different data sets aneurysm, foot and Perlin noise from left to right. As discussed in the main work, the differences across the data sets are marginal.
        }
        \label{fig:appendix_volumerender_complete}
    \end{figure*}

    \begin{figure*}
        \centering
        \begin{tikzpicture}[ampersand replacement=\&]
            \matrix [outer sep=0pt] {
                \node[]{data scaling method:}; \&
                \draw[color=resampling, very thick] (0.1,-0.1) -- (0.3,0.1);
                \draw[color=resampling, very thick] (0.0,0.0) -- (+0.4,0.0) node[right,black] {resampling}; \&
                \draw[color=replication, very thick] (0.2,-0.1) -- (0.2,0.1);
                \draw[color=replication, very thick] (0.0,0.0) -- (+0.4,0.0) node[right,black] {replication}; \&
                \draw[color=extent, very thick] (0.3,-0.1) -- (0.1,0.1);
                \draw[color=extent, very thick] (0.0,0.0) -- (+0.4,0.0) node[right,black] {extent}; \\
            };
        \end{tikzpicture}
        \vspace{0.5em}

        \begin{subfigure}[h]{0.33\textwidth}%
            \centering
            aneurysm
            \vspace{2pt}

            \includegraphics[width=\textwidth]{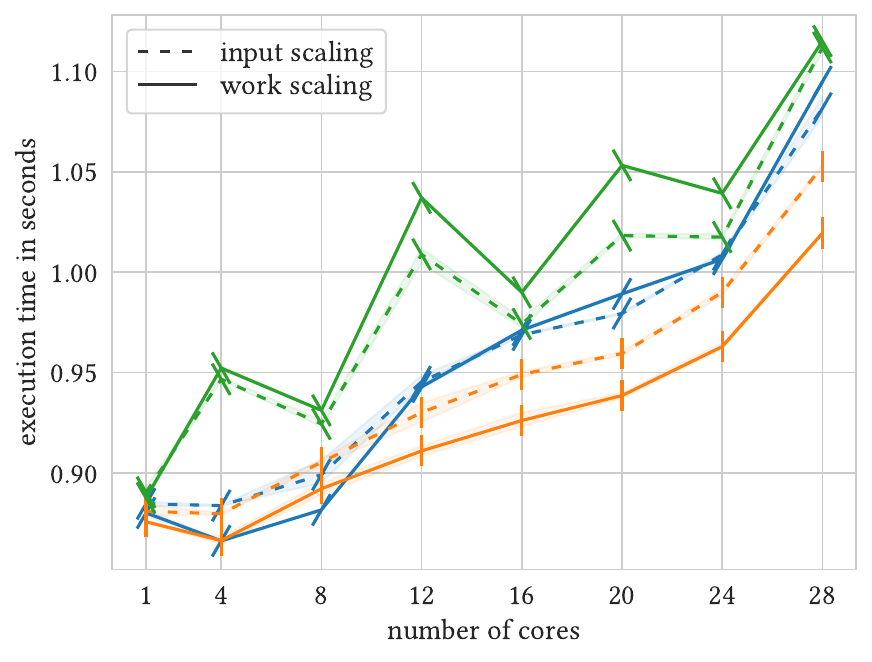}
            \vspace{-1cm}
            \caption{}\label{fig:mc_aneurism_work_vs_input}
        \end{subfigure}
        \begin{subfigure}[h]{0.33\textwidth}%
            \centering
            foot
            \vspace{2pt}

            \includegraphics[width=\textwidth]{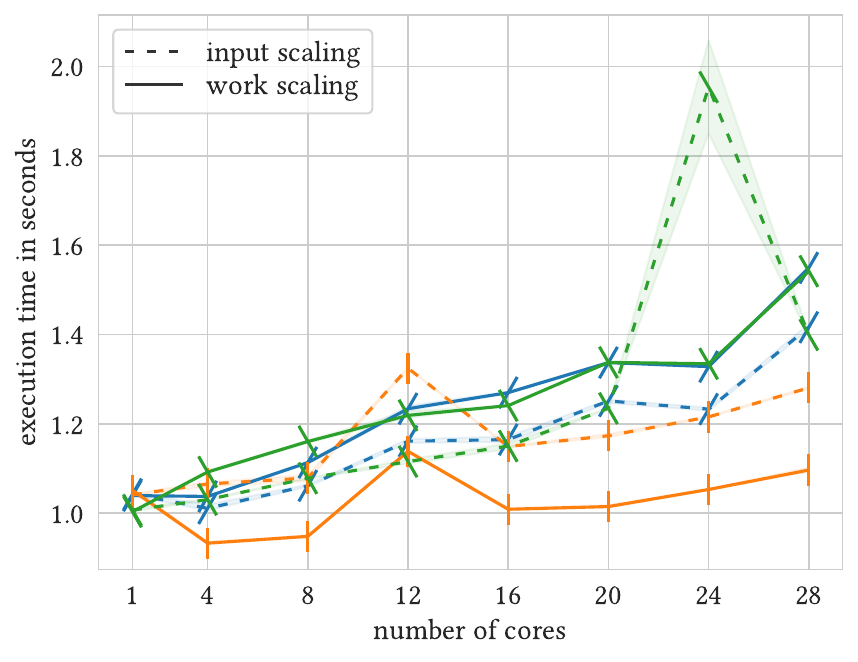}
            \vspace{-1cm}
            \caption{}\label{fig:mc_bones_work_vs_input}
        \end{subfigure}
        \begin{subfigure}[h]{0.33\textwidth}%
            \centering
            Perlin noise
            \vspace{2pt}

            \includegraphics[width=\textwidth]{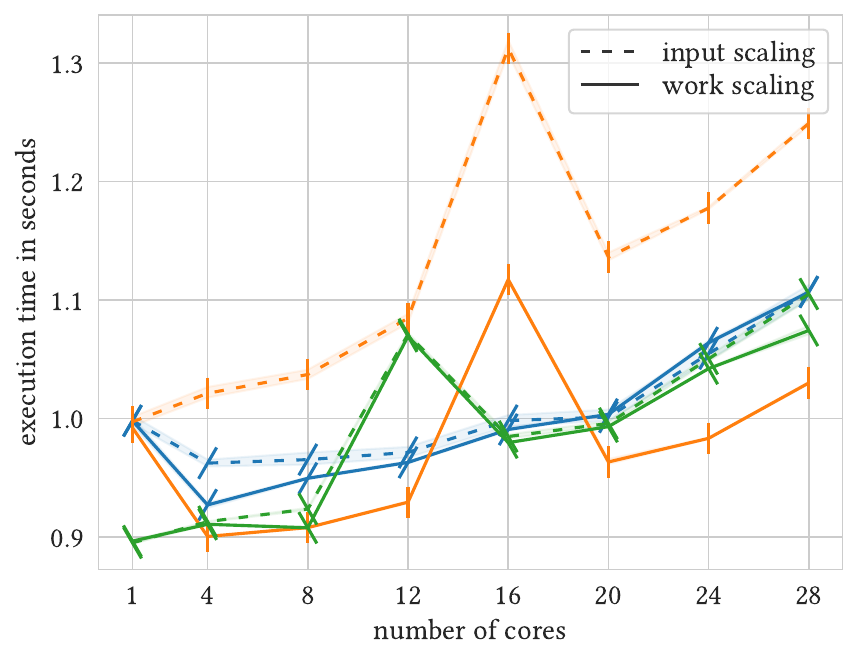}
            \vspace{-1cm}
            \caption{}\label{fig:mc_pn_work_vs_input}
        \end{subfigure}
        \begin{subfigure}[h]{0.33\textwidth}%
            \centering
            \includegraphics[width=\textwidth]{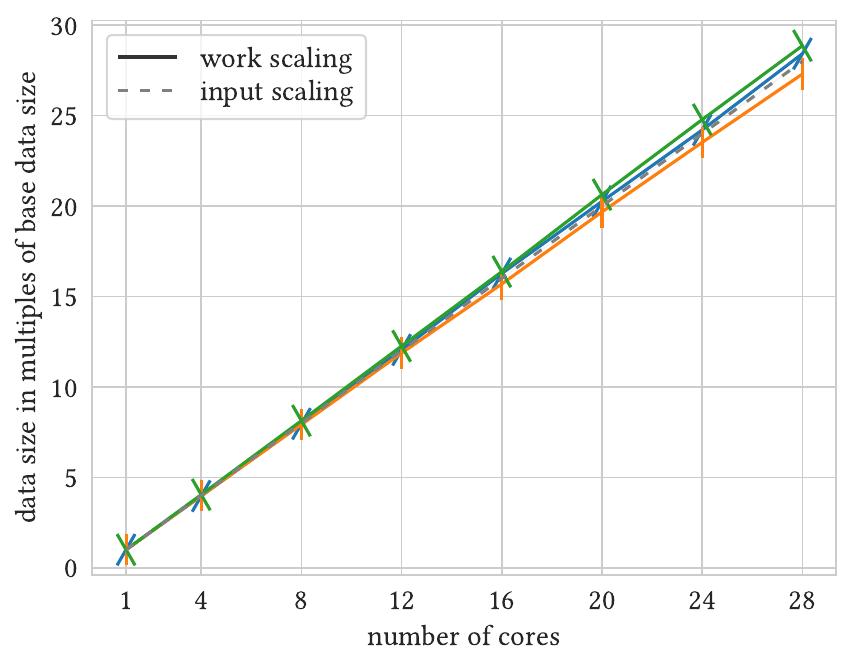}
            \vspace{-1cm}
            \caption{}\label{fig:mc_aneurism_sf}
        \end{subfigure}
        \begin{subfigure}[h]{0.33\textwidth}%
            \centering
            \includegraphics[width=\textwidth]{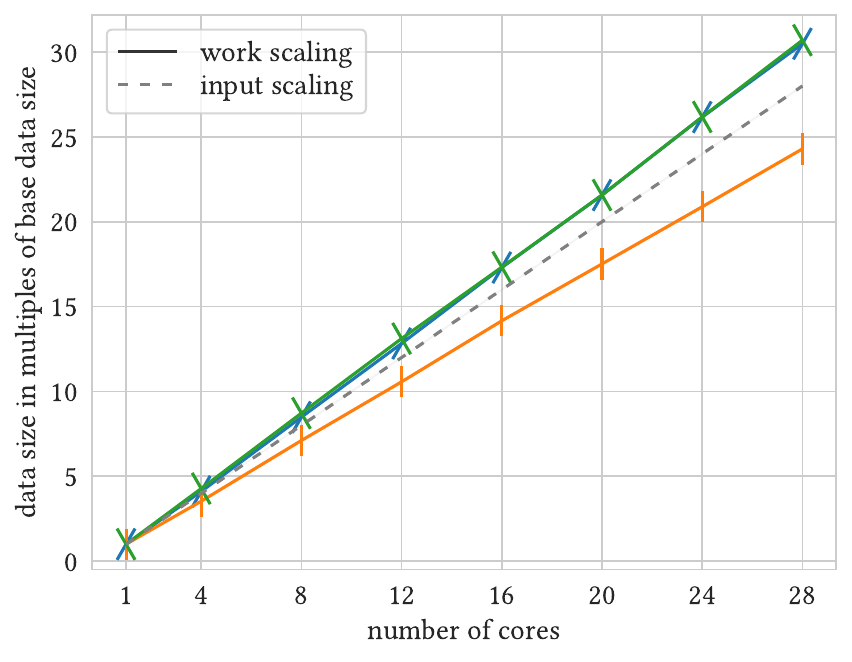}
            \vspace{-1cm}
            \caption{}\label{fig:mc_bones_sf}
        \end{subfigure}
        \begin{subfigure}[h]{0.33\textwidth}%
            \centering
            \includegraphics[width=\textwidth]{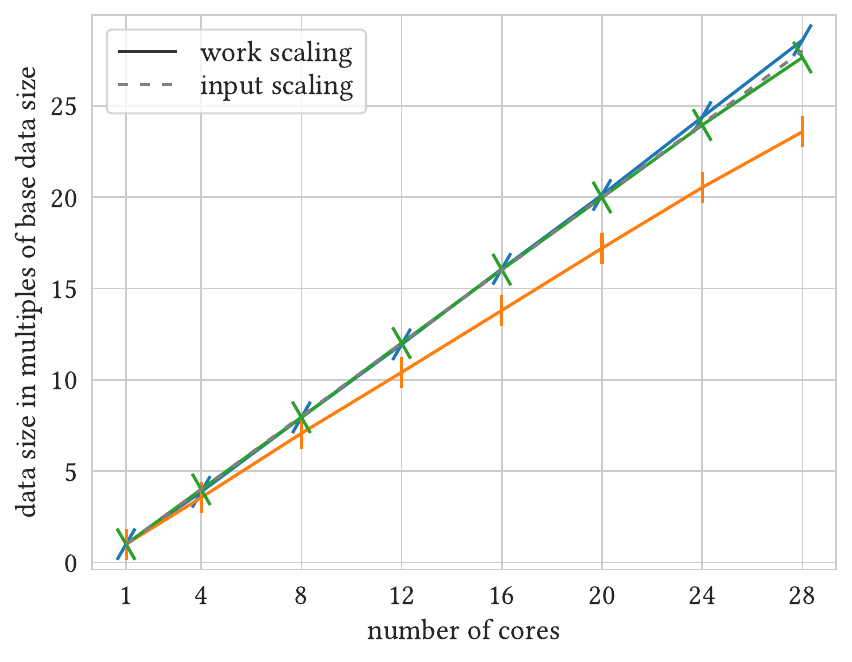}
            \vspace{-1cm}
            \caption{}\label{fig:mc_pn_sf}
        \end{subfigure}
        \begin{subfigure}[h]{0.33\textwidth}%
            \centering
            \includegraphics[width=\textwidth]{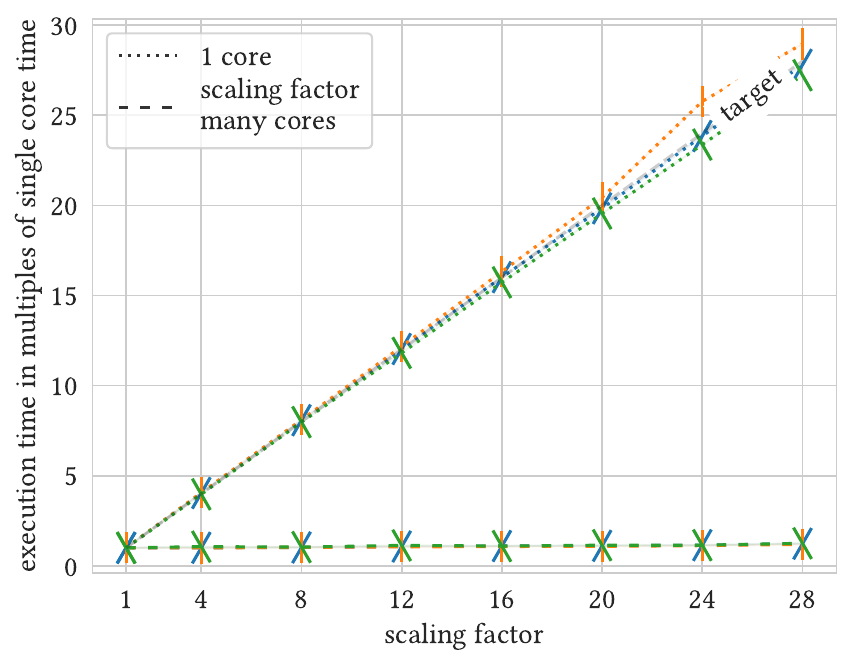}
            \vspace{-1cm}
            \caption{}\label{fig:mc_aneurism_input_scaling}
        \end{subfigure} 
        \begin{subfigure}[h]{0.33\textwidth}%
            \centering
            \includegraphics[width=\textwidth]{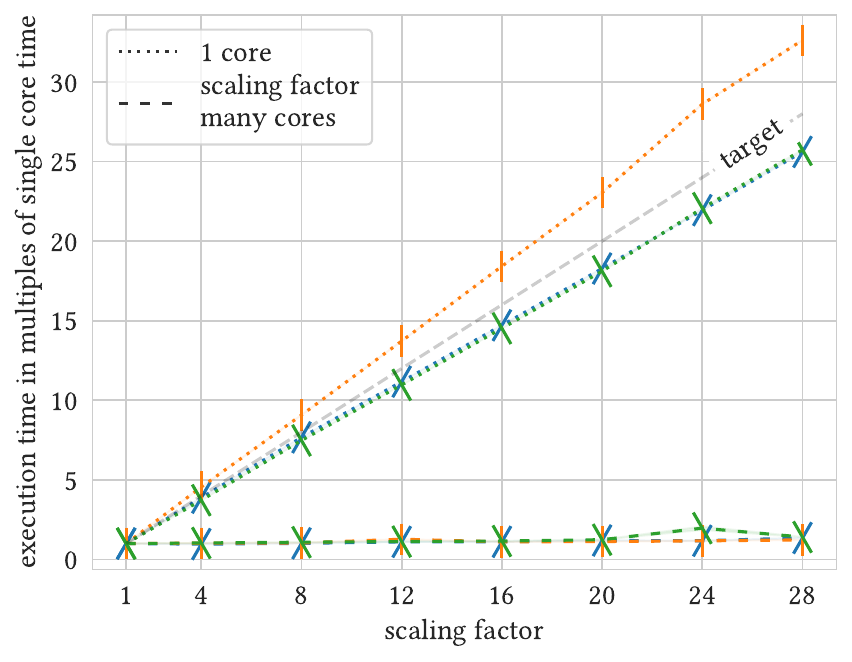}
            \vspace{-1cm}
            \caption{}\label{fig:mc_bones_input_scaling}
        \end{subfigure}
        \begin{subfigure}[h]{0.33\textwidth}%
            \centering
            \includegraphics[width=\textwidth]{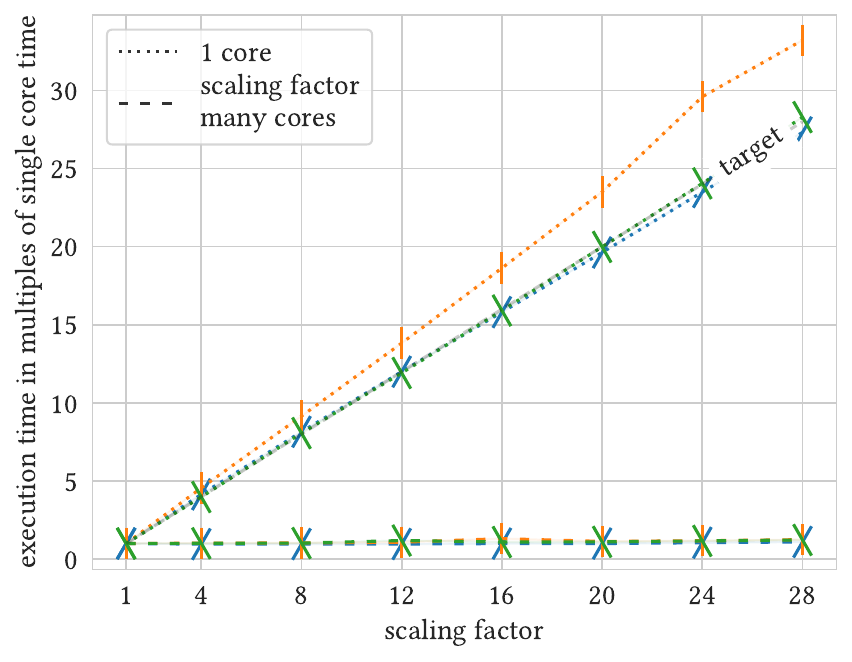}
            \vspace{-1cm}
            \caption{}\label{fig:mc_pn_input_scaling}
        \end{subfigure} 
        \caption{Plots showing \emph{work scaling} vs. \emph{input scaling} (top row), the corresponding scaling factor (middle row) and the work approximation of input-size scaling via execution time on a single core (bottom row) for iso-contour extraction via marching cubes. The columns correspond to the different data sets aneurysm, foot and Perlin noise from left to right. In contrast to the other results, marching cubes is parallelized by running one MPI rank per core.
        One can see that \emph{extent} and \emph{replication} exhibit irregular behavior in the form of execution time peaks. Presumably, this depends on the data segmentation for distributing the data across MPI ranks.
        Here \emph{Work scaling} executions are not always more consistent. However, the relative deviation in execution time for all methods is already low in comparison to other observed methods.
        }
        \label{fig:appendix_contour_marchingcubes}
    \end{figure*}

    \begin{figure*}
        \centering
        \begin{tikzpicture}[ampersand replacement=\&]
            \matrix [outer sep=0pt] {
                \node[]{data scaling method:}; \&
                \draw[color=resampling, very thick] (0.1,-0.1) -- (0.3,0.1);
                \draw[color=resampling, very thick] (0.0,0.0) -- (+0.4,0.0) node[right,black] {resampling}; \&
                \draw[color=replication, very thick] (0.2,-0.1) -- (0.2,0.1);
                \draw[color=replication, very thick] (0.0,0.0) -- (+0.4,0.0) node[right,black] {replication}; \&
                \draw[color=extent, very thick] (0.3,-0.1) -- (0.1,0.1);
                \draw[color=extent, very thick] (0.0,0.0) -- (+0.4,0.0) node[right,black] {extent}; \\
            };
        \end{tikzpicture}
        \vspace{0.5em}

        \begin{subfigure}[h]{0.33\textwidth}%
            \centering
            aneurysm
            \vspace{2pt}

            \includegraphics[width=\textwidth]{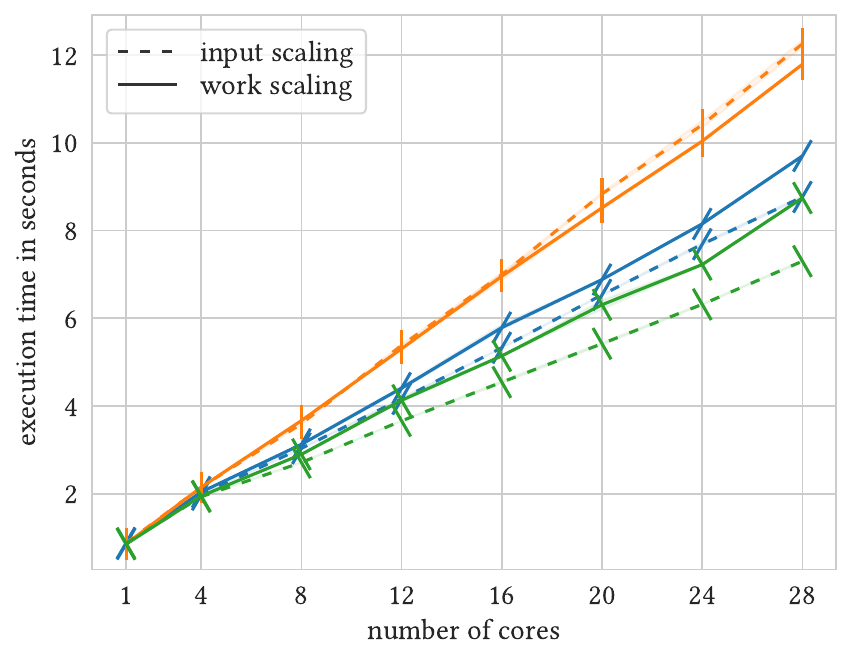}
            \vspace{-1cm}
            \caption{}\label{fig:ttkct_aneurism_work_vs_input}
        \end{subfigure}
        \begin{subfigure}[h]{0.33\textwidth}%
            \centering
            foot
            \vspace{2pt}

            \includegraphics[width=\textwidth]{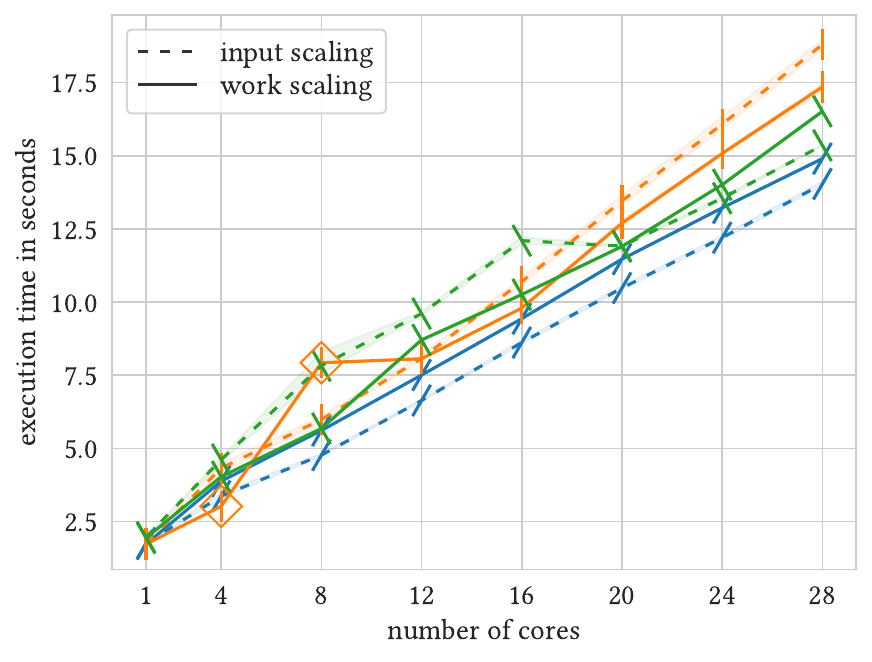}
            \vspace{-1cm}
            \caption{}\label{fig:ttkct_bones_work_vs_input}
        \end{subfigure}
        \begin{subfigure}[h]{0.33\textwidth}%
            \centering
            Perlin noise
            \vspace{2pt}

            \includegraphics[width=\textwidth]{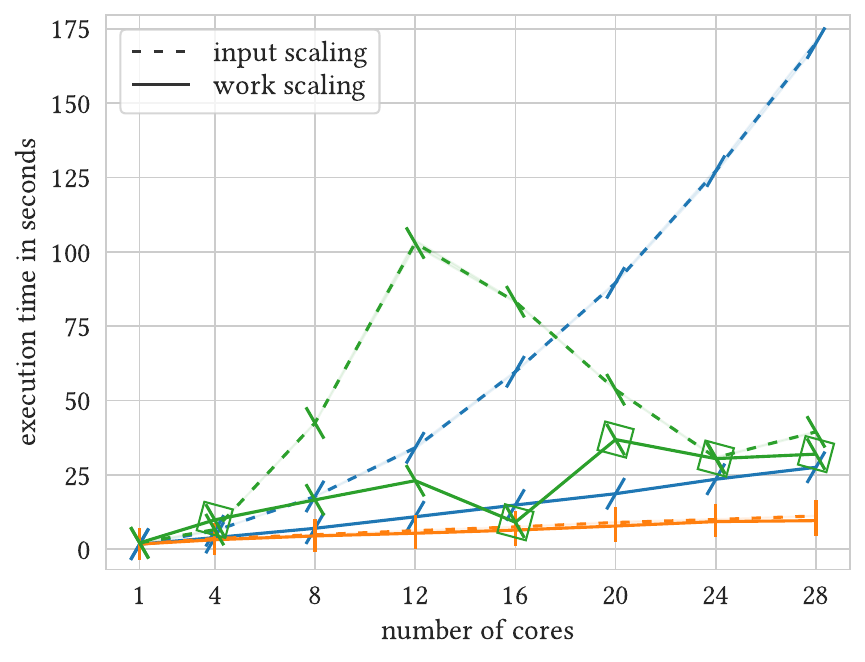}
            \vspace{-1cm}
            \caption{}\label{fig:ttkct_pn_work_vs_input}
        \end{subfigure}
        \begin{subfigure}[h]{0.33\textwidth}%
            \centering
            \includegraphics[width=\textwidth]{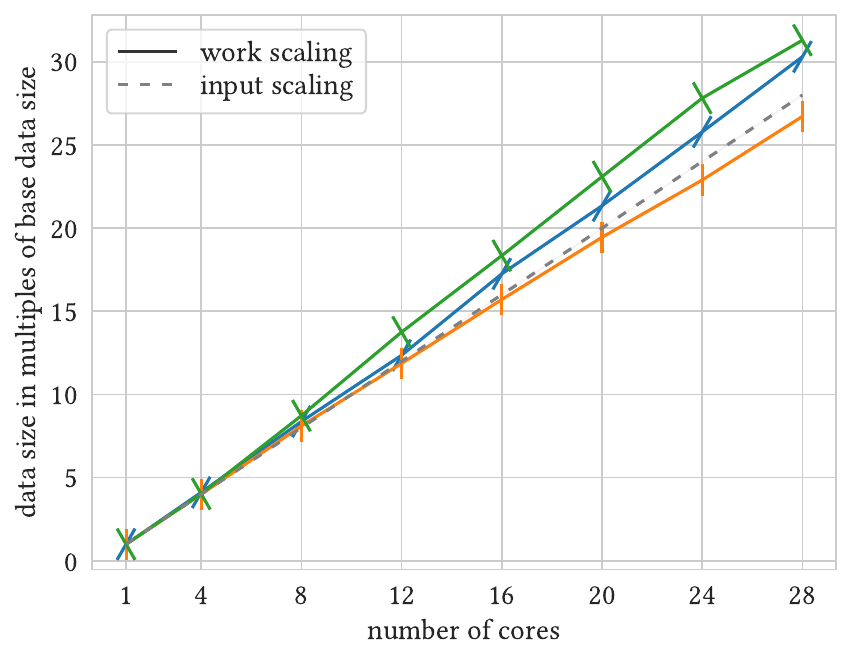}
            \vspace{-1cm}
            \caption{}\label{fig:ttkct_aneurism_sf}
        \end{subfigure}
        \begin{subfigure}[h]{0.33\textwidth}%
            \centering
            \includegraphics[width=\textwidth]{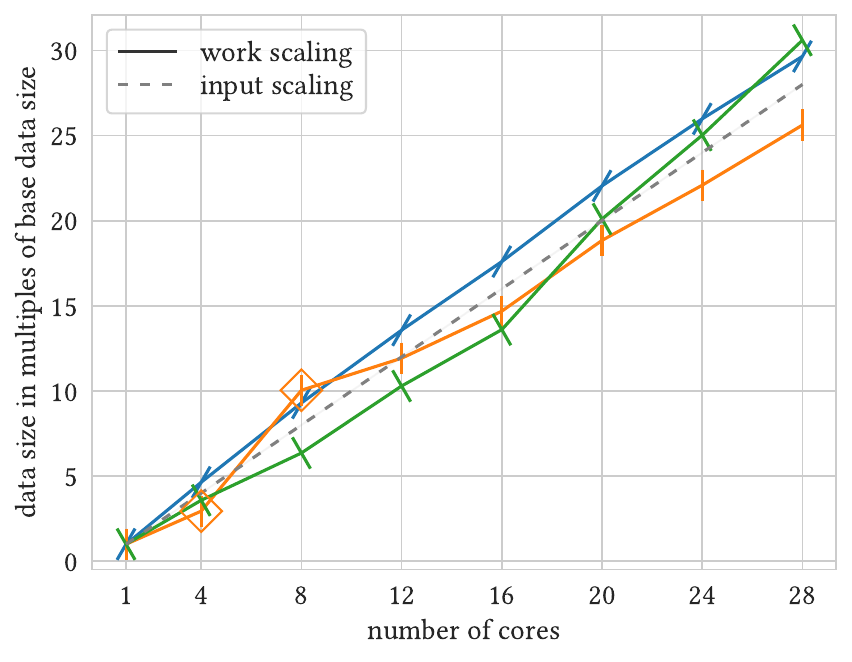}
            \vspace{-1cm}
            \caption{}\label{fig:ttkct_bones_sf}
        \end{subfigure}
        \begin{subfigure}[h]{0.33\textwidth}%
            \centering
            \includegraphics[width=\textwidth]{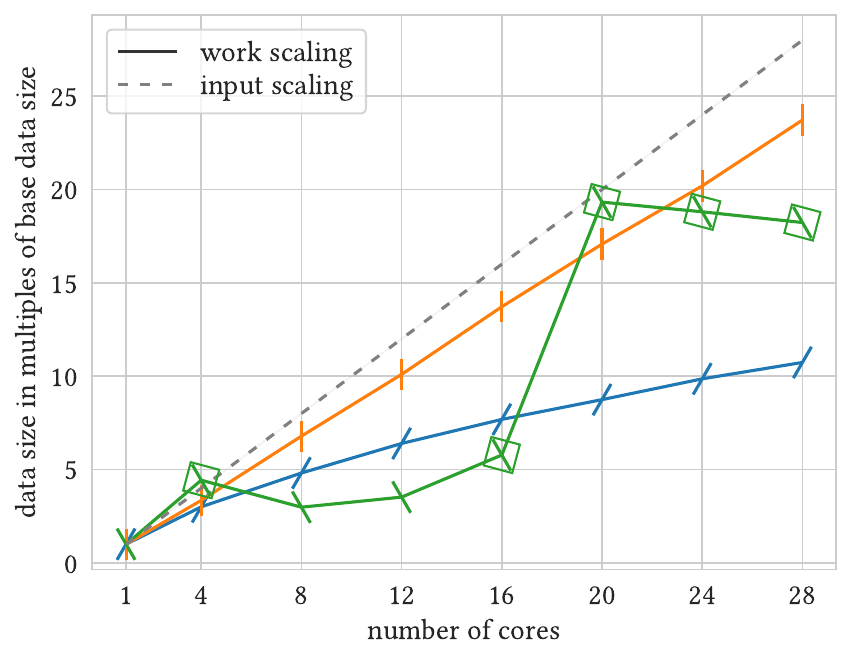}
            \vspace{-1cm}
            \caption{}\label{fig:ttkct_pn_sf}
        \end{subfigure}
        \begin{subfigure}[h]{0.33\textwidth}%
            \centering
            \includegraphics[width=\textwidth]{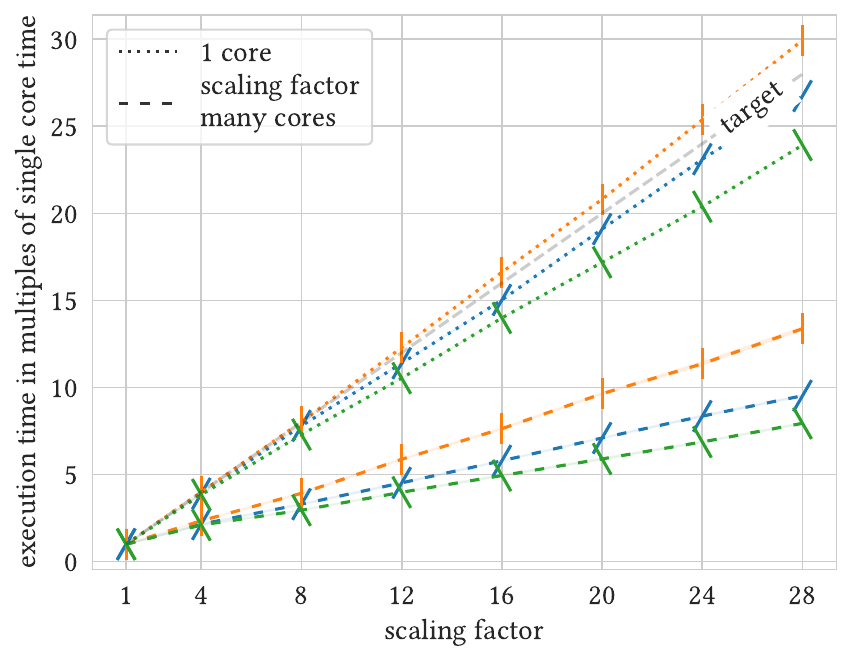}
            \vspace{-1cm}
            \caption{}\label{fig:ttkct_aneurism_input_scaling}
        \end{subfigure} 
        \begin{subfigure}[h]{0.33\textwidth}%
            \centering
            \includegraphics[width=\textwidth]{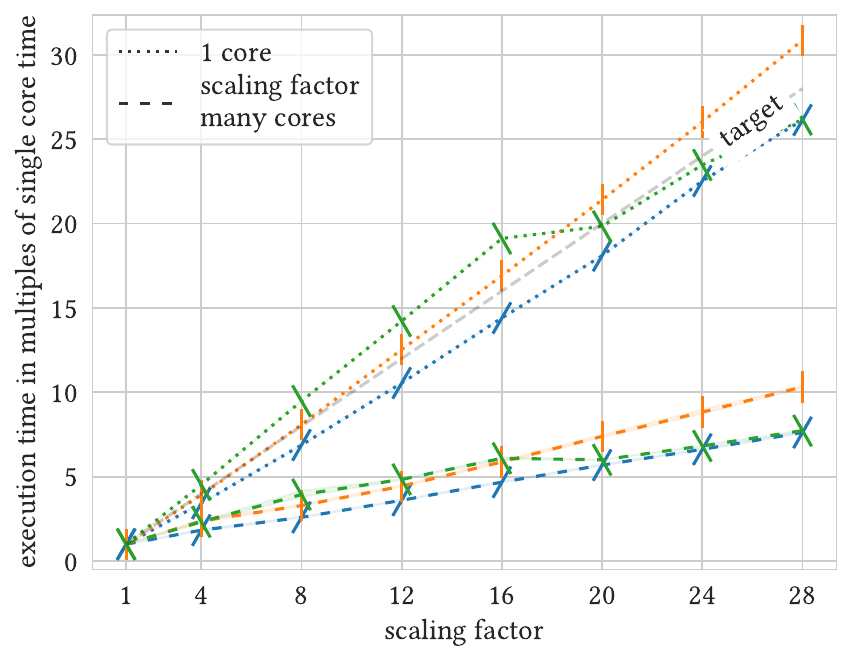}
            \vspace{-1cm}
            \caption{}\label{fig:ttkct_bones_input_scaling}
        \end{subfigure}
        \begin{subfigure}[h]{0.33\textwidth}%
            \centering
            \includegraphics[width=\textwidth]{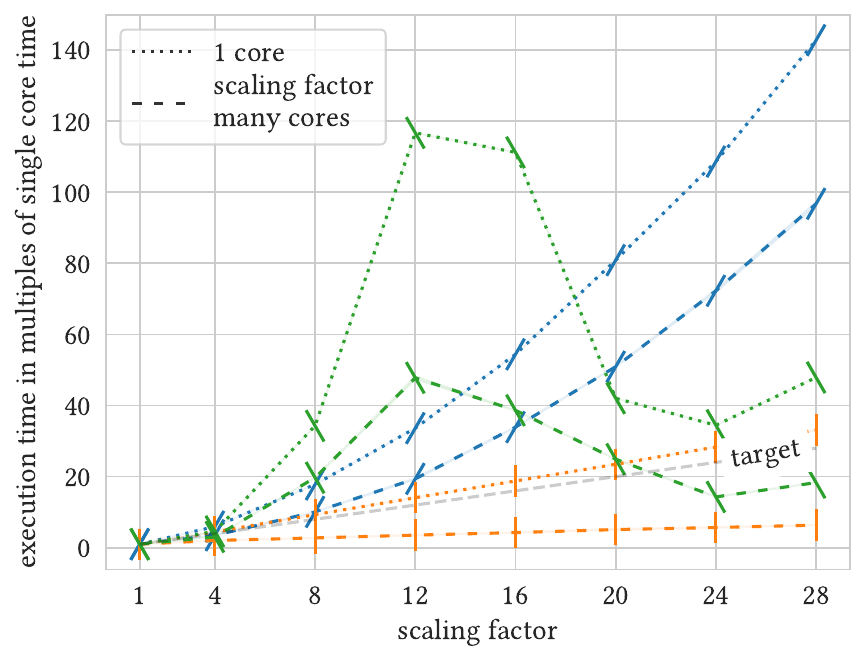}
            \vspace{-1cm}
            \caption{}\label{fig:ttkct_pn_input_scaling}
        \end{subfigure} 
        \caption{Plots showing \emph{work scaling} vs. \emph{input scaling} (top row), the corresponding scaling factor (middle row) and the work approximation of input-size scaling via execution time on a single core (bottom row) for augmented contour tree computation with the FTM algorithm by \cite{gueunet2017} in TTK. The columns correspond to the different data sets aneurysm, foot and Perlin noise from left to right.
        For the foot and aneurysm data set we observe only small deviations in the introduced work with \emph{input scaling} from the ideal one-to-one ratio of work to cores. The \emph{work scaled} weak scaling runs are only slightly more consistent.
        Larger differences between scaling methods can be observed in the Perlin noise data set. The execution time for \emph{extent} scaling peaks at $12$ cores and then declines while \emph{resampling} leads to a super-linear increase in execution time.
        The FTM algorithm subdivides the data for parallel processing with respect to the field on which the contour tree is computed. Thus, stitching of the local trees and potential load imbalances highly depend on the composition of the data set, which was also identified by \cite{gueunet2017}.
        }
        \label{fig:appendix_contourtree_ttk}
    \end{figure*}

\end{document}